\documentclass[useAMS,usenatbib,onecolumn,showpacs,superscriptaddress,floatfix,nofootinbib,noeprint,prd]{revtex4-2}

\usepackage{graphicx}
\usepackage{epsf}
\usepackage{epsfig}
\usepackage{amssymb,amsmath}
\usepackage[usenames]{color}
\usepackage{amssymb}
\usepackage{times}
\usepackage{mathrsfs}
\usepackage{hyperref}
\hypersetup{
  colorlinks=true,        
  linkcolor=blue,         
  citecolor=cyan,         
}
\usepackage{float}

\usepackage[T1]{fontenc}
\usepackage[utf8]{inputenc}
\usepackage{babel}
\usepackage[font=small,labelfont=bf]{caption}

\newcommand{\ie}{{i.e.,}~}

\renewcommand{\u}{\boldsymbol{u}}
\newcommand{\halb}{\frac{1}{2}}

\renewcommand{\d}{\mathbf{D}}

\renewcommand{\u}{\boldsymbol{u}}
\newcommand{\w}{\boldsymbol{w}}

\newcommand{\F}{\boldsymbol{F}}
\renewcommand{\S}{\mathbf{S}}
\newcommand{\f}{\boldsymbol{f}}

\newcommand{\A}{\boldsymbol{A}}
\newcommand{\B}{\boldsymbol{B}}

\newcommand{\upp}{}

\allowdisplaybreaks

\begin{document}

\title[A strongly hyperbolic first--order BSSNOK system]{A monolithic first--order BSSNOK formulation of the Einstein--Euler equations and its solution with path-conservative finite difference CWENO schemes}

\date{\today}
\label{firstpage}

\author{Michael Dumbser}
\affiliation{
Laboratory of Applied Mathematics, University of Trento,
Via Mesiano 77, 38123 Trento, Italy
}

\author{Olindo Zanotti}
\affiliation{
Laboratory of Applied Mathematics, University of Trento,
Via Mesiano 77, 38123 Trento, Italy
}

\author{Gabriella Puppo}
\affiliation{
Dipartimento di Matematica, University of Roma La Sapienza,
Piazza Aldo Moro 5, 00185 Roma,
Italy 
}

\begin{abstract}
We present a new, monolithic first--order (both in time and space) BSSNOK formulation of the coupled Einstein--Euler equations. The entire system of hyperbolic PDEs is solved in a completely unified manner via one single numerical scheme applied to both the 
conservative sector of the matter part
and to the first--order strictly non--conservative sector of the spacetime evolution. The coupling between matter and space-time is achieved via algebraic source terms.  
The numerical scheme used for the solution of the new monolithic first order formulation is a path-conservative central WENO (CWENO) finite difference scheme,
with suitable insertions to account for the presence of the non--conservative terms. 
By solving several crucial tests of numerical general relativity, including a stable neutron star, Riemann problems in relativistic matter with shock waves and the stable long-time evolution of single and binary puncture black holes up and beyond the binary merger, 
we show that our new CWENO scheme,
introduced two decades ago for the compressible Euler equations of gas dynamics, 
can be successfully applied also to numerical general relativity, solving all equations at the same time with one single numerical method. In the future the new monolithic approach proposed in this paper may become an attractive alternative to traditional methods that couple central finite difference schemes with Kreiss-Oliger dissipation for the space-time part with totally different TVD schemes for the matter evolution and which are currently the state of the art in the field.
\end{abstract}

\maketitle


\section{Introduction}


Since the detection of GW150914~\cite{Abbott2016}, Gravitational Waves (GWs) astronomy has experienced a fundamental boost~\cite{Bailes2021}, with 93 sources detected up to March 2024\footnote{\url{https://gwosc.org/}}. 
The accurate numerical modeling of gravitational waveforms
has become a key ingredient to extract signals from detector noises and to understand the properties of astrophysical
sources, specifically binary systems \cite{Jimenez2017,Abbott2019}.
In this respect, numerical general relativity has been providing an invaluable contribution to the scientific progress (see, among the others,~\cite{EventHorizon2022a,EventHorizon2022b,Aranguren2023,Camilletti2022,Palenzuela2022,Camilletti2024,Breschi2024,Topolski2024}).

The two main pillars of any successful numerical simulation are represented by an appropriate mathematical formulation of the equations that need to be solved
and by the availability of an accurate and robust numerical scheme.
 
Concerning the mathematical formulation of the Einstein field equations, 
although many of them have been proposed over the years, both as first--order or second--order PDEs systems, 	
two of them are the most adopted in numerical relativity codes today:	 $i)$
the generalized harmonic formulation (GH), which is usually written as a first order (in space derivatives) PDE system~\cite{Lindblom2006}, with very important scientific results, including the first stable numerical simulation
of a binary black hole merger presented in~\cite{Pretorius2005a}, and $ii)$ the
second--order (in space derivatives) BSSNOK formulation (with its variants represented by CCZ4/Z4c) within the 3+1 splitting of spacetime~\cite{Baumgarte99,Alcubierre:2008,Baumgarte2010}. 
In this article we are concerned only with the latter approach.
The scientific research on  first--order formulations of the BSSNOK system has been somewhat left aside during the last decade or more, with only a few exceptions in relatively recent years~\citep{Bona-and-Palenzuela-Luque-2005:numrel-book,Brown2012,Dumbser2017strongly,Dumbser2020GLM}.  

Concerning the numerical scheme, on the other hand, it is fair to say that,
in spite of considerable efforts carried out in the last decade 
for adopting innovative numerical schemes in the solution of the Einstein equations, 
most  of the existing numerical relativity codes still rely on rather traditional central finite differencing with artificial dissipation of the Kreiss-Oliger type, in combination with Runge--Kutta time integration.
Of course there are exceptions to this trend. For example, there is 
a active field of research to extend discontinuous Galerkin (DG) schemes to 
full general relativity~\cite{Teukolsky2015,Miller2017,Kidder2017,Dumbser2017strongly,Hebert2018,Tichy2021,Tichy2023,DumbserZanottiGaburroPeshkov2023,DumbserZanottiPeshkov2024,Deppe2024};
or to develop spectral methods both for the initial value problem and for the time evolution of the Einstein equations~\cite{Boyle2007,Boyle2007b,Duez2008,Scheel2009,Grandclement2009,Buchman2012,Szilagyi2014};
or to go beyond Runge--Kutta schemes for the time integration by migrating to so--called 
ADER schemes~\cite{Zanotti2015,Zanotti2015d,Zanotti2016,ADERGRMHD,Dumbser2018conformal,Gaburro2021PNPMLimiter}.
In spite of these progresses, if we list the most popular numerical relativity codes that perform actual calculations of waveforms from binary mergers
(Einstein--Toolkit\footnote{\url{http://einsteintoolkit.org/}}~\cite{Loffler2012}, 
LazEv\footnote{\url{https://ccrg.rit.edu/content/software/lazev}}~\cite{Zlochower2005,Lousto2023}, BAM\footnote{\url{http://data.cardiffgravity.org/bam-catalogue/}}~\cite{Bruegmann2008,Husa_2008, Thierfelder2011},
GRChombo\footnote{\url{https://www.grchombo.org/}}~\cite{Clough_2015,Andrade2021},
AMReX\footnote{\url{https://amrex-codes.github.io/amrex/}}~\cite{Peterson_2023}, Nmesh~\cite{Tichy2023}, SACRA\footnote{\url{https://www.aei.mpg.de/1095648/SACRA-description}}\cite{Yamamoto2009,Kiuchi2017}, 
SpEC\footnote{\url{https://www.black-holes.org/code/SpEC.html}}~\cite{Kidder2000,Duez2008,Haas2016},
SPHINCS\_BSSN~\cite{Rosswog2021},
SENR/NRPy~\cite{Ruchlin2018}, MHDueT\footnote{\url{http://mhduet.liu.edu/}}\cite{Palenzuela2018,Palenzuela2021}),
GR-Dendro\footnote{\url{https://paralab.github.io/Dendro-GR/}}~\cite{Milinda2019,Milinda2023})
we observe that, apart from Nmesh, SpEC and GR-Dendro, all of them still use finite difference schemes for the Einstein sector of the spacetime evolution.

Now, if we combine our considerations about the formulation of the Einstein equations, which rests
on the second--order BSSNOK system, and the choice of
a suitable numerical scheme, which rests mainly on traditional finite differencing, 
we see that the lack of a reliable first--order formulation 
represents a serious obstacle for the application of the whole class of advanced numerical schemes
of Godunov-type
proposed in the last three decades (or more) for the solution of nonlinear first--order hyperbolic PDE systems. 
As an additional complication, when the 
Einstein equations are cast in first--order form, they naturally come up with non--conservative terms and curl-type involutions,
which may give the (erroneous) impression of being untractable by Godunov--type methods\footnote{For an alternative mixed  elliptic-hyperbolic formulation see also \cite{Cordero-Carrion2008,Cordero-Carrion2010}.}.
On the contrary, since
all the characteristic fields arising from the Einstein equations are linearly degenerate, 
discontinuities cannot develop, and therefore non--conservative terms in the PDE system can be taken into account
in a  relatively easy way. 

In light of the above comments, 
assuming  that finite difference numerical schemes are likely to remain the preferred choice by
the NR community for a certain amount of time onward, 
due to their simplicity,
in this paper we resume a central WENO (CWENO) finite difference numerical scheme,
proposed in a different context originally by \cite{Levy1999,Levy2000,Levy2001}, that may contribute
to a substantial improvement with respect to the present standard in a few specific aspects:
\begin{itemize}
\item It relies on a hyperbolic \emph{first--order} but \emph{non--conservative} version  of the BSSNOK formulation.
\item It applies unmodified both to the spacetime (Einstein) and to the matter (Euler) evolution, achieving high order of accuracy in both space and time.
\end{itemize}
Our investigation follows some very recent progress made in  \cite{Balsara2023,Balsara2024b,Balsara2024c},
who have opened a rather  innovative line of research, by 
introducing a new class of finite difference WENO schemes 
for first--order non--conservative systems. Because such schemes are quite general
and not limited to NR, we address the interested reader to \cite{shuosher1,shuosher2,shu_efficient_weno,Levy2000,Levy2001,WENOZ,cravero2018cweno,AOWENO1,AOWENO2,AOWENO3}, where all possible information
about these new methods can be found.


The structure of the paper is as follows. In Section~\ref{sec:second} we present the new first order version of the BSSNOK formulation of the coupled Einstein-Euler equations.
In Section~\ref{sec:cweno} we describe the new numerical
scheme that can be used in a monolithic way for the full system of the Einstein--Euler PDEs. 
In Section~\ref{sec:tests} we show a number of benchmark results
to demonstrate the correctness of both the formulation and the numerical
solver. Finally, the conclusions are summarized in Section
\ref{sec:conclusions}.

We work in a geometrized set of units, in which the speed of light and
the gravitational constant are set to unity, \ie $c=G=1$. Greek indices
run from $0$ to $3$, Latin indices run from $1$ to $3$ and we use the
Einstein summation convention of repeated indices.

\section{The BSSNOK system}
\label{sec:second}

\subsection{The standard second order formulation}
Many numerical codes solving the Einstein--Euler equations within the $3+1$ formalism adopt the so--called second order 
BSSNOK formulation, which, due to its popularity, does not require an extended discussion and it can be briefly described as follows
(for more details see ~\cite{Alcubierre:2008,Baumgarte2010,Gourgoulhon2012,Rezzolla_book:2013}). 
The spatial metric $\gamma_{ij}$ of the spacelike hypersurfaces is rescaled conformally according to
\begin{equation}
\tilde\gamma_{ij}=\psi^{-4}\gamma_{ij}=e^{-4\phi}\gamma_{ij}\,,
\end{equation}
where $\tilde\gamma_{ij}$ has unit determinant, such that $\psi=\gamma^{1/12}$, where $\gamma$ denotes as usual the determinant of $\gamma_{ij}$.
Hence, the spatial metric $\gamma_{ij}$, with only 6 independent components,
generates in fact 7 evolved quantities, namely the factor $\phi$, related to the determinant of  $\gamma_{ij}$ as $\phi=\ln\psi=\frac{1}{12}\ln\gamma$, and the 
rescaled unit determinant metric $\tilde\gamma_{ij}$.
Similarly, the symmetric extrinsic curvature $K_{ij}$, with only 6 independent components,
generates in fact 7 evolved quantities, namely the trace $K$ of  $K_{ij}$, and the rescaled trace-free tensor
\begin{equation}
\tilde{A}_{ij}=\psi^{-4}A_{ij}=e^{-4\phi}(K_{ij}-\frac{1}{3}\gamma_{ij}K)\,.
\end{equation}
Overall, the (redundant) 17 evolved quantities within the BSSNOK framework obey the following system of PDEs, which is of first order in time and of mixed first and second order in space:  
\newpage 
\begin{eqnarray}
\label{dtgamma}
  (\partial_t -{\cal L}_{\beta})~ \tilde\gamma_{ij}
  &=& - {2\,\alpha}\,\tilde A_{ij}\,,
\\
\label{dtphi}
  (\partial_t -{\cal L}_{\beta})~ \phi
  &=& - \frac{1}{6}\alpha\,K\,,
\\
\label{dtKij}
   (\partial_t - {\cal L}_{\beta})~\tilde A_{ij} &=& e^{-4\phi}\bigg[-\upp\nabla_i\upp\nabla_j\alpha
    + \alpha\,   R_{ij} 
    - 8\pi \alpha S_{ij}\,\bigg]^{TF}
+ \alpha \left(   K\,\tilde A_{ij}   - 2\tilde A_{im}\tilde A^m_j  \right)	\,,	
\\
\label{dtK}
  (\partial_t -{\cal L}_{\beta})~ K
  &=& -\upp\nabla_i\upp\nabla^i\alpha + \alpha\left( \tilde A_{ij}\tilde A^{ij} + \frac{1}{3}K^2  +4\pi(E+S)  \right)\,,
	\\
\label{dtGammai}
  (\partial_t -{\cal L}_{\beta})~ \tilde\Gamma^i&=&\tilde\gamma^{jk}\partial_j\partial_k \beta^i + \frac{1}{3}\tilde\gamma^{ij}\partial_j\partial_k\beta^k
	-2\tilde A^{ij}\partial_j \alpha  \\
	\nonumber
	&+&2\alpha\left(\tilde\Gamma^i_{jk}\tilde A^{jk} + 6\tilde A^{ij}\partial_j\phi-\frac{2}{3}\tilde\gamma^{ij}\partial_j K  -8\pi e^{4\phi} M^i \right)
\end{eqnarray}
where, according to the standard notation of the $3+1$ formalism,
$\alpha$ is the lapse, $\beta^i$ is the shift, $R_{ij}$ is the purely spatial Ricci tensor,
$S_{ij}$ is the purely spatial part of the energy momentum tensor,  $S$ is its trace and the superscript $TF$ stands for the trace-free part of a tensor. Moreover, ${\cal L}_{\beta}$ denotes the Lie derivative
along the shift vector.
We also recall that Eq.~\eqref{dtKij} and \eqref{dtK} incorporate the Hamiltonian constraint, while Eq.~\eqref{dtGammai} incorporates
the momentum constraint. In terms of the BSSNOK variables these constraints can be written as
\begin{eqnarray}
\label{eqn.adm1}
H &=& R - \tilde A_{ij} \tilde A^{ij} + \frac{2}{3}K^2 - 16\pi E = 0\,, \\
\label{eqn.adm2}
M^i &=& \partial_j\tilde A^{ij}+\tilde\Gamma^i_{jk}\tilde A^{jk}+6\tilde A^{ij}\partial_j\phi-\frac{2}{3}\tilde\gamma^{ij}\partial_j K  - 8 \pi S^i = 0\,,
\end{eqnarray}
and these quantities will be monitored during a numerical simulation. In particular, we will 
check the normalized $L^2$ errors of the above constraints, computed for a generic quantity $q$ as
\begin{equation}
	\langle L_q^2 \rangle=\sqrt{\frac{\int_\Omega\, q^2\,d^3x}{\int_\Omega\,d^3x}}\,\hspace{1cm}\textrm{with}\hspace{1cm} q \in \left\{ H, \, M^i \right\}\,.
\end{equation}
%

\subsection{The new first--order version}
\label{sec:first}

A promising first--order formulation of BSSNOK was proposed by \cite{Brown2012}, though it seems
that second--order BSSNOK formulations are still the most popular choice
in large scale simulations of astrophysical sources.
Here, as already done in the first--order CCZ4 formulation of 
\cite{Dumbser2017strongly}, we introduce 30 auxiliary variables containing first derivatives of the metric terms, namely
\begin{align}
\label{eq:Auxiliary}
A_k := \partial_k\ln\alpha = \frac{\partial_k \alpha }{\alpha}\,, \qquad
B_k^{\,\,i} := \partial_k\beta^i\,, \qquad
D_{kij} := \frac{1}{2}\partial_k\tilde\gamma_{ij}\,. \qquad
P_k := \frac{\partial_k \psi}{\psi} = \partial_k \phi = \frac{1}{12}\partial_k\ln \gamma
\end{align}
Since the auxiliary variables are gradients of primary evolution quantities, they must remain curl-free for all times if they are initially curl-free (this is obvious either via the Schwarz theorem on the symmetry of second derivatives, or alternatively because the curl of a gradient is zero). Hence,  
we have the following second-order ordering constraints induced by \eqref{eq:Auxiliary}: 
\begin{equation}
	\label{eq:curlfree}
	\mathcal{A}_{lk}   = \partial_l A_k - \partial_k A_l = 0, \quad
	\mathcal{B}_{lk}^i = \partial_l B_k^{\,\,i} - \partial_k B_l^{\,\,i} = 0, \quad
	\mathcal{D}_{lkij} = \partial_l D_{kij} - \partial_k D_{lij} = 0, \quad
	\mathcal{P}_{lk}   = \partial_l P_k - \partial_k P_l = 0. 
\end{equation}
Since $\tilde{A}_{ij}$ is trace-free one has $\tilde{\gamma}^{ij} \tilde{A}_{ij} = 0$ and therefore another differential constraint 
\begin{equation}
 \mathcal{C}_k =  
 \partial_k \tilde{\gamma}^{ij}  \tilde{A}_{ij} + 
  \tilde{\gamma}^{ij} \partial_k \tilde{A}_{ij} = 0
\end{equation}
that can be used in the first order version of BSSNOK. 
In this paper we have found it not necessary to apply specific techniques to ensure a curl-free evolution, as done, for instance, in \cite{Dumbser2020GLM} for the first order CCZ4 equations. On the contrary, we have found it crucial, for strong hyperbolicity, to insert the curl-free term proportional to $\mu$ 	on the left hand side of Eq.~\eqref{eqn.B}. This is also discussed in the list of comments below Eq.~\eqref{eqn.nablanablaalpha}. 
With the aid of \eqref{eq:Auxiliary}, the system~\eqref{dtgamma}--\eqref{dtGammai}, 
augmented by the matter part and by the gauge conditions, 
can be written as a monolithic first--order BSSNOK formulation of the Einstein-Euler system, \ie

\clearpage 
\begin{align}
\label{eqn.rho}
&\partial_t (\sqrt{\gamma}D)+\partial_i\left[\sqrt{\gamma}(\alpha v^i D - \beta^i D)\right]=0\,,\\ 
\label{eqn.S}
&\partial_t (\sqrt{\gamma}S_j)+\partial_i\left[\sqrt{\gamma}(\alpha S^i_{\,\,j} - \beta^i S_j)\right]=\sqrt{\gamma}\left[\alpha S^{ik}D_{jik}  +
S_i  B_j^{\,\,i} - \alpha E A_j   \right]\,,\\ 
\label{eqn.E}
&\partial_t (\sqrt{\gamma}E)+\partial_i\left[\sqrt{\gamma}(\alpha S^i - \beta^i E)\right]=\sqrt{\gamma}\left[\alpha S^{ij}e^{4\phi}\left(\tilde A_{ij} + \frac{1}{3}\tilde\gamma_{ij}K\right) - \alpha S^j A_j  \right]\,,\\ 
\label{eqn.gamma}
&\partial_t\tilde\gamma_{ij} - \beta^k\partial_k\tilde\gamma_{ij}=\tilde\gamma_{ik} B_{j}^{\,\,k}  + \tilde\gamma_{kj} B_{i}^{\,\,k} -\frac{2}{3}\tilde\gamma_{ij}
B_{k}^{\,\,k} - 2\alpha \tilde A_{ij}\,,\\
\label{eqn.phi}
&\partial_t \phi  - \beta^k\partial_k \phi= \frac{1}{6}B_{k}^{\,\,k}  -\frac{1}{6}\alpha K  \,,\\
\label{eqn.Kij}
&\partial_t \tilde A_{ij} - \beta^k \partial_k \tilde A_{ij}  + \alpha e^{-4\phi} \left( \partial_{(i} {A}_{j)} -\frac{1}{3}\tilde\gamma_{ij}\tilde\gamma^{mn}\partial_{(m} {A}_{n)}\right)
-\alpha e^{-4\phi} \bigg[ (R_{ij})^{TF}_{ncp}      \bigg] = \nonumber \\
&  \qquad \qquad 
\tilde A_{ik} B_j^{\,\,k} + \tilde A_{kj} B_i^{\,\,k} -\frac{2}{3}\tilde A_{ij}B_k^{\,\,k} 
- \alpha e^{-4\phi} \left[ A_i A_j -  \Gamma^k_{ij} A_k -\frac{1}{3}\tilde\gamma_{ij}\tilde\gamma^{mn}(A_m A_n -  \Gamma^k_{mn} A_k)  \right] +\nonumber\\
&  \qquad \qquad 
+\alpha e^{-4\phi} \bigg[ (R_{ij})^{TF}_{src}      \bigg] 
- 8\pi\alpha e^{-4\phi} \left(S_{ij}-\frac{1}{3}e^{4\phi}\tilde\gamma_{ij}\,S\right)  
 +  \alpha ( K \tilde A_{ij} - 2  \tilde A_{il} \tilde\gamma^{lm} \tilde A_{mj} ) \,, \\
\label{eqn.K}
&\partial_t K - \beta^k \partial_k K + \alpha e^{-4\phi}\tilde\gamma^{ij}\partial_{(i} A_{j)} 
= -\alpha e^{-4\phi}\tilde\gamma^{ij}  \left( A_i A_j  -\Gamma^k_{ij} A_k \right) + \alpha \left(\tilde A_{ij}\tilde A^{ij} + \frac{1}{3}K^2 + 4\pi(E+S)  \right)\,,\\
\label{eqn.slicing}
&\partial_t \ln\alpha - \beta^k{\partial_k\ln\alpha} = -g(\alpha)\alpha(K-K_0)\,,\\
\label{g-driver1}
&\partial_t \beta^i - b\, s\,\beta^k\partial_k \beta^i  = \frac{3}{4}\,s\,b^i, \\
\label{g-driver2}
&\partial_t b^i - b\, s\,(\beta^k\partial_k b^i - \beta^k\partial_k \tilde\Gamma^i) = s(\partial_t \tilde\Gamma^i - \eta b^i)\,,\\
\label{eqn.Gammai}
&  \partial_t\tilde\Gamma^i -s\,\bigg[ \beta^k\partial_k \tilde\Gamma^i
+\tilde\gamma^{jk}\partial_{(j} B_{k)}^{\,\,i}
 + \frac{1}{3}\tilde\gamma^{ij}\partial_{(j} B_{k)}^{\,\,k}  
-\frac{4}{3}\alpha\tilde\gamma^{ij}\partial_j K 
 \bigg]
 =\nonumber\\ 
& \qquad \qquad s\,\bigg[\frac{2}{3}\tilde\Gamma^i B_k^{\,\,k} - \tilde\Gamma^k B_k^{\,\,i}
	-2 \alpha \tilde A^{ij} A_j   + 
	2\alpha\left(\tilde\Gamma^i_{jk}\tilde A^{jk} + 6\tilde A^{ij}P_j  -8\pi e^{4\phi} M^i \right)\bigg]\,,\\
\label{eqn.A}
& \partial_t A_{i} - {\beta^k \partial_k A_i} + \alpha g(\alpha) \left( \partial_i K - \partial_i K_0  \right)
=-\alpha A_i \left( K - K_0  \right) \left( g(\alpha) + \alpha g^\prime(\alpha)  \right)
  +   B_i^{\,\,k} ~A_{k} 
\\
\label{eqn.B}
& \partial_t B_k^{\,\,i}  - s\left[  \frac{3}{4} \partial_k b^i  
+ b\left(\beta^m\partial_m B_k^{\,\,i}  \right)
- 
\mu\, \alpha^2  \gamma^{ij} \gamma^{nl} \left( \partial_k D_{ljn} - \partial_l D_{kjn} \right)  
\right]
=		s\,b\, B_m^{\,\,i} B_k^{\,\,m} \,,
\\
\label{eqn.D}
&
\partial_t D_{kij} - {\beta^m \partial_m D_{kij}} 
         - \frac{1}{2} {\tilde\gamma}_{mi} \partial_{(k} {B}_{j)}^{\,\,m}
         - \frac{1}{2} {\tilde\gamma}_{mj} \partial_{(k} {B}_{i)}^{\,\,m}
         + \frac{1}{3} {\tilde\gamma}_{ij} \partial_{(k} {B}_{m)}^{\,\,m}
				 +  \alpha \partial_k {\tilde A}_{ij} = B_k^{\,\,m} D_{mij} + B_j^{\,\,m} D_{kmi} + B_i^{\,\,m} D_{kmj}  \nonumber \\ 
&	\qquad \qquad 	-\frac{2}{3}B_m^{\,\,m} D_{kij} - \alpha A_k  {\tilde A}_{ij}
+
\frac{1}{3}\alpha \tilde\gamma_{ij}\left[ \tilde\gamma^{nm} \partial_k \tilde A_{nm} + \tilde A_{nm} \partial_k \tilde\gamma^{nm} \right]
\,,\\
\label{eqn.P}
&
\partial_t P_i - \beta^k\partial_k P_i+\frac{1}{6}\alpha\partial_i K-\frac{1}{6}\partial_{(i} B_{k)}^{\,\,k}=P_k B_i^{\,\,k} - \frac{1}{6}\alpha K A_i 
\end{align}

We also  list a few expressions and identities that are useful when writing the above PDE system in non--conservative form:
\begin{eqnarray}
{\gamma} &=& \textnormal{det}( {\gamma}_{ij} )=e^{12\phi}\,, \\
\partial_k {\tilde\gamma}^{ij} & = &  - 2 {\tilde\gamma}^{in} {\tilde\gamma}^{mj} D_{knm}\,, \\
%
\tilde\Gamma_{ij}^k &=&  {\tilde\gamma}^{kl} \left( D_{ijl} + D_{jil} - D_{lij} \right)\,, \\
\Gamma_{ij}^k &=& \tilde\Gamma_{ij}^k + 2( \delta^k_i P_j + \delta^k_j P_i ) - 2 \tilde\gamma^{km}P_m \tilde\gamma_{ij} \,, \\
{\tilde\Gamma}^i & = &  {\tilde\gamma}^{jk}\, {\tilde\Gamma}^i_{jk} = \tilde\gamma^{im}\tilde\gamma^{jk}\partial_j \tilde\gamma_{mk} = -\partial_j  \tilde\gamma^{ij} \,,\\ 
\tilde\Gamma_{ijk} &=&  \left( D_{kij} + D_{jik} - D_{ijk} \right)\,, \\
D_{kij}&=& \frac{1}{2}(\tilde\Gamma_{ijk}+\tilde\Gamma_{jik})  \,,\\
\label{eqn.dchr}
\partial_k \tilde\Gamma_{ij}^m & = &   - 2 {\tilde\gamma}^{mn} {\tilde\gamma}^{pl} D_{knp}  \left( D_{ijl} + D_{jil} - D_{lij} \right)
												             + {\tilde\gamma}^{ml} \left( \partial_{(k} {D}_{i)jl} + \partial_{(k} {D}_{j)il} - \partial_{(k} {D}_{l)ij} \right) , \\ %
\partial_k \Gamma_{ij}^m & = & \partial_k \tilde\Gamma_{ij}^m + 2 \delta^m_i \partial_k P_j + 2 \delta^m_j \partial_k P_i + 4\tilde\gamma^{mr}\tilde\gamma^{ns}D_{krs}P_n\tilde\gamma_{ij} 
-2\tilde\gamma_{ij}\tilde\gamma^{mn}\partial_k P_n - 4\tilde\gamma^{mn}P_n D_{kij}
\,,\\
 R^m_{\,\,ikj} & = & \partial_k \Gamma^m_{ij} - \partial_j \Gamma^m_{ik} + \Gamma^m_{lk}\, \Gamma^l_{ij}  - \Gamma^m_{lj}\,\Gamma^l_{ik} , \\
 \label{Rij}
 R_{ij} & = &   R^k_{\,\,ikj}=\tilde R_{ij} + R^\phi_{ij} \,, \\
 \tilde R_{ij} & = & -\tilde\gamma^{lm}\partial_lD_{mij}+\tilde\gamma_{k(i}\partial_{j)}\tilde\Gamma^k+\tilde\Gamma^k\tilde\Gamma_{(ij)k}+\tilde\gamma^{lm}(2\tilde\Gamma^k_{l(i}\tilde\Gamma_{j)km}
   + \tilde\Gamma^k_{im}\,\tilde\Gamma_{klj})  \,, \\
 R^\phi_{ij} & = & -2\partial_i P_j -2\tilde\gamma_{ij}\tilde\gamma^{kn}\partial_k P_n + 4P_i P_j-4\tilde\gamma_{ij}\tilde\gamma^{kn}P_k P_n + 2 \tilde\Gamma^k_{ij}P_k +2\tilde\gamma_{ij}\tilde\gamma^{kr}\tilde\Gamma^n_{kr}P_n \,, \\
 R & = &  \gamma^{ij}\, R_{ij}=e^{-4\phi}\tilde\gamma^{ij}\, R_{ij}\,, \\
\label{eqn.nablanablaalpha}
\upp\nabla_i \upp\nabla_j \alpha &=& \alpha A_i A_j - \alpha\, \Gamma^k_{ij} A_k + \alpha \partial_{(i} {A}_{j)}\,.
\end{eqnarray}
\begin{itemize}
	\item Eq.~\eqref{eqn.slicing} is the gauge condition for the lapse, which can be either the harmonic one, setting $g(\alpha)=1$, or the {\emph 1+log} one, setting $g(\alpha)=2/\alpha$.
	\item Eqs.~\eqref{g-driver1}--\eqref{eqn.Gammai} all together form the well-known {\emph{ Gamma--driver}}, which constitutes the gauge condition for the shift and it can
	be activated or not through the parameter $s$, either 1 or 0.
	The extra parameter $b$ in Eq.~\eqref{g-driver1} and  Eq.~\eqref{g-driver2}, either 1 or 0, is used to switch the convection term on or off in the evolution of $\beta^i$ and $b^i$.
	\item The coefficient $\mu$ in Eq.~\eqref{eqn.B} can be used to modulate the insertion of the ordering
	constraints into the system, specifically the terms $\gamma^{ij} \gamma^{nl}\left( \partial_k D_{ljn} - \partial_l D_{kjn} \right)$, 
	 which are zero at the continuous level due to the symmetry of second derivatives, see Eq. \eqref{eq:curlfree}. These terms are crucial to obtain a strongly hyperbolic system. 
	\item The extra term proportional to $ \left[  \tilde\gamma^{jk} \partial_i \tilde A_{jk} + \tilde A_{jk} \partial_i \tilde\gamma^{jk}   \right]$ on the right hand side of Eq.~\eqref{eqn.D}, 
	is added to take into account that $\tilde A_{jk}$ is trace free, see also \cite{Dumbser2017strongly}, and is crucial to obtain a strongly hyperbolic system.
	\item The terms $(R_{ij})^{TF}_{ncp}$ and $(R_{ij})^{TF}_{src}$ in Eq.~\eqref{eqn.Kij} contain
	the non--conservative products and the purely algebraic factors, respectively, that can be extracted from
	Eq.~\eqref{Rij} when all the partial derivatives of the Christoffel symbols are expanded.
\end{itemize}
The definition of the matter quantities $D$, $E$, $S_i$, $S_{ij}$ for a non--dissipative fluid 
in  Eq.~\eqref{eqn.rho}--\eqref{eqn.E} is standard nowadays and can be found, for instance, in 
\cite{Rezzolla_book:2013}. Throughout this paper we are assuming a simple ideal gas equation of state
$p=\rho\epsilon(\gamma-1)$, where $\epsilon$ is the specific internal energy, while $\gamma$ is the adiabatic index. 
Of course, present day NR simulations adopt much more realistic equations of state, especially for
neutron star binaries (see, among the others, \cite{Takami2014,Dexheimer2019,Weigh2019}), but this is not the focus of this work, and we have therefore set up the
simplest physical conditions. For the sake of completeness we recall the definitions of the conservative variables and of the stress tensor in terms of the primitive (physical) variables: 
\begin{equation}
	D   = \rho W, \quad
	S_i = \rho h W^2 v_i, \quad 
	E   = \rho h W^2 - p, \quad
	S_{ij} = \rho h W^2 v_i v_j + p\, \gamma_{ij} , \quad  
	h   = 1 + \epsilon + \frac{p}{\rho}, \quad   
	W   = \frac{1}{\sqrt{1-v^i v_i}},	
\end{equation}
with $h$ the specific enthalpy, $\rho$ the rest mass density, $v_i$ the fluid velocity and $W$ the Lorentz factor. For an efficient and robust conversion from conservative to primitive variables, even including vacuum, see \cite{DumbserZanottiGaburroPeshkov2023}. The new first order BSSNOK system proposed above is strongly hyperbolic and in  Appendix \ref{app.eigenstructure} we provide the list of eigenvalues and eigenvectors for a reduced subset of the full system.

\section{The numerical method}
\label{sec:cweno}

\subsection{Central WENO reconstruction}
\label{sec:CWENO}

Central WENO (CWENO) schemes were introduced more than 20 years ago in a series of papers by \cite{Levy1999,Levy2000,Levy2001}. Here we provide
a step-by-step presentation of this approach, showing how it can be practically implemented within the framework of conservative WENO finite difference methods \cite{shuosher1,shuosher2,shu_efficient_weno,Balsara2024c}. 
We consider for simplicity the $x$-direction, while the extension to 
a 3D implementation can be easily obtained after repeating the same reconstruction also in the $y$ and $z$ directions. 
We assume the discrete solution of Eq.~\eqref{eqn.pde.mat.preview} below to be given by the \textit{point values}  $\mathbf{u}_{i,j,k}=\mathbf{u}(x_i,y_j,z_k)$, with the points located in the centers of the logically equidistant control volumes, i.e. 
$x_i = \halb ( x_{i-\halb} + x_{i+\halb})$ and $\Delta x = x_{i+\halb} - x_{i-\halb}$. More precisely, we have
\begin{equation}
	 x_i = x_L + \halb \Delta x + (i-1) \Delta x, 
	 \qquad 
	 \Delta x = \frac{x_R - x_L}{\textnormal{IMAX}} 
\end{equation} 
with $x_L$ and $x_R$ the left and right boundaries of the computational domain and $\textnormal{IMAX}$ the number of cells in the $x$ direction. Similar definitions apply to 
$y_j$ and $z_k$. 
\begin{enumerate}
%
\item   
In each direction, 
there is a large stencil $\mathcal{S}_{opt}$ composed of $N+1$ points and given by
$\mathcal{S}_{opt} = \left\{ i-r, \cdots, i-1, i, i+1, \cdots, i+r \right\}$ with $r=N/2$ the half stencil size and $N$ the degree of the polynomial to be reconstructed on such a stencil.
 In addition, there are a number of sub-stencils for the construction of lower degree polynomials, which we list below according to the value of $N$:
\begin{itemize}
	\item  For $N=2$
\begin{equation}	
    \mathcal{S}_{L} = \left\{ i-1, i \right\},\hspace{6.3cm}      
	\mathcal{S}_{R} = \left\{ i, i+1 \right\}\,. \hspace{1.55cm}  
\end{equation}
	\item For  $N>2$
\begin{equation}	
	\hspace{-1cm}
	\mathcal{S}_{L} = \left\{ i-2, i-1, i \right\},\hspace{1cm}   
	\mathcal{S}_{C} = \left\{ i-1, i, i+1 \right\},  \hspace{1cm}   
	\mathcal{S}_{R} = \left\{ i, i+1, i+2 \right\}\,.
\end{equation}

\end{itemize}
%
\item A generic polynomial of degree $M=M(k)$ on any stencil $\mathcal{S}_k$ is expressed in terms of a set of polynomial basis functions  $\psi_m(x)$ and associated degrees of freedom $\hat u_k^m$ as follows:
\begin{equation}
	P^M_k(x) = \sum \limits_{m=0}^M \psi_m(x) \, \hat u^m_k\,, \qquad \textrm{with}\qquad k \in \left\{ opt, L, C, R \right\}\,.
\end{equation}
Throughout this paper we will use the simple Taylor monomials $\psi_m(x) = x^m$ as basis functions. 
The polynomials $P_k^M(x)$ are obtained by \textit{reconstruction}, i.e. by interpreting the
point values $\mathbf{u}_{i,j,k}$ of the finite difference scheme as \textit{cell averages}, see \cite{shuosher2}. The associated reconstruction equations (which are \textit{not} interpolation equations, see \cite{shuosher2,shu_efficient_weno}) for a generic polynomial of degree $M$ computed on an associated stencil $\mathcal{S}_k$ read
\begin{equation}
	\frac{1}{x_{j+\halb}-x_{j-\halb}} \int \limits_{x_{j-\halb}}^{x_{j+\halb}} P_k^M(x) dx = u_j, \qquad
	\forall j \in \mathcal{S}_k.  
\end{equation}
In practice, the polynomials to be reconstructed, according to the value of $N$, are chosen as:
\begin{itemize}
	\item  For $N=2$ \hspace{2cm}
	 $P^2_{opt}(x)$\,, \hspace{2cm} $P^1_L(x)$\,, \hspace{2cm} $P^1_R(x)$\,.
	\item For $N>2$  \hspace{2cm}
	$P^N_{opt}(x)$\,, \hspace{2cm} $P^2_L(x)$\,, \hspace{2cm} $P^2_C(x)$\,, \hspace{2cm} $P^2_R(x)$\,.
\end{itemize}
%
\item After that, we assume that the polynomial $P^M_{opt}(x)$, with  $M \in \left\{ 2,N\right\}$ is split in terms of the poynomials over the sub-stencils, plus one additional
(yet unknown) $P^M_0(x)$ polynomial, i.e.
\begin{itemize}
	\item  For $N=2$
\begin{equation} 
	\label{eqn.optimalorderl1} 
\hspace{-2cm}	P^2_{opt}(x) = \lambda_0 P^2_0(x) + \lambda_L P^1_L(x) + \lambda_R P^1_R(x)\,.
\end{equation}
	\item For  $N>2$
\begin{equation} 
	\label{eqn.optimalorderl2}
	P^N_{opt}(x) = \lambda_0 P^N_0(x) + \lambda_L P^2_L(x) + \lambda_C P^2_C(x) + \lambda_R P^2_R(x)\,.
\end{equation}		
\end{itemize}		
The linear weights $\lambda_0$,  $\lambda_C$, $\lambda_L$ and $\lambda_R$ 
 in a CWENO scheme can be chosen arbitrarily. Throughout this paper we typically set 
$\lambda_0 = 10^8$, $\lambda_C = 10^4$, and $\lambda_L = \lambda_R = 1$, following \cite{dumbser2017central}, though different choices have also been considered,
depending on the test, as specified later.
\item 
Consequently, from \eqref{eqn.optimalorderl1} and \eqref{eqn.optimalorderl2} we now compute an \textit{auxiliary} polynomial $P^N_0(x)$ as 
\begin{itemize}
	\item  For $N=2$
\begin{equation}	
	\label{aux-pol1}
	\hspace{-2cm}
	P^2_0(x) = \frac{1}{\lambda_0} \left( 
	P^2_{opt}(x) - \lambda_L P^1_L(x)  - \lambda_R P^1_R(x) \right)\,.
\end{equation}	
	\item For  $N>2$
\end{itemize}		
\begin{equation}	
	\label{aux-pol2}
	\hspace{1cm}
	P^N_0(x) = \frac{1}{\lambda_0} \left( 
	P^N_{opt}(x) - \lambda_L P^2_L(x) - \lambda_C P^2_C(x) - \lambda_R P^2_R(x) \right)\,. 
\end{equation}
%
\item
Finally, a non-linear WENO reconstruction is performed in terms of the polynomial
$P^N_0(x)$ and of the lower degree polynomials built over the existing sub-stencils.
In particular, we compute for all
polynomials of degree $M(k)$ the nonlinear WENO oscillation indicators as usual in the following manner 
\begin{equation}
	\sigma_k = \sum_{\alpha=1}^{M} \int \limits_{x_{i-\halb} }^{x_{i+\halb }} \left( \frac{\partial^\alpha P^{M}_k(x)}{\partial x^\alpha} \right)^2 \Delta x^{2 \alpha-1} \, dx,\qquad \textrm{with}\qquad k \in \left\{ 0, L, C, R \right\}.
\end{equation}
Tab.~\ref{tab.cweno} shows the main values of the degree polynomials associated to their stencils.
The nonlinear WENO weights ${\omega}_k$ are then computed as usual as 
\begin{equation}
	\label{WENOwr}
	{\omega}_k = \frac{\tilde{\omega}_k}{\sum_m \tilde{\omega}_m}, 
	\qquad 
	\textnormal{ with } 
	\qquad 
	\tilde{\omega}_k = \frac{\lambda_k}{\left( \sigma_k + \epsilon \right)^r}\,, 
\end{equation}
where we typically set $r=4$ and $\epsilon=10^{-7}$.
The final CWENO reconstruction polynomial then reads 
\begin{equation}
	w_h(x) = \sum_k \omega_k P^{M}_k(x). 
\end{equation} 
\end{enumerate}
\begin{table}[!t]
	\caption{Main parameters of the CWENO polynomials.} 
	\renewcommand{\arraystretch}{1.0}
	\begin{center}
		\begin{tabular}{cc|cccc}
			\hline
			\hline
			scheme order & $N$       & $M(0)$& $M(L)$  & $M(C)$  & $M(R)$  \\
			\hline
			3            & 2         & 2     & 1      & \        & 1    \\
			\hline
			5            & 4         & 4     & 2      & 2        & 2      \\
			\hline
			7            & 6         & 6     & 2      & 2        & 2      \\
			\hline 
			9            & 8         & 8     & 2      & 2        & 2      \\
			\hline
			\hline 
		\end{tabular}
	\end{center}
	\label{tab.cweno}
\end{table}
The reconstructed states on the left and right interface of each grid point are given component-wise and in a dimension-by-dimension fashion as  
\begin{equation}
	w^{\mp}_{i \pm \halb,j,k} = w_h \left( x_{i\pm\halb}\right) 
\end{equation}
so that the corresponding high order accurate approximation of the first spatial derivative can be obtained as
\begin{equation}
	\partial_x \u_{i,j,k} = \frac{	\w^-_{i + \halb,j,k} -	\w^+_{i - \halb,j,k}}{\Delta x},
\end{equation} 
which is a very interesting and peculiar feature of conservative WENO schemes, see \cite{shuosher2,shu_efficient_weno}. 
The same can be done also for the discrete derivatives in $y$ and $z$ direction so that we finally get the discrete spatial gradient $\nabla \u_{i,j,k} = \left( \partial_x \u_{i,j,k}, \partial_y \u_{i,j,k}, \partial_z \u_{i,j,k} \right)^T$ to be used in the scheme \eqref{semi-discrete-scheme} shown later. 
The CWENO reconstruction presented above for the point values of the states $\u_{i,j,k}$ is also applied to the point values of the fluxes, i.e. for $\f^x_{i,j,k}$ in the $x$ direction, $\f^y_{i,j,k}$ in the $y$ direction and $\f^z_{i,j,k}$ in the $z$ direction.

\subsection{CWENO finite difference discretization of the Einstein-Euler system}

The full Einstein-Euler  
equations~\eqref{eqn.rho}--\eqref{eqn.D} for the coupled evolution of matter and spacetime form a first--order PDE system in which 
the matter sector enters as a purely conservative sub--system with algebraic sources, while
the Einstein sector is purely non--conservative. On the overall it can be written as
\begin{equation}
	\label{eqn.pde.mat.preview}
	\frac{\partial \u }{\partial t} 
	+ \frac{\partial \f^i}{\partial x_i} +
	\B^i(\u) \frac{\partial \u}{\partial x_i} 
	= \S(\u),
	\qquad 
	\textnormal{or, equivalently,}
	\qquad 
	\frac{\partial \u }{\partial t} + \nabla \cdot \F(\u)
	+ \B(\u) \cdot \nabla \u  
	= \S(\u)\,,
\end{equation}
where $\u$ is the state vector, composed of 63 dynamical variables\footnote{More specifically:  
5 for the matter part, 
11 for the lapse, the shift vector, the metric components and the scalar $\phi$, 
7 for $\tilde A_{ij}$ and the scalar $K$,
3 for $b^i$, 
3 for $\tilde\Gamma^i$ and 
1 for $K_0$; 
plus the auxiliary variables needed for the first order reduction: 
3 for $A_i$, 
9 for $B_k^{\,i}$, 
18 for $D_{kij}$ and 
3 for $P_i$.
In comparison, the classical second order BSSNOK system \cite{Alcubierre:2008} has only 30 evolution variables, while the well-known first order generalized harmonic (GH) formulation \cite{Lindblom2006} of the full Einstein-Euler equations requires 55 variables, namely 50 for the Einstein sector and 5 variables for the matter part.},
while $\F=\left(\f^x, \f^y, \f^z \right)^T$ is the flux vector. 
A finite difference WENO scheme for the vacuum Einstein equations, namely with $T^{\mu\nu}=0$ and which can successfully cope with the presence of non--conservative terms, has been recently proposed by \cite{Balsara2024c}. Here we follow a similar approach, with a few modifications.
Still in semi-discrete form, a path-conservative finite difference scheme for the discretization of \eqref{eqn.pde.mat.preview}, reads 

\vspace{6mm} 

\begin{eqnarray}
\label{semi-discrete-scheme}
\frac{d \u_{i,j,k} }{d t} &=& 
 - \frac{\f^x_{i+\halb,j,k}-\f^x_{i-\halb,j,k}}{\Delta x}   
 - \frac{\f^y_{i,j+\halb,k}-\f^y_{i,j-\halb,k}}{\Delta y}   
 - \frac{\f^z_{i,j,k+\halb}-\f^z_{i,j,k-\halb}}{\Delta z}  
- \B(\u_{i,j,k}) \cdot \nabla \u_{i,j,k}  +  \nonumber \\
&& 
- \frac{\d^x_{i+\halb,j,k}+\d^x_{i-\halb,j,k}}{\Delta x}  -
 \frac{\d^y_{i,j+\halb,k}+\d^y_{i,j-\halb,k}}{\Delta y}  -
 \frac{\d^z_{i,j,k+\halb}+\d^z_{i,j,k-\halb}}{\Delta z}   + \S(\u_{i,j,k} ).
\end{eqnarray}
The calculation of the
discrete spatial gradient $\nabla \u_{i,j,k} = \left( \partial_x \u_{i,j,k}, \partial_y \u_{i,j,k}, \partial_z \u_{i,j,k} \right)^T$
is obtained through the CWENO reconstruction,
as described in Sect.~\ref{sec:CWENO}.
The calculation of the numerical fluxes $\f^x_{i\pm\halb,j,k}$, $\f^y_{i,j\pm\halb,k}$ and
$\f^z_{i,j,k\pm\halb}$ is obtained through
an approximate Riemann solver, for which we provide two possibilities
\begin{itemize}
\item 
	a Rusanov-type (local Lax-Friedrichs) Riemann solver (written here along the $x$-direction), \ie
\begin{equation}
	\label{Rusanov}
	\f^x_{i+\halb,j,k}=\frac{1}{2}\left(\f^{x,-}_{i + \halb,j,k} + \f^{x,+}_{i + \halb,j,k}   \right)-\frac{1}{2}\lambda^{\max}_{i + \halb,j,k}\left( \w^{+}_{i + \halb,j,k} - \w^{-}_{i + \halb,j,k} \right)\,,
\end{equation}	
\item
an HLL Riemann solver (written here along the $x$-direction), \ie
\begin{equation}
	\label{HLL}
	\f^x_{i+\halb,j,k}=\frac{\lambda^{+}_{i + \halb,j,k}\f^{x,-}_{i + \halb,j,k} - \lambda^{-}_{i + \halb,j,k}\f^{x,+}_{i + \halb,j,k}    + \lambda^{-}_{i + \halb}\lambda^{+}_{i + \halb,j,k}\left( \w^{+}_{i + \halb,j,k} - \w^{-}_{i + \halb,j,k} \right)}{\lambda^{+}_{i + \halb,j,k} - \lambda^{-}_{i + \halb,j,k} }\,,
\end{equation}
\end{itemize}
where
\begin{eqnarray}
	\lambda^{-}_{i + \halb,j,k}&=&	\textnormal{min}\left( 0.0, \textnormal{min}\left(\tilde{\lambda}(\w^{-}_{i + \halb,j,k}),\tilde{\lambda}(\w^{+}_{i + \halb,j,k})\right) \right) \\
	\lambda^{+}_{i + \halb,j,k}&=&  \textnormal{max}\left( 0.0, \textnormal{max}\left(\tilde{\lambda}(\w^{-}_{i + \halb,j,k}),\tilde{\lambda}(\w^{+}_{i + \halb,j,k})\right) \right)\\
	\lambda^{\max}_{i + \halb,j,k}&=&\textnormal{max}\left( |\tilde{\lambda}(\w^{-}_{i + \halb,j,k})|,|\tilde{\lambda}(\w^{+}_{i + \halb,j,k})|\right)
\end{eqnarray}
are obtained from the eigenvalues $\tilde\lambda$ of the system matrix $\A^x(\u) = \partial \f^x / \partial \u + \B^x(\u)$ computed at the interface as
\begin{equation}
	\tilde{\A}^x_{i+\halb,j,k}=\A^x(\bar \w)\,, \qquad 
\bar\w=\frac{1}{2}\left(\w^{+}_{i + \halb,j,k}+\w^{-}_{i + \halb,j,k}\right)\,.	
\end{equation}
Note that one only needs the eigenvalues, not the matrix itself. 
Eq.~\eqref{Rusanov} and \eqref{HLL} should be regarded as  {\emph{four--state flux formulas}}, where each of  the four entries
$\w^{-}_{i + \halb,j,k}$, $\w^{+}_{i + \halb,j,k}$, $\f^{x,-}_{i + \halb,j,k}$ and 
$\f^{x,+}_{i + \halb,j,k}$ are \emph{reconstructed} via the CWENO strategy,
as described in Sect.~\ref{sec:CWENO}. The whole
procedure must of course be repeated for the calculation
of $\f^y_{i,j\pm\halb,k}$ and
$\f^z_{i,j,k\pm\halb}$.
According to \cite{AOWENO3,Balsara2024c} the three fractions involving the $\d^{x,y,z}$ terms in  Eq.~\eqref{semi-discrete-scheme} are equivalent to the  
jump terms in path-conservative finite volume schemes \cite{Castro2006,Pares2006,Castro2008,CastroPardoPares} and they are given as follows: 
\begin{itemize}
	\item Rusanov--type scheme:
\begin{eqnarray}
\label{Rusanov-diss}
\d^x_{i + \halb,j,k} & = & \frac{1}{2}  \tilde{\B}^x_{i+\halb,j,k} \cdot   
\left( \w^{+}_{i + \halb,j,k} - \w^{-}_{i + \halb,j,k} \right)  \,,\\ 
\end{eqnarray}
\item HLL--type scheme:
\begin{eqnarray}
\label{HLL-diss}
\d^x_{i+\halb,j,k}  &=&  -\frac{\lambda^{-}_{i + \halb}}{\lambda^{+}_{i + \halb}-\lambda^{-}_{i + \halb}}  \tilde{\B}^x_{i+\halb,j,k} \cdot \left( \w^{+}_{i + \halb,j,k} - \w^{-}_{i + \halb,j,k} \right)\,, \nonumber \\ 
\d^x_{i-\halb,j,k}  &=&  \phantom{-} \frac{\lambda^{+}_{i - \halb}}{\lambda^{+}_{i - \halb}-\lambda^{-}_{i - \halb}}  \tilde{\B}^x_{i-\halb,j,k} \cdot \left( \w^{+}_{i - \halb,j,k} - \w^{-}_{i - \halb,j,k} \right).
\end{eqnarray}
\end{itemize}
The approximation of the terms 
$\tilde{\B}^x_{i+\halb,j,k}$ in \eqref{Rusanov-diss} and \eqref{HLL-diss}
is achieved through a simple midpoint rule as
\begin{equation}
	\tilde{\B}^x_{i+\halb,j,k}=\B^x(\bar \w)\,, \qquad 
	\bar\w=\frac{1}{2}\left(\w^{+}_{i + \halb,j,k}+\w^{-}_{i + \halb,j,k}\right)\,.
\end{equation}
At first glance, one may not have expected such path-conservative jump terms in the numerical scheme and in fact they do not appear in a traditional central finite difference scheme.
However, as shown by \cite{Balsara2024c}, in the context of WENO finite difference schemes they have the practical 
effect of increasing  the order of accuracy by one with very little extra cost, and furthermore also substantially increase the robustness of the final numerical method 
and should therefore be retained. 

Similar expressions for the path-conservative jump terms must also be produced for the $y$ and $z$ directions, respectively.

\noindent 
According to \cite{shuosher2} the WENO approximation of the discrete gradient at the cell-center can be written via the divided differences of the boundary-reconstructed states as 
\begin{equation}
	\nabla \u_{i,j,k} = \left( 
	\begin{array}{c}
		(\w^{-}_{i + \halb,j,k} - \w^{+}_{i - \halb,j,k})/\Delta x \\
		(\w^{-}_{i,j + \halb,k} - \w^{+}_{i,j - \halb,k})/\Delta y \\
		(\w^{-}_{i,j,k + \halb} - \w^{+}_{i,j,k - \halb})/\Delta z 
	\end{array} 
	\right).
\end{equation}
This concludes the description of the spatial discretization used in this paper. 

\noindent
Finally, the nonlinear ODE system resulting from \eqref{semi-discrete-scheme}
is discretized in time via high order classical or TVD Runge--Kutta schemes, see \cite{shuosher1,shuosher2,shu2}.

\paragraph*{Grid stretching.}
\label{sec:grid}
The choice of coordinates is largely determined by the initial conditions of the problem to be solved.
Most of the  3D simulations of NR adopt initial conditions that are conformally flat, and 
for this reason the corresponding coordinates are called Cartesian. Our simple finite difference code does not yet incorporate Adaptive Mesh Refinement (AMR), but it implements a coordinate transformation 
from the logical Cartesian grid to the \emph{physical domain}, with the aim of producing a high resolution (but uniform) grid in the inner part, while producing a stretched non-uniform grid in the outer part, with an outer border much farther away. In practice, along each direction
we perform a mapping from the logical coordinate $\xi_i$ to the physical coordinate $x_i$ simply as
\begin{equation}
	\label{coordmap}
	x_i(\xi_i)= \left\{
	\begin{array}{llll}
		\xi_i &   {\rm if} & |\xi_i|\leq \xi_i^c \,, \\
		
		\frac{\xi_i}{|\xi_i|}\left(a|\xi_i|^3 + b|\xi_i|^2 + c|\xi_i| + d\right) &   {\rm if} & |\xi_i|> \xi_i^c \,, 
	\end{array} \right.
\end{equation}
where $\xi_i^c$ sets the border of the inner uniform grid,
while the coefficients $a$, $b$, $c$ and $d$ are selected in such a way to obtain
a smooth matching of the cubic with the inner linear branch.
The same mapping is applied along all coordinate directions. 
It is also worth stressing that, though not equivalent to AMR in terms
of adaptability, the usage of a grid stretching like \eqref{coordmap}, or similar ones, reduces the need for AMR in many circumstances. An extensive usage of coordinate mapping in the context of numerical relativity has been presented in \cite{Hemberger2013}.

\paragraph*{Non-reflecting boundary conditions.}
\label{sec:bc}
Our code allows for two different strategies to obtain non-reflecting boundary conditions.
Primarily, we use Sommerfeld boundary conditions for the generic quantity $Q$ as \cite{Zhang2025}
\begin{equation}
	\partial_t Q + \frac{r}{x_i}\partial_i Q=-\frac{1}{r}(Q-Q_0)\vert_{\partial\Omega}\,,
\end{equation}
where $r=\sqrt{x^2+y^2+z^2}$, while $Q_0$ denotes the initial conditions. As an alternative, more pragmatic, approach, we have used a so-called \emph{sponge layer} boundary condition, which performs a weighted average among the numerical solution and the initial condition at the grid border, resulting in an absorption of waves. In practice, in a thin shell at the border we modify the solution according to
\begin{equation}
	Q\vert_{\partial\Omega}= (1-k)Q\vert_{\partial\Omega} + kQ_0\vert_{\partial\Omega} \,,
\end{equation}
where $k$ is a small number chosen typically as $k=0.1$.

\paragraph*{Well--balancing.}
\label{sec:wb}
For stationary  problems, our numerical scheme can easily be
made {\emph {well--balanced}},  by simply subtracting a discrete version
of the equilibrium solution from the discretized time-dependent PDE system.
In this way, an initial equilibrium can be maintained stationary up to machine precision
virtually for ever. 
We recall, however, that well--balancing (WB) is not a necessary requirement of our numerical scheme. It is instead an \textit{additional feature} that is advantageous only for stationary solutions
and it can be activated to increase the accuracy of the computation. 
In this paper we have used it only for the TOV star in Sect.~\ref{sec:TOV} to highlight its potential advantages.

The interested reader is referred to \cite{PareschiRey,berberich2021high,DumbserZanottiGaburroPeshkov2023}, where he/she can find all the practical details for the simple and straightforward implementation of a well--balanced scheme in a PDE system of mixed
conservative$/$non--conservative form. For more theoretical background on well-balanced numerical methods for hyperbolic PDE, see \cite{Bermudez1994,leveque1998balancing,gosse2001well,audusse2004fast,BottaKlein,Castro2006,Pares2006,Castro2008} and references therein.

\section{Numerical tests}
\label{sec:tests}

In this Section we present a large set of canonical tests for numerical general relativity, obtained with  
our new high order path-conservative finite difference CWENO schemes applied to the novel first order BSSNOK formulation of the coupled Einstein-Euler system. There are of course plenty of such tests
that have been proposed over the year, some of which already quite standardized~\cite{Alcubierre2004,Babiuc_2008}.
Due to lack of space, we had to perform a selection, as outlined below. For all computational tests shown in this section we have used the classical fourth order Runge-Kutta scheme to integrate the semi-discrete CWENO method in time. Moreover, in several of the tests below we monitor the Einstein constraints of Eq.~\eqref{eqn.adm1}--\eqref{eqn.adm2}, referring to them for simplicity simply as $H$, $M1$, $M2$, $M3$.

\subsection{Linearized gravitational wave }
\label{sec:linearized-grav-wave}

We start with a simple  wave perturbation of the flat Minkowski spacetime~\cite{Alcubierre2004} for which the metric is given by
\begin{equation} 
	\label{eqn.lw.metric}
	d s^2 = - d t^2 + d x^2 + (1+b)\,d y^2 + (1-b)\,d z^2, \quad \textnormal{with} \quad b = \epsilon \sin \left( 2 \pi (x-t) \right)\,.
\end{equation} 
All the metric terms are directly deduced from~\eqref{eqn.lw.metric}, 
where, for instance, the extrinsic curvature follows from 
$K_{ij} = \partial_t  \gamma_{ij} /(2\alpha)$. Moreover, since we choose $\epsilon$ to be
$\epsilon = 10^{-8}$, the overall dynamics is linear and the terms depending on $\epsilon^2$ can be neglected. 
We use the \emph{harmonic gauge condition}, while the \emph{gamma--driver} can be turned off, i.e. $s=0$.
\begin{figure}[!htbp]
	\includegraphics[width=0.45\textwidth]{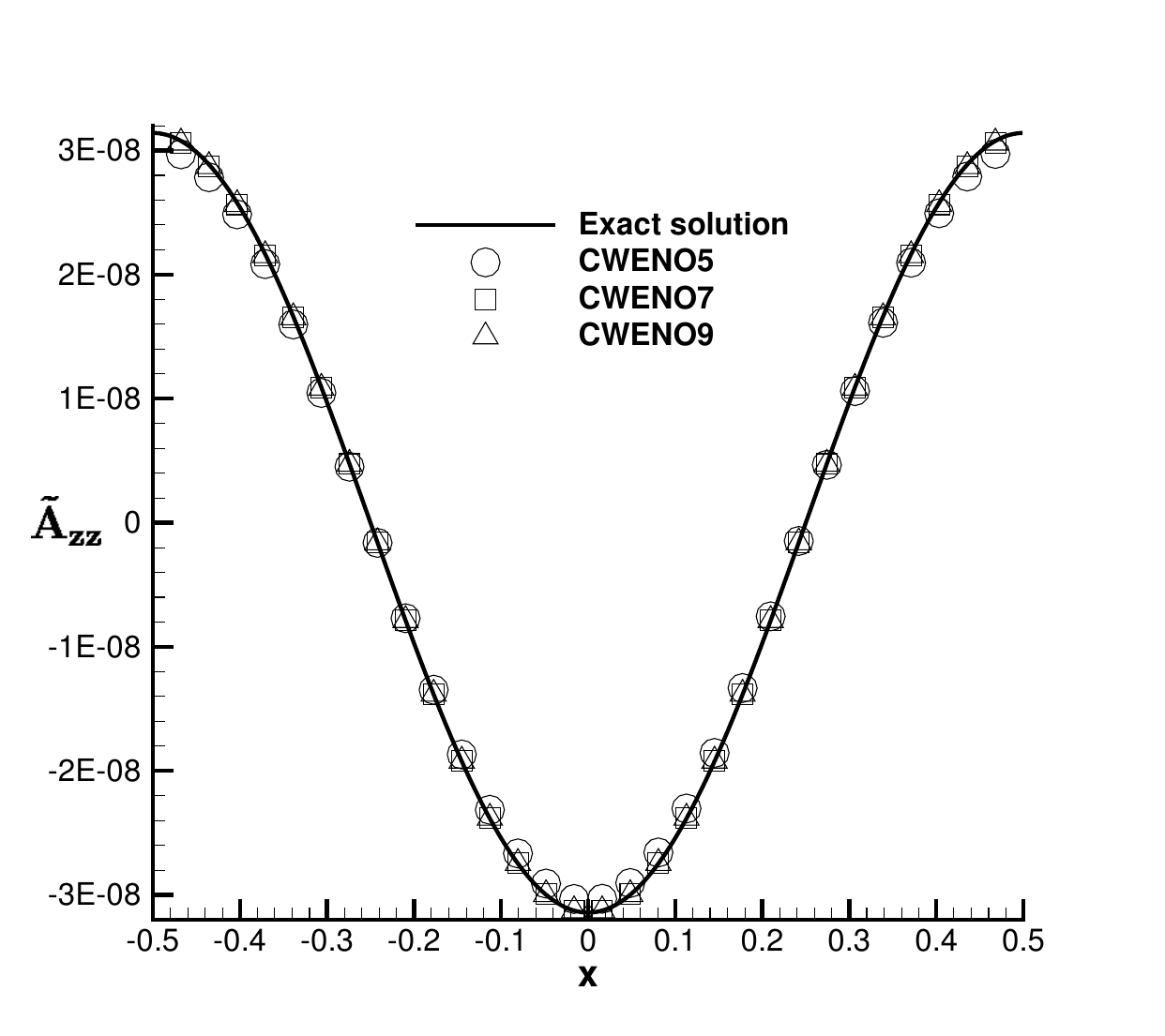}
	\includegraphics[width=0.45\textwidth]{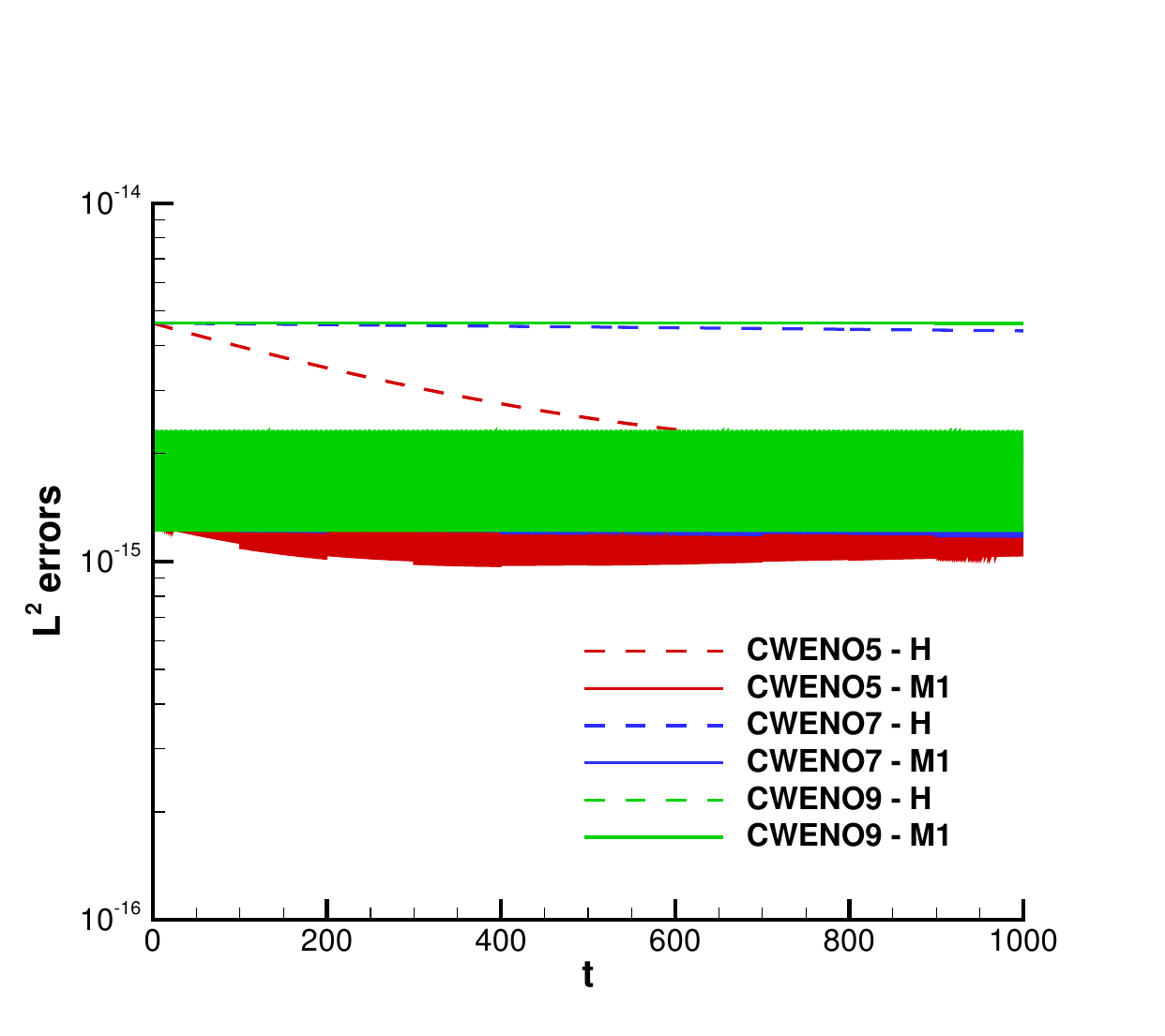}
		\caption{Linearized gravitational wave test solved with different version of the CWENO scheme.
		Left panel: $\tilde A_{zz}$ component of the extrinsic curvature at the final time, compared to the exact solution. Right panel: time evolution of the normalized Einstein constraints.}
		\label{fig.linearized-grav-wave}
\end{figure}
Matter is absent in this test.  
The computational domain is  given by the rectangle $\Omega = [-0.5,0.5]\times[-0.05,0.05]$, which is discretized using  $32 \times 4$ grid-points, and adopting
periodic boundary conditions in both directions.

In Figure~\ref{fig.linearized-grav-wave} we report the results of the calculation comparing the performance of our CWENO schemes at different orders. 
In the left panel we show the profiles of $\tilde A_{zz}$ at $t=1000$,
compared to the exact solution. Apart from the CWENO5 scheme, which shows an appreciable deviation with respect to the solid line, the behavior of the solutions
obtained with CWENO7 and CWENO9 is very accurate.
In the right panel we monitor the evolution of the Einstein constraints, which remain close to machine precision all along the simulation.

\subsection{The robust stability test}
\label{sec:stability-test}

\begin{figure}[!htbp]
	\begin{center}
		\begin{tabular}{cc}
			\includegraphics[width=0.45\textwidth]{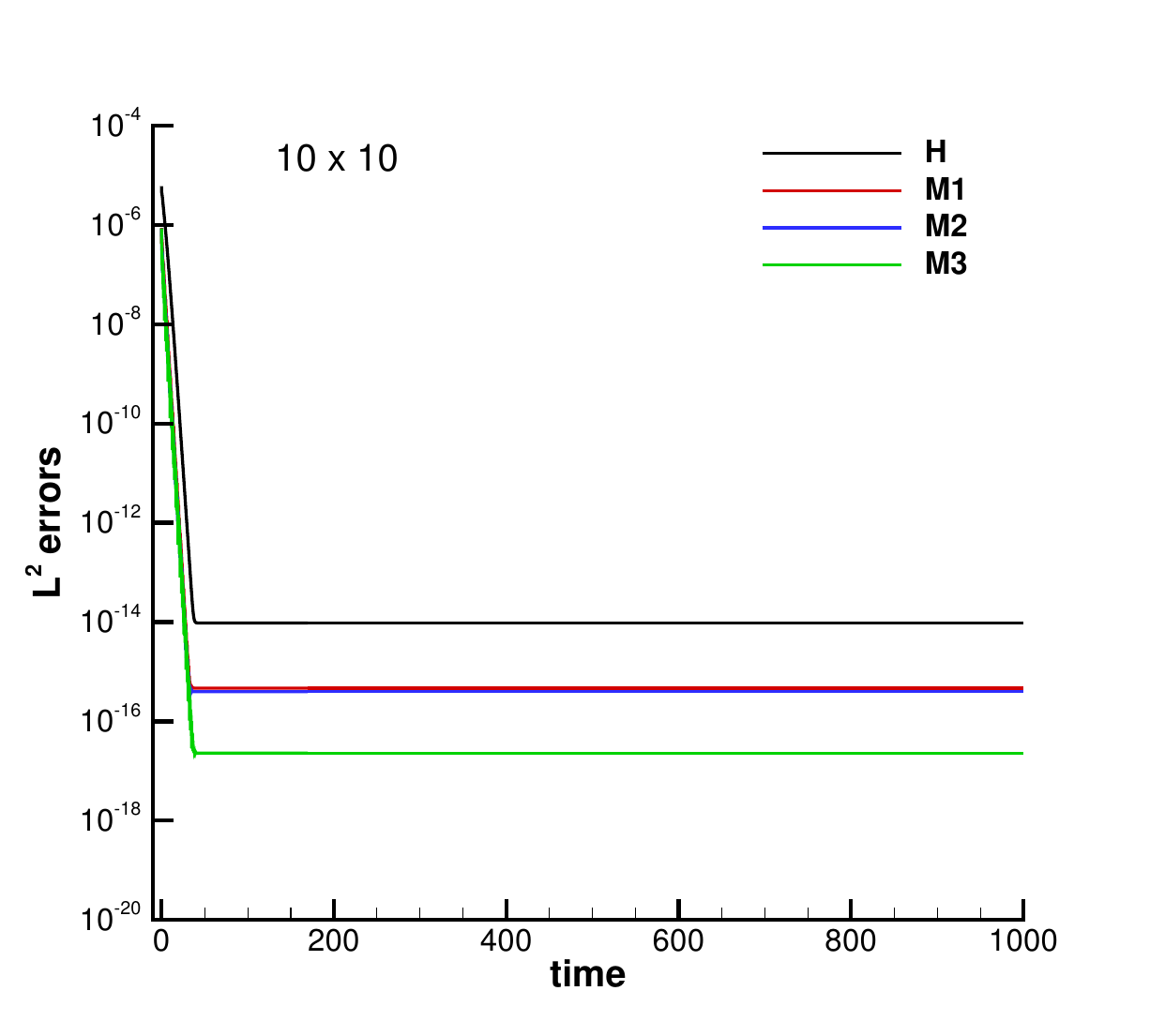} &
			\includegraphics[width=0.45\textwidth]{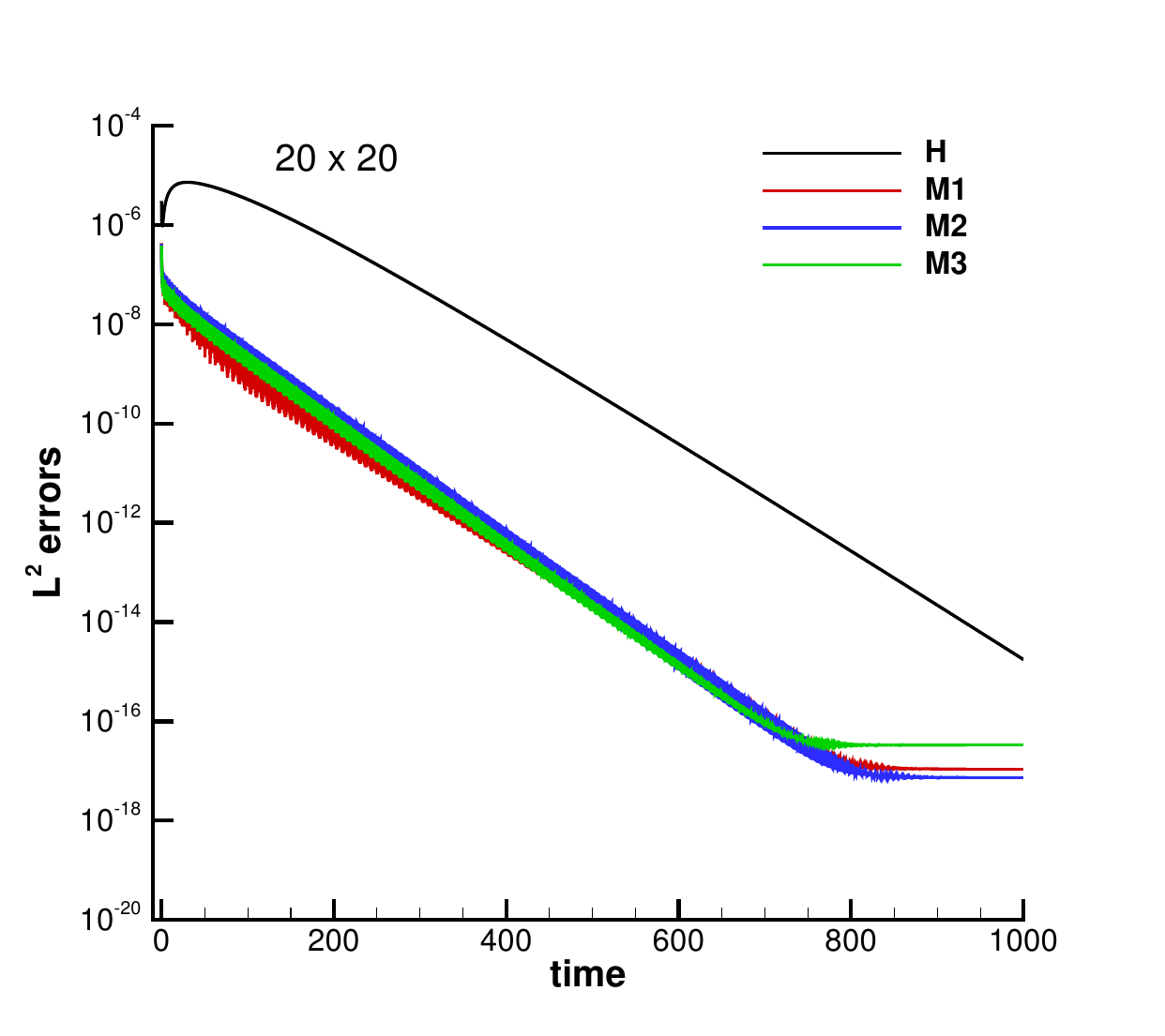} \\
			\includegraphics[width=0.45\textwidth]{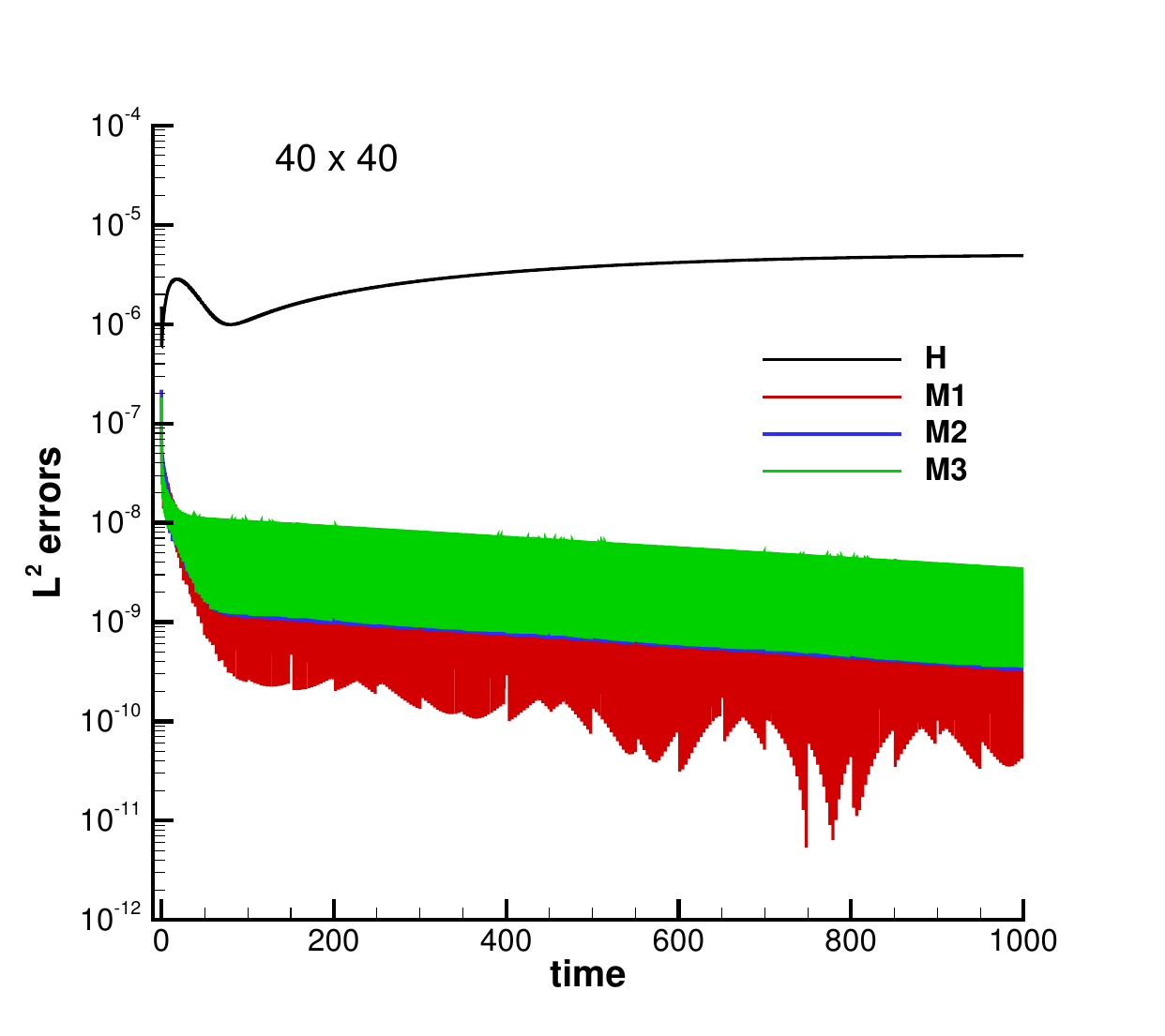} &
			\includegraphics[width=0.45\textwidth]{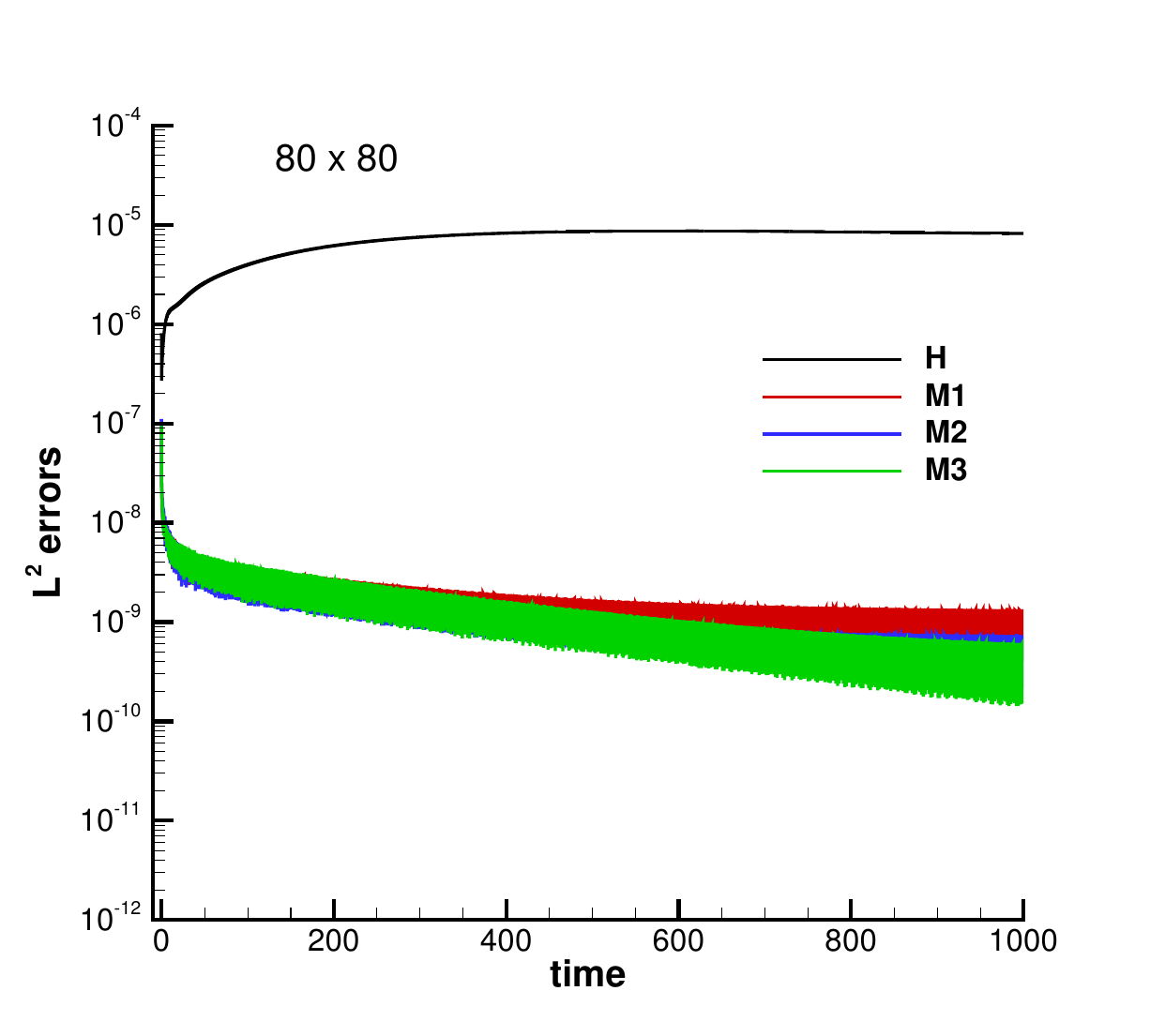} \\
		\end{tabular}
		\caption{Robust stability test case  with a random initial perturbation of amplitude
			$10^{-7}/\rho^2$ in all quantities on a sequence of successively
			refined meshes on the unit square in 2D.
			The {\emph {gamma--driver}} shift condition, $1+\log$ slicing and CWENO5
			scheme have been used. Top left: $10\times10$ elements. Top right: $20\times20$ elements. 
			Bottom left: $40\times40$ elements. Bottom right: $80\times80$ elements. }
		\label{fig.robstab}
	\end{center}
\end{figure}
The {\emph {robust stability test}} introduced in \cite{Alcubierre2004}
 is used to
highlight potential unstable and exponentially growing modes in the solution, and it is therefore a 
procedure to verify the hyperbolicity of the PDE system in a pragmatic and empirical way. 
It is performed on the two dimensional domain $\Omega = [-0.5;0.5]\times[-0.5;0.5]$
by perturbing a flat  Minkowski spacetime
with a random  perturbation with  amplitude $\pm 10^{-7}/\varrho^2$.
The parameter $\varrho$, which is used to scale the perturbation, affects also the resolution of the grid, which is composed of $10 \varrho \times 10 \varrho$ gridpoints.
As customary for this test, the {\emph {gamma--driver}} shift condition is activated.

Fig.~\ref{fig.robstab} shows the results of our calculations, by reporting the evolution of the 
normalized Einstein constraints in a battery of tests where the  CWENO5 version of the scheme is adopted.
These numbers and the corresponding 
results are certainly satisfactory, even if not of the same quality as those reported
in Fig. 4 of
 \cite{DumbserZanottiGaburroPeshkov2023}, where a DG scheme was used in combination with the Z4 formulation of the Einstein equations~~\cite{Bona:2003qn,Bona-and-Palenzuela-Luque-2005:numrel-book}.
%

\subsection{The gauge wave}
\label{sec:gauge-wave}

\begin{figure}[!htbp]
	\includegraphics[width=0.45\textwidth]{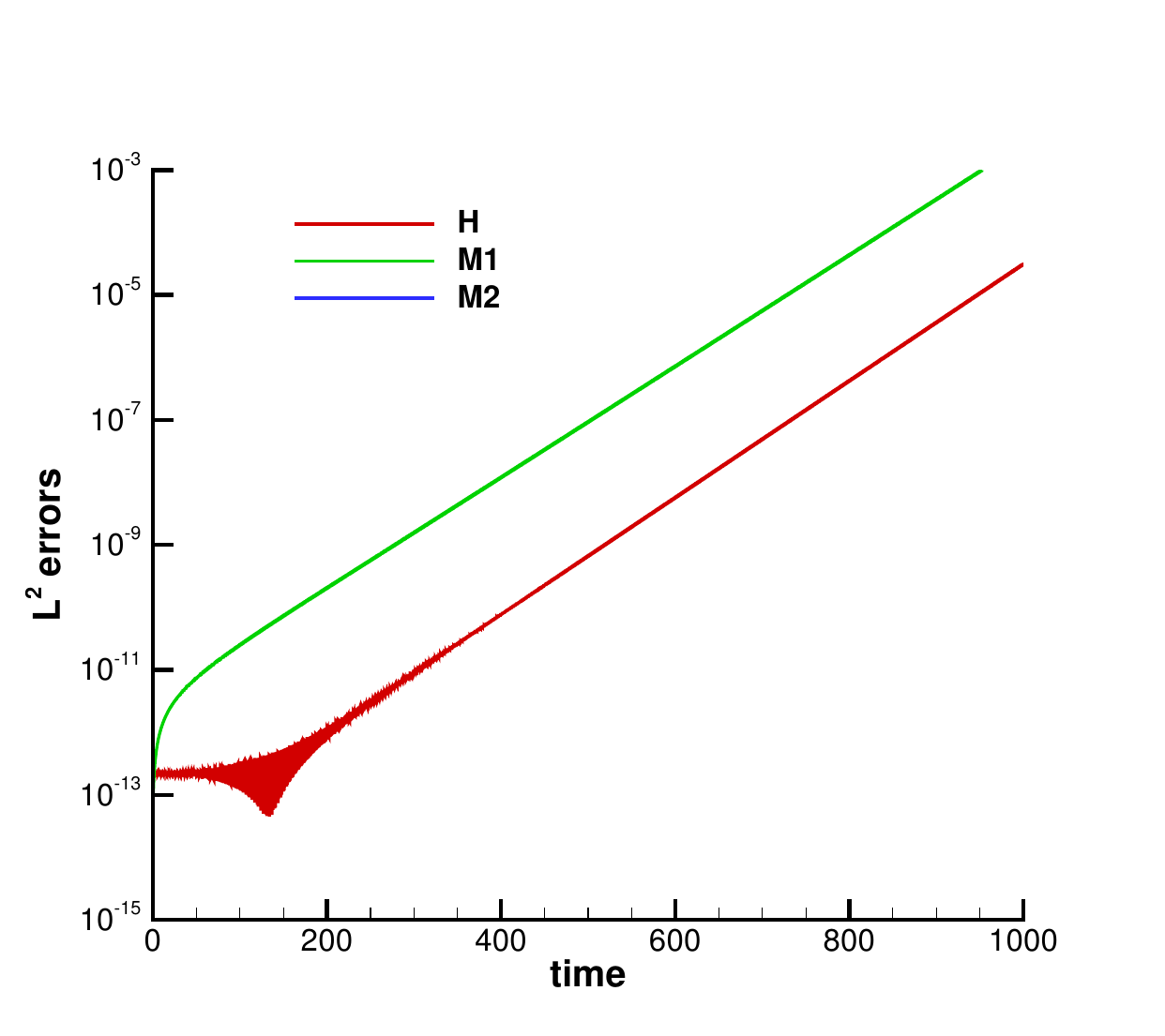}
	\includegraphics[width=0.45\textwidth]{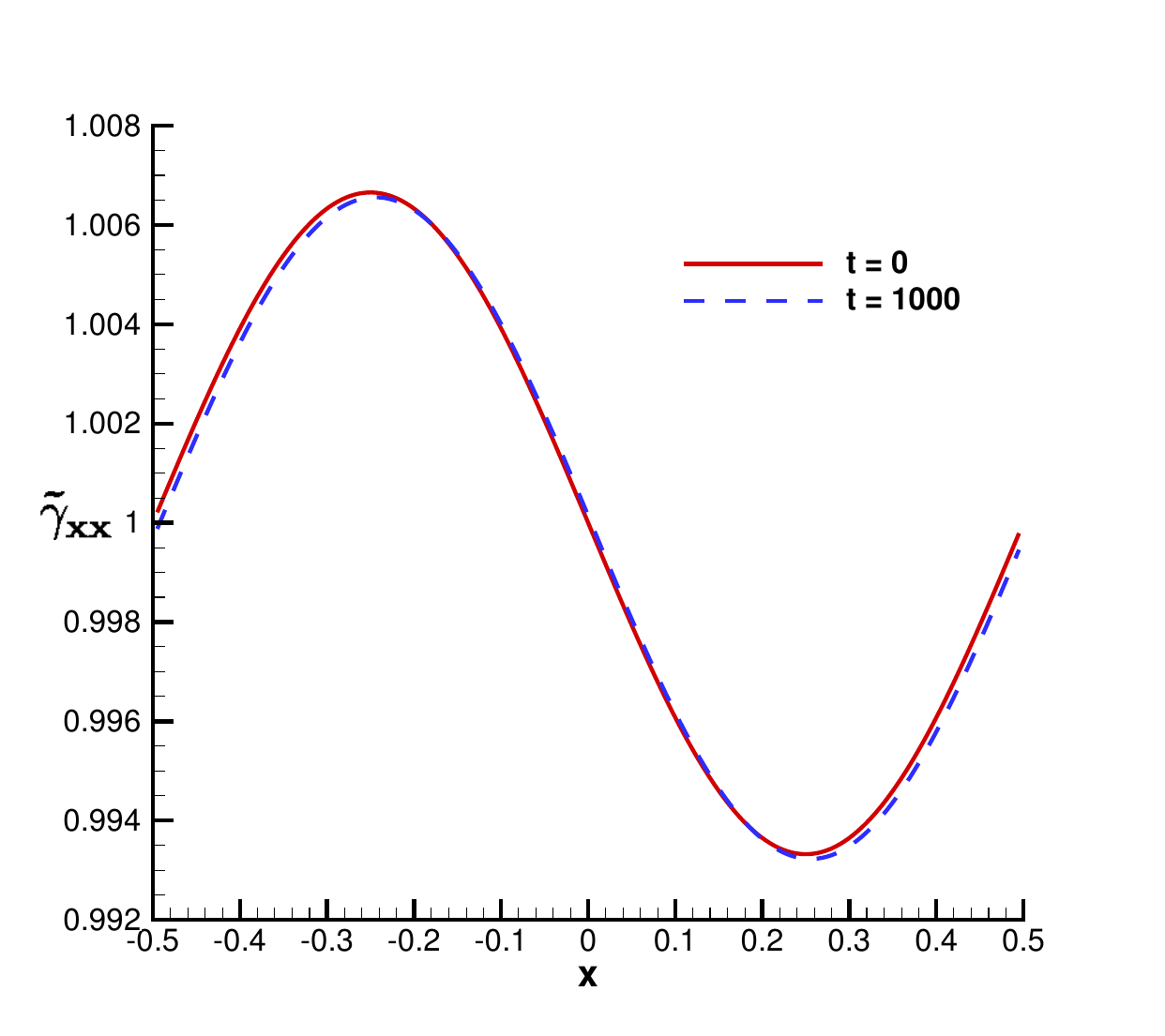}
	\caption{Solution of the gauge wave test at $t=1000$ with $A=0.01$ using the CWENO7 scheme. 
		Left panel: Evolution of the Einstein constraints.
		Right panel: profile of the metric term $\tilde{\gamma}_{xx}$ at the final time compared to the initial condition. 	  
	}
	\label{fig.gaugewavetest-smallA}
\end{figure}

The so--called \textit{gauge wave test}, taken from~\cite{Alcubierre2004}, is traditionally a challenging test for the BSSNOK formulation of the Einstein equations
(see~\cite{Alic:2011a,Brown2012}).
We found that this is still true even with our first--order BSSNOK implementation.
We recall that, on the contrary, rather successful performances with this test were reported by  
\cite{Dumbser2017strongly} in their first--order reformulation of the CCZ4 system
and more recently by \cite{DumbserZanottiGaburroPeshkov2023} with an undamped first--order version of the 
pure Z4 formulation.
In this test the metric is obtained from a simple coordinate transformation performed in the Minkowski spacetime and it is given by 
\begin{equation}
	\label{eqn.gw.metric}
	d s^2 = - H(x,t) \, d t^2 + H(x,t) \, d x^2 + d y^2 + d z^2, \quad \text{where} \quad H(x,t) = 
	1-A\,\sin \left( 2\pi(x-t) \right)\,,
\end{equation}
which is in fact equivalent to a sinusoidal gauge wave of amplitude $A$ propagating along the $x$-axis.
A \emph{harmonic gauge condition} is required along with periodic boundary conditions.

We have first run this test with a small wave amplitude $A=0.01$ over
a rectangular domain of size  $\Omega = [-0.5,0.5] \times [-0.02, 0.02]$ using 
a CWENO7 scheme.  The grid is composed of $256\times 4$ elements, uniformly distributed, and the final time is $t=1000$.
As expected, the gauge wave test manifests several pathologies when evolved 
within the BSSNOK formulation over a long timescale, independently of the numerical scheme adopted. 
The problem is well known in the literature and has been documented by several authors over the years (see, among others \cite{Alcubierre2004,Tichy2009,Brown2012}). On the contrary, we recall that this test can be solved successfully within the Z4 or the GH formulations~\cite{DumbserZanottiGaburroPeshkov2023,Babiuc2006,Boyle2007}.
We have confirmed this behaviour, which is clearly in-printed in the exponential growth of the Einstein constraints, as reported in the left panel
of Figure~\ref{fig.gaugewavetest-smallA}. In the right panel we can appreciate the mismatch of the final solution at $t=1000$ with respect to the initial conditions,
taking the metric function $\tilde{\gamma}_{xx}$ as a representative quantity. 

In spite of these deficiencies, it is still possible to perform a numerical convergence analysis based on this test case, provided the final time is short enough in such a way that 
the pathologies highlighted above do not have the time to spoil the numerical properties of the scheme. Tab.~\ref{tab.conv1} contains the result of this analysis for a gauge wave
with $A=0.1$ at the final time $t=1$ and it essentially confirms the 
nominal orders of convergence of the scheme. 

\begin{table}
\caption{Numerical convergence results for the gauge wave
    with $A=0.1$ at a final time of $t=1$.  The $L^2$
    errors and the corresponding observed convergence order are reported for
    the variables $K$, $\phi$ and $\alpha$.}
\begin{center}
\begin{tabular}{ccccccc}
\hline
  $N$ & ${L^2}$ error $K$ & $\mathcal{O}(K)$  & ${L^2}$ error $\phi$ & $\mathcal{O}(\phi)$  & ${L^2}$ error $\alpha$ & $\mathcal{O}(\alpha)$  \\
\hline
  \multicolumn{7}{c}{CWENO3}   \\
\hline
$32$   & 3.984E-04 &     & 7.963E-06 &      & 4.778E-05 &      \\
$64$   & 5.111E-05 & 3.0 & 1.007E-06 & 3.0  & 6.039E-06 & 3.0  \\
$128$  & 6.418E-06 & 3.0 & 1.261E-07 & 3.0  & 7.566E-07 & 3.0  \\
$256$  & 9.042E-07 & 2.8 & 1.919E-08 & 2.8  & 1.151E-07 & 2.7  \\
\hline
  \multicolumn{7}{c}{CWENO5}   \\
\hline
$32$   & 1.046E-05 &     & 1.291E-07 &      & 7.746E-07 &      \\
$64$   & 3.380E-07 & 5.0 & 4.135E-09 & 5.0  & 2.481E-08 & 5.0  \\
$128$  & 1.066E-08 & 5.0 & 1.302E-10 & 5.0  & 7.809E-10 & 5.0  \\
$256$  & 3.341E-10 & 5.0 & 4.079E-12 & 5.0  & 2.448E-11 & 5.0  \\
\hline
  \multicolumn{7}{c}{CWENO7}   \\
\hline
$32$   & 3.681E-06  &     & 3.591E-08 &      & 2.155E-07 &      \\
$64$   & 5.294E-07  & 6.7 & 5.076E-09 & 6.8  & 3.046E-08 & 6.8  \\
$128$  & 4.536E-09  & 6.9 & 4.330E-11 & 6.9  & 2.598E-10 & 6.9  \\
$256$  & 4.475E-11  & 6.7 & 4.525E-13 & 6.6  & 2.715E-12 & 6.6  \\
\hline
\end{tabular}
\end{center}
\label{tab.conv1}
\end{table}

\subsection{Special relativistic Riemann Problems in the Cowling approximation }
\label{sec:SR}
As a first test involving the matter part of the PDE system, we consider a few canonical Riemann problems in the
Cowling approximation~\cite{Cowling41}, namely after neglecting the evolution of the spacetime, which is assumed
to be flat Minkowski.
We have selected a sample of three Riemann problems, whose  
initial conditions are reported in Table~\ref{tab.RP.ic}.
Each of them has been solved using a CWENO3 scheme, a resolution of $512$ points
and WENO parameters of Eq.~\eqref{eqn.optimalorderl2}--\eqref{WENOwr}, that, contrary to the
standard values declared in Sect.~\ref{sec:CWENO}, are given by $\lambda_0=10^8$, $\lambda_C=10^4$, $\lambda_L=1$, $\lambda_R=1$, $r=10$, $\epsilon=10^{-14}$.

\begin{figure}
{\includegraphics[angle=0,width=5.5cm,height=5.0cm]{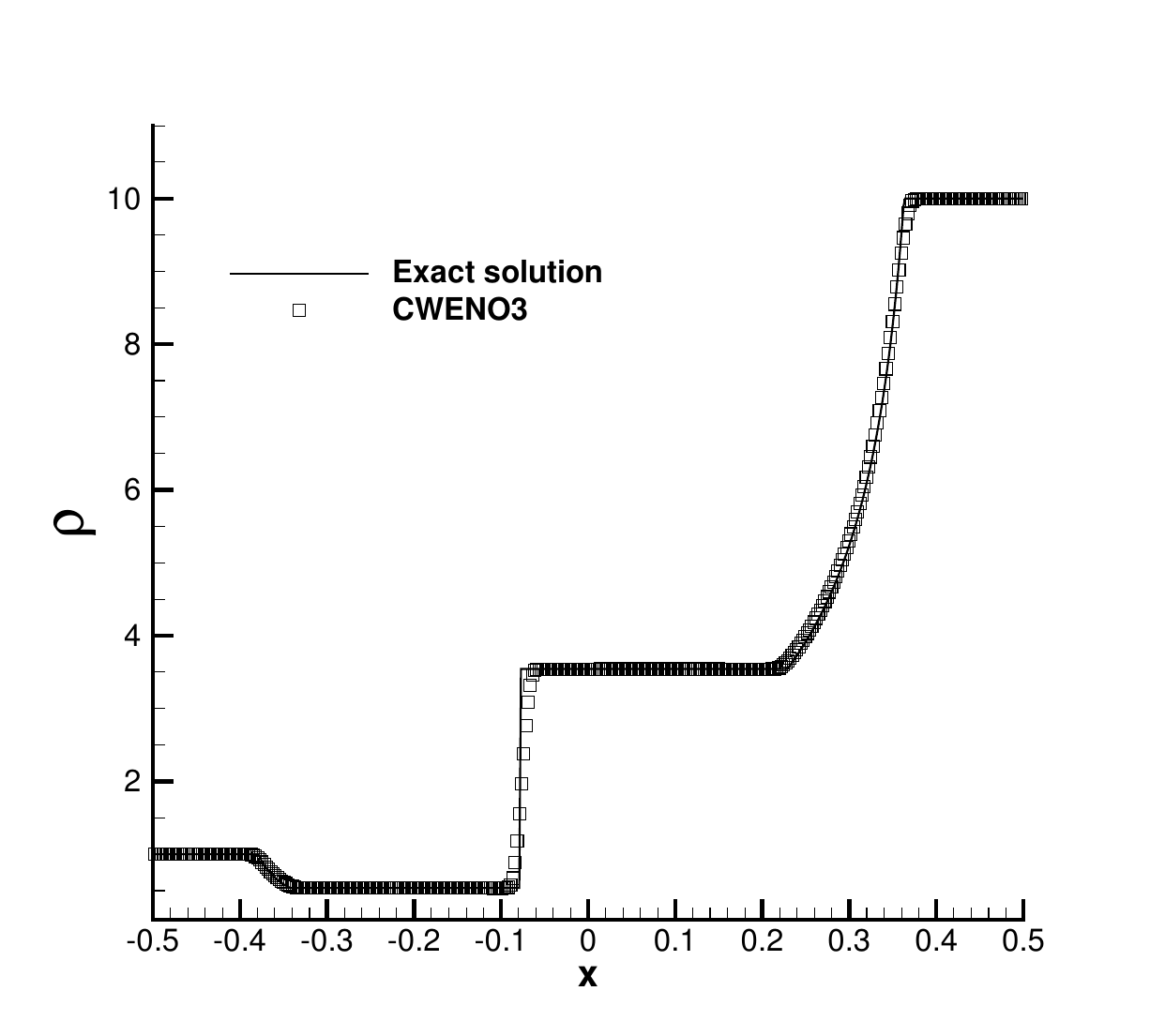}}
{\includegraphics[angle=0,width=5.5cm,height=5.0cm]{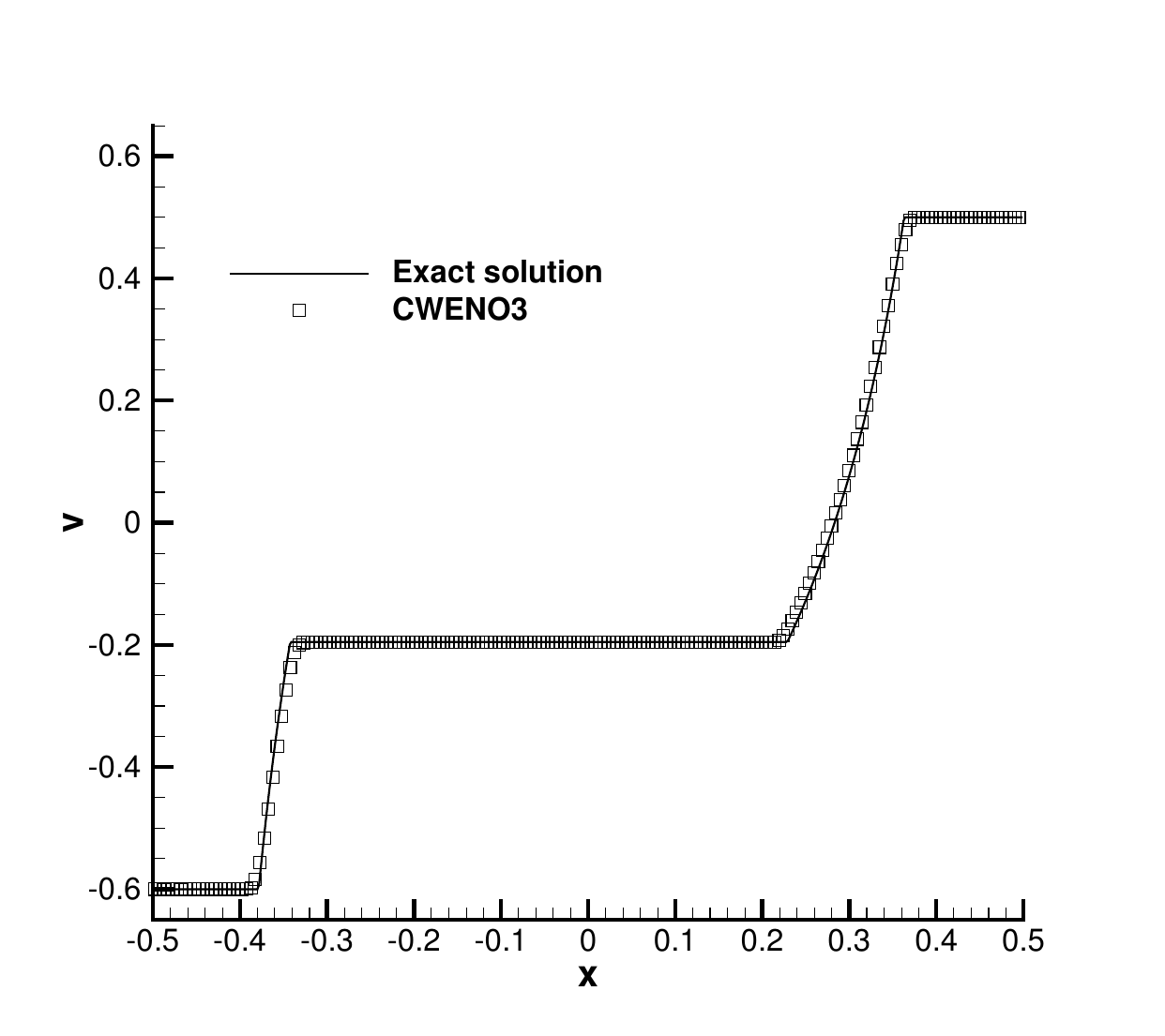}}
{\includegraphics[angle=0,width=5.5cm,height=5.0cm]{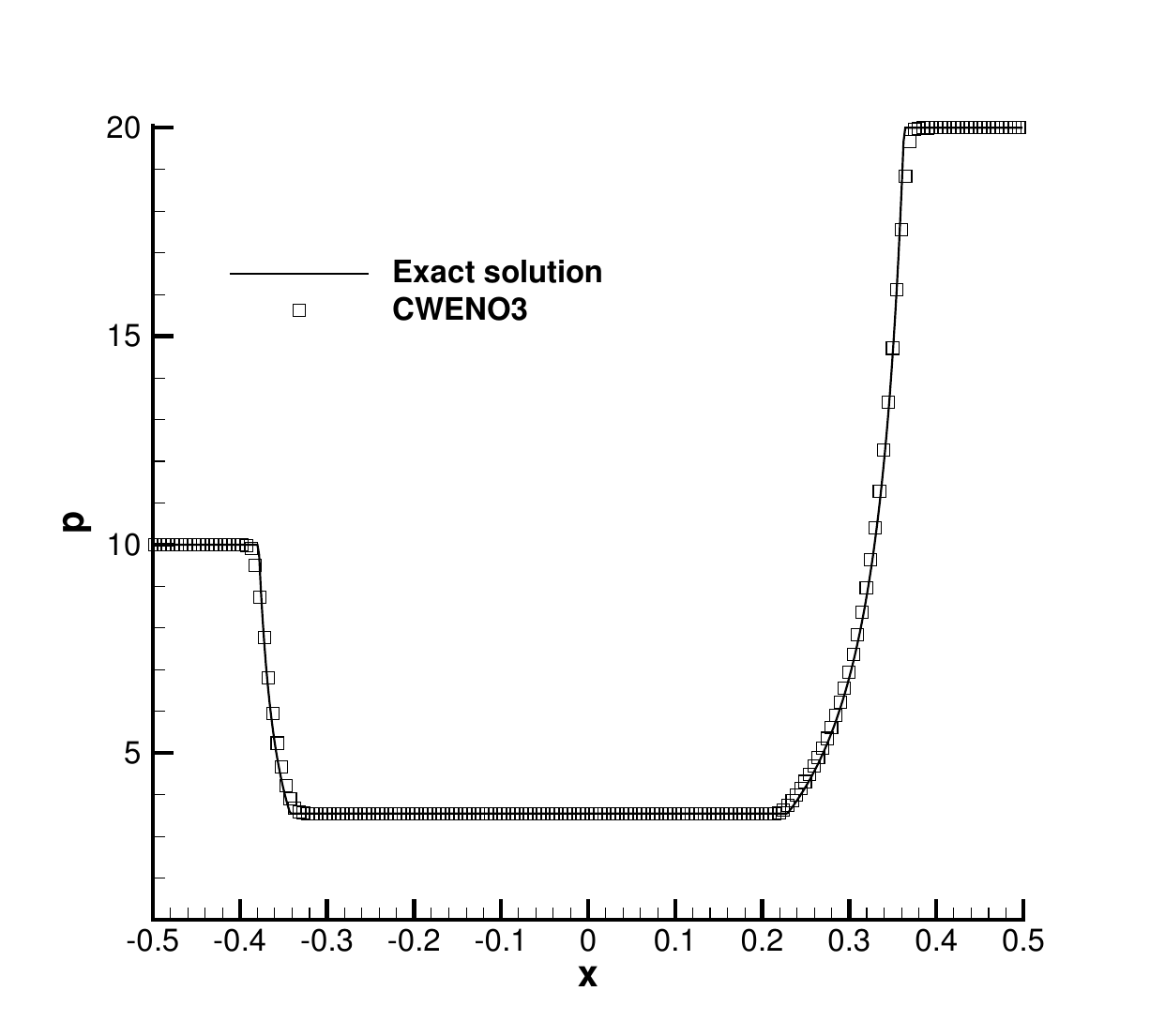}}
\caption{Solution of Riemann Problem 1  at time $t=0.4$.}
\label{fig:shock-tube-2R}
\end{figure}

\begin{figure}
{\includegraphics[angle=0,width=5.5cm,height=5.0cm]{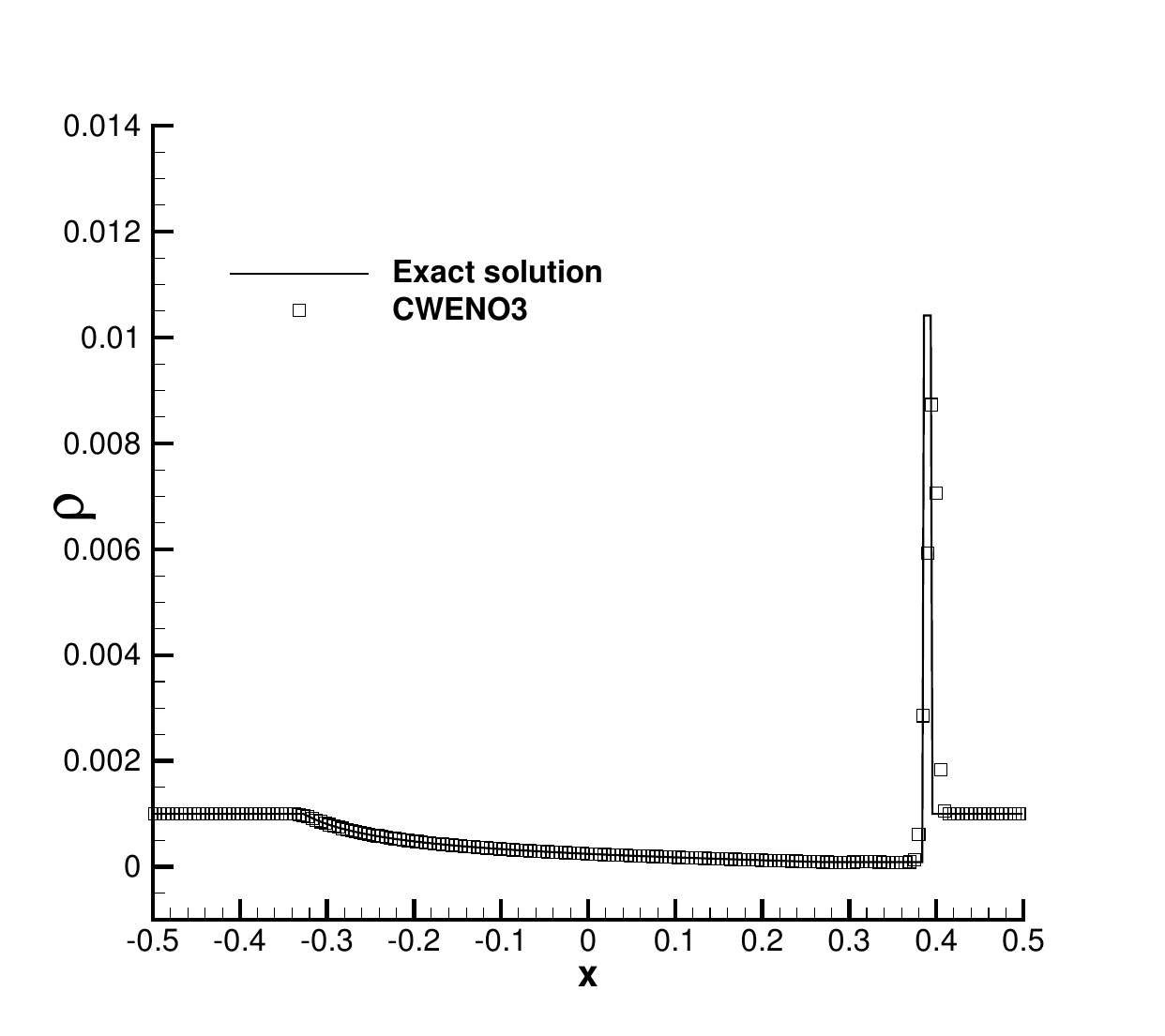}}
{\includegraphics[angle=0,width=5.5cm,height=5.0cm]{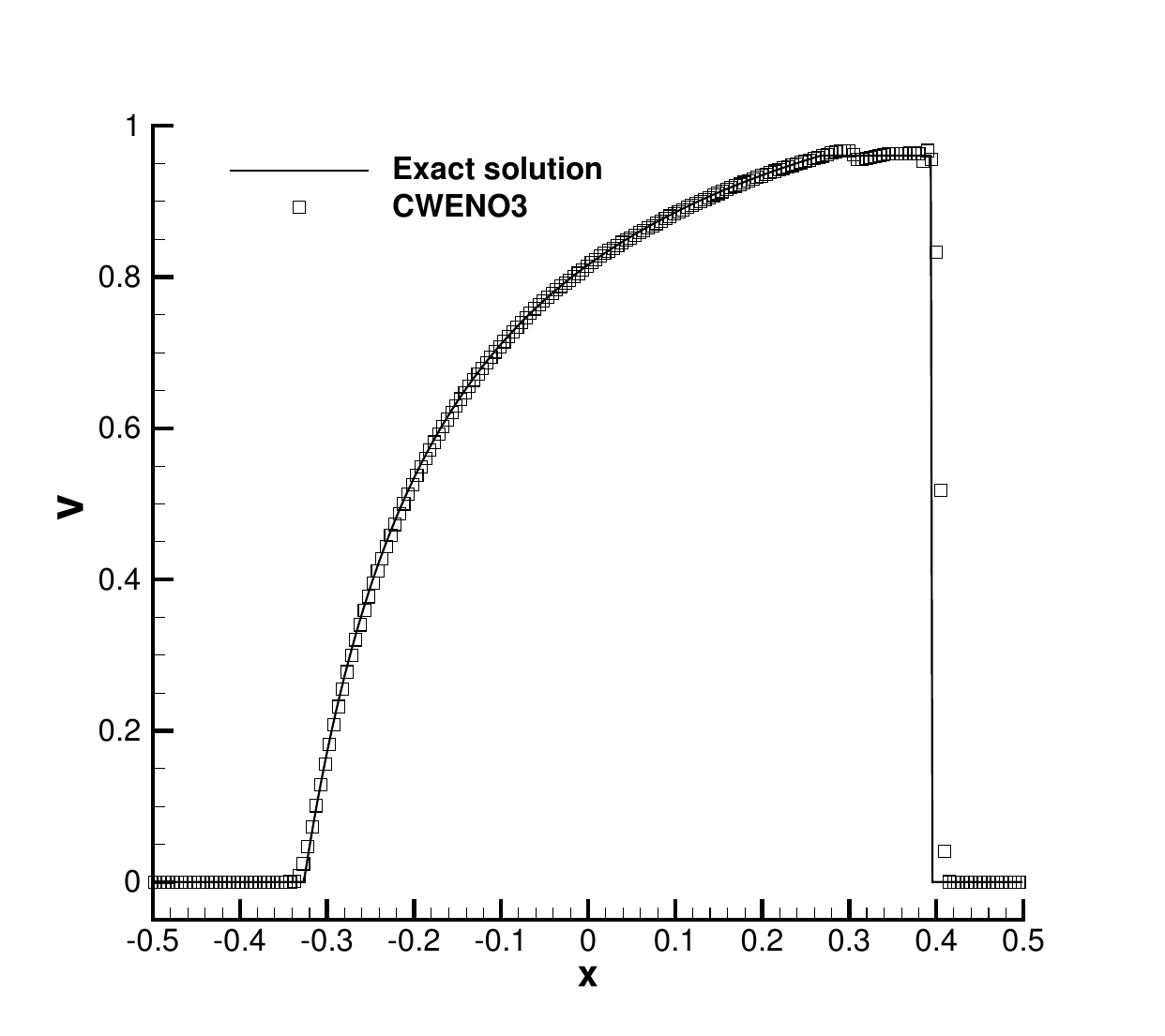}}
{\includegraphics[angle=0,width=5.5cm,height=5.0cm]{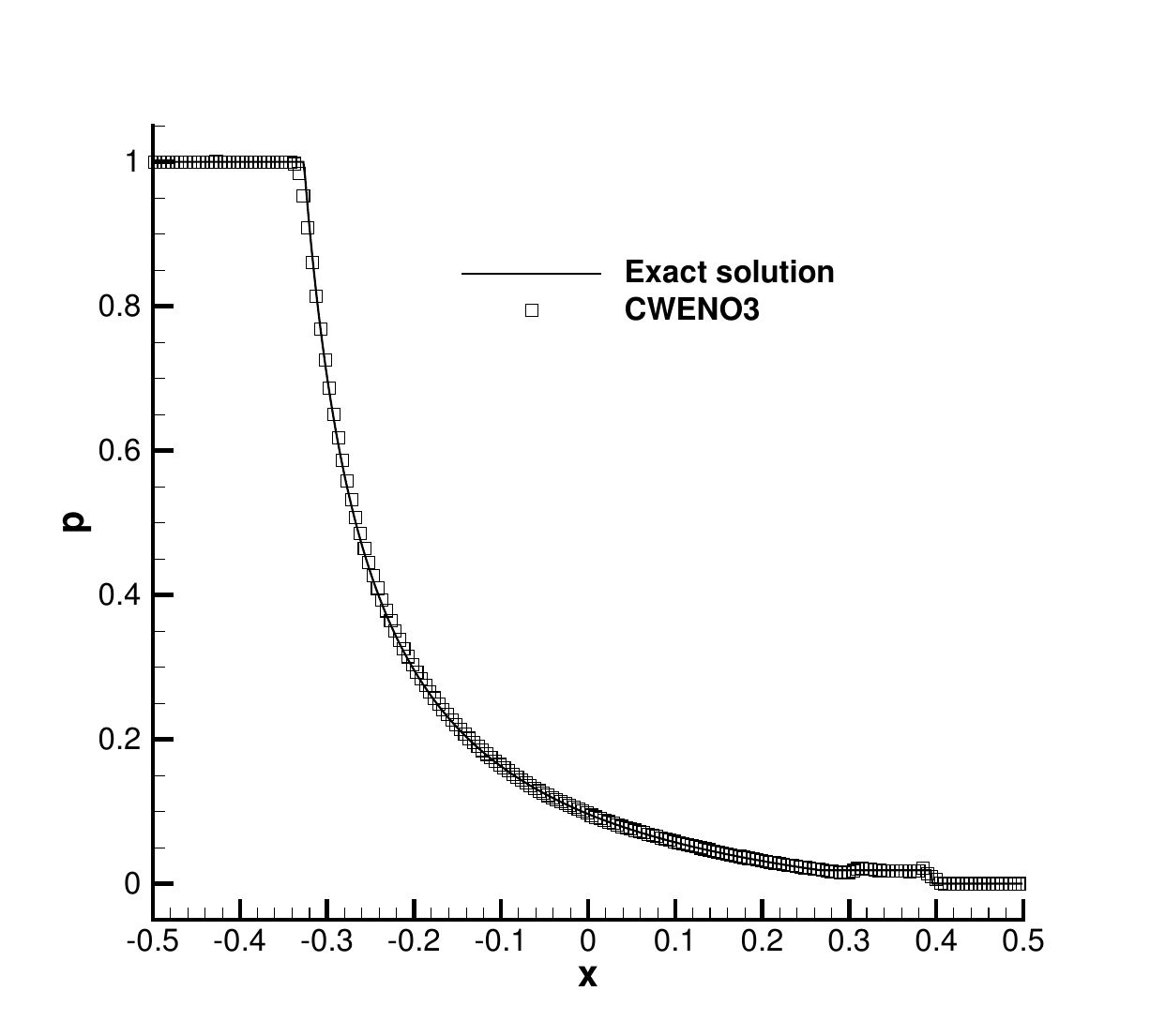}}
\caption{Solution of  Riemann Problem 2 at time $t=0.4$.}
\label{fig:shock-tube-RS}
\end{figure}

\begin{figure}
{\includegraphics[angle=0,width=5.5cm,height=5.0cm]{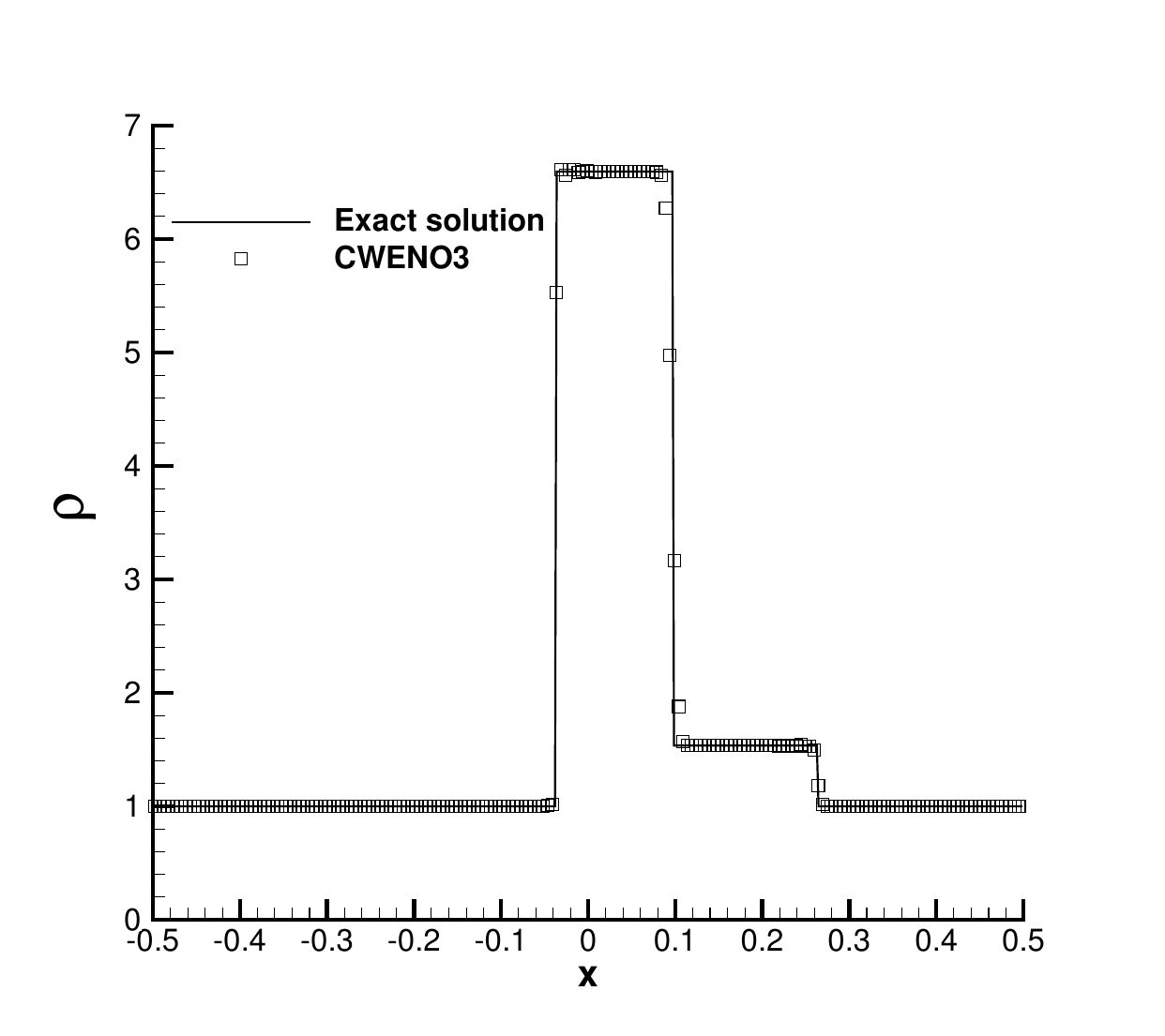}}
{\includegraphics[angle=0,width=5.5cm,height=5.0cm]{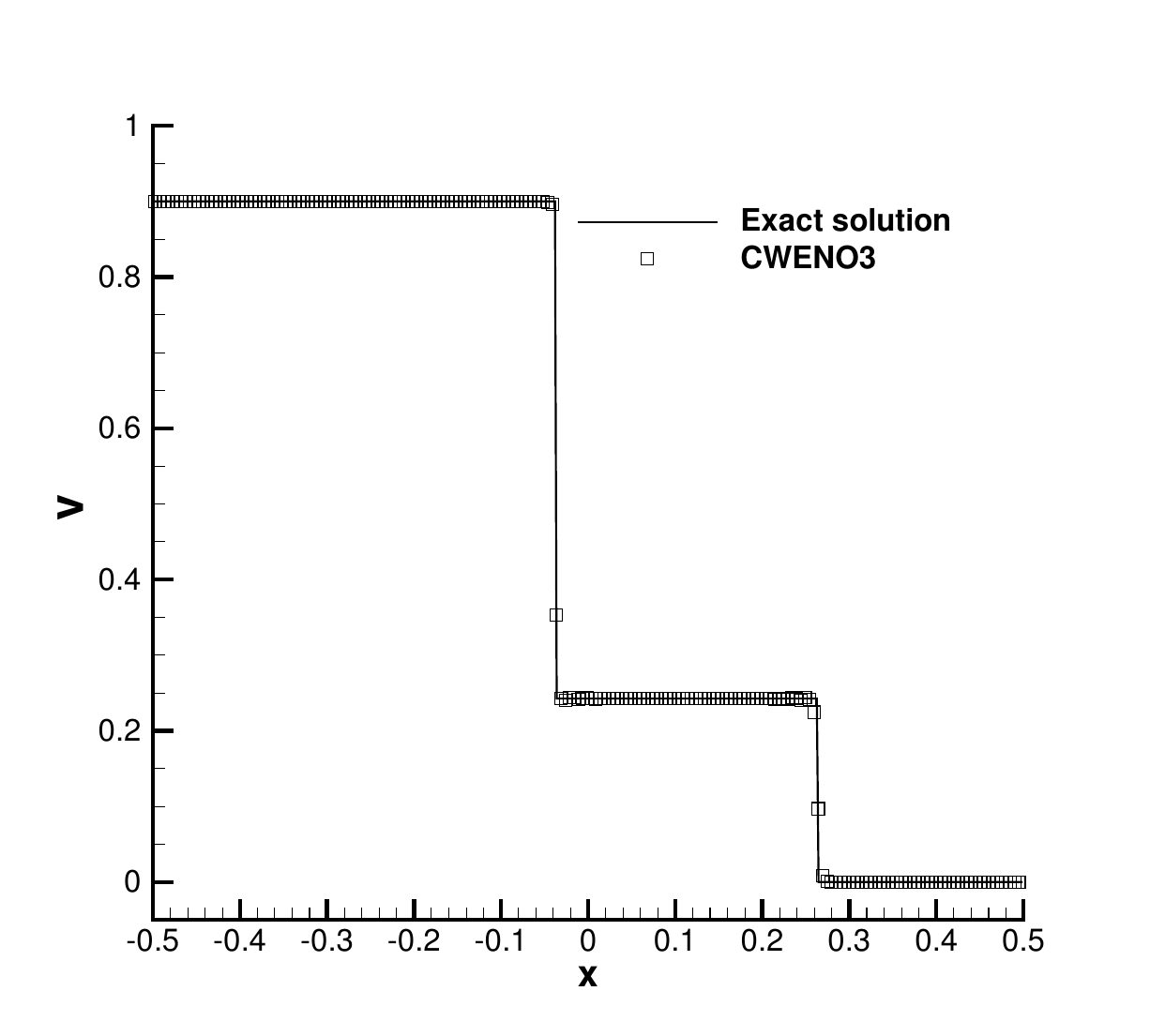}}
{\includegraphics[angle=0,width=5.5cm,height=5.0cm]{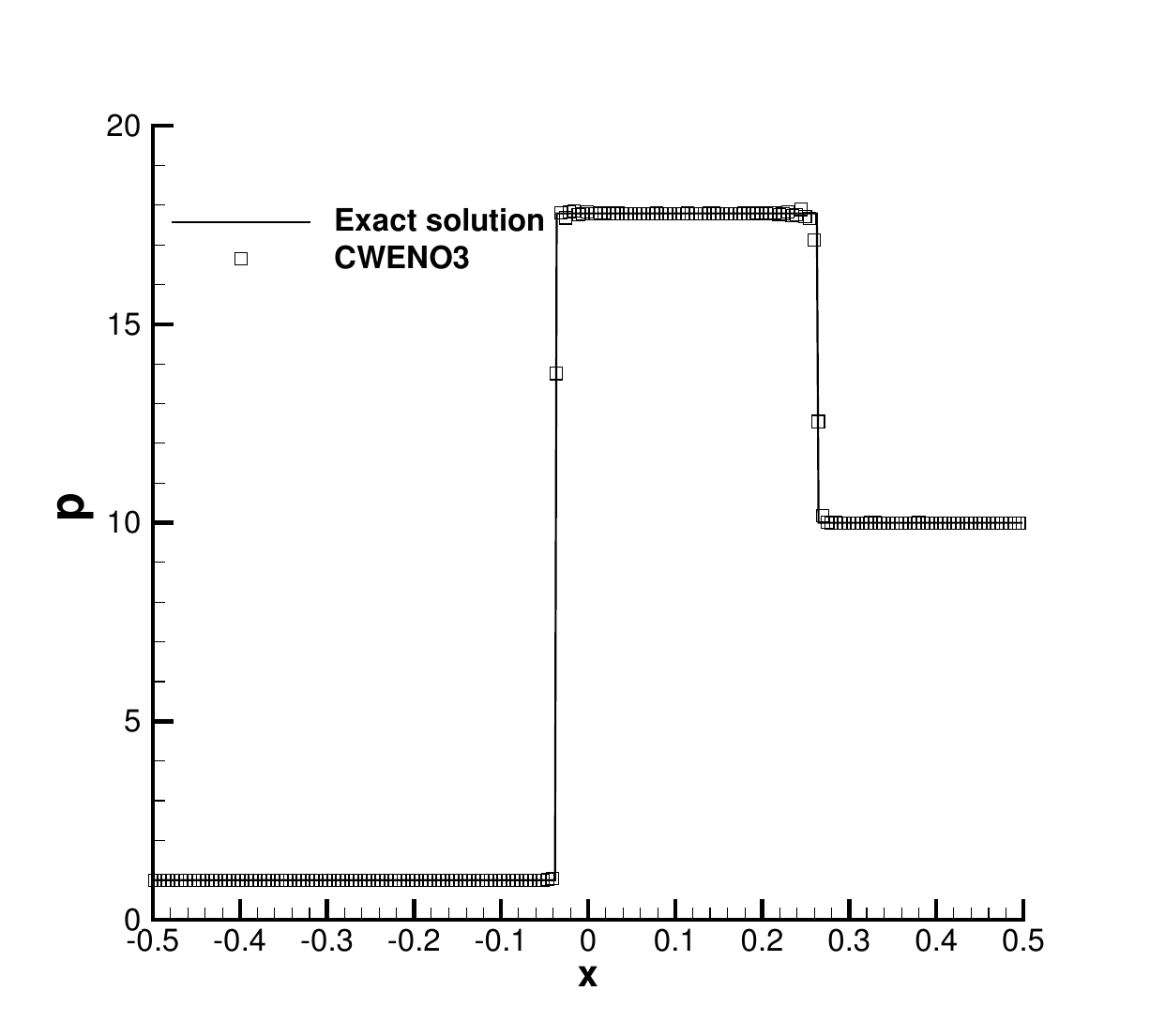}}
\caption{Solution of  Riemann Problem 3 at time $t=0.4$.}
\label{fig:shock-tube-2S}
\end{figure}

The corresponding solutions  are shown in Fig.~\ref{fig:shock-tube-2R}--\ref{fig:shock-tube-2S}, and they are schematically described as
\begin{itemize}
\item Riemann Problem 1, already considered by \cite{Mignone2005}, produces a rarefaction wave propagating to the left, a contact discontinuity and  a second rarefaction wave propagating to the right.
\item Riemann Problem 2, already considered by \cite{Radice2012a}, produces a rarefaction wave propagating to the left, a contact discontinuity and  a shock wave propagating to the right.
\item Riemann Problem 3, already considered by \cite{Zhang2006}, produces a shock wave propagating to the left, a contact discontinuity and  a second shock wave propagating to the right.
\end{itemize}
All together, these calculation prove the ability of the numerical scheme in treating strong discontinuities in the relativistic regime.
\begin{table}[!b]
\caption{Initial left (L) and right (R) states of the Riemann problems. 
  The last column reports the final time $t_f$ of the simulation.} 
\vspace{0.5cm}
\renewcommand{\arraystretch}{1.0}
\begin{center}
\begin{tabular}{ccccccccc}
\hline
\hline
 Problem  & $\rho_L$  & $v_L$ & $p_L$  & $\rho_R$  & $v_R$ & $p_R$      & $\gamma$ & $t_f$ \\
\hline
1         & 1         & -0.6  & 10     & 10        & 0.5   &  20        & 5/3      & 0.4   \\
\hline
2         & $10^{-3}$ & 0     & 1      & $10^{-3}$ & 0     &  $10^{-5}$ & 5/3      & 0.4   \\
\hline
3         & 1         & 0.9   & 1      & 1         & 0     &  10        & 4/3      & 0.4   \\
\end{tabular}
\end{center}
\label{tab.RP.ic}
\end{table}

%
%
%
\subsection{Michel accretion }
\label{sec:Michel}

\begin{figure}[!htbp]
	\begin{center}
	{\includegraphics[angle=0,width=7.3cm,height=7.3cm]{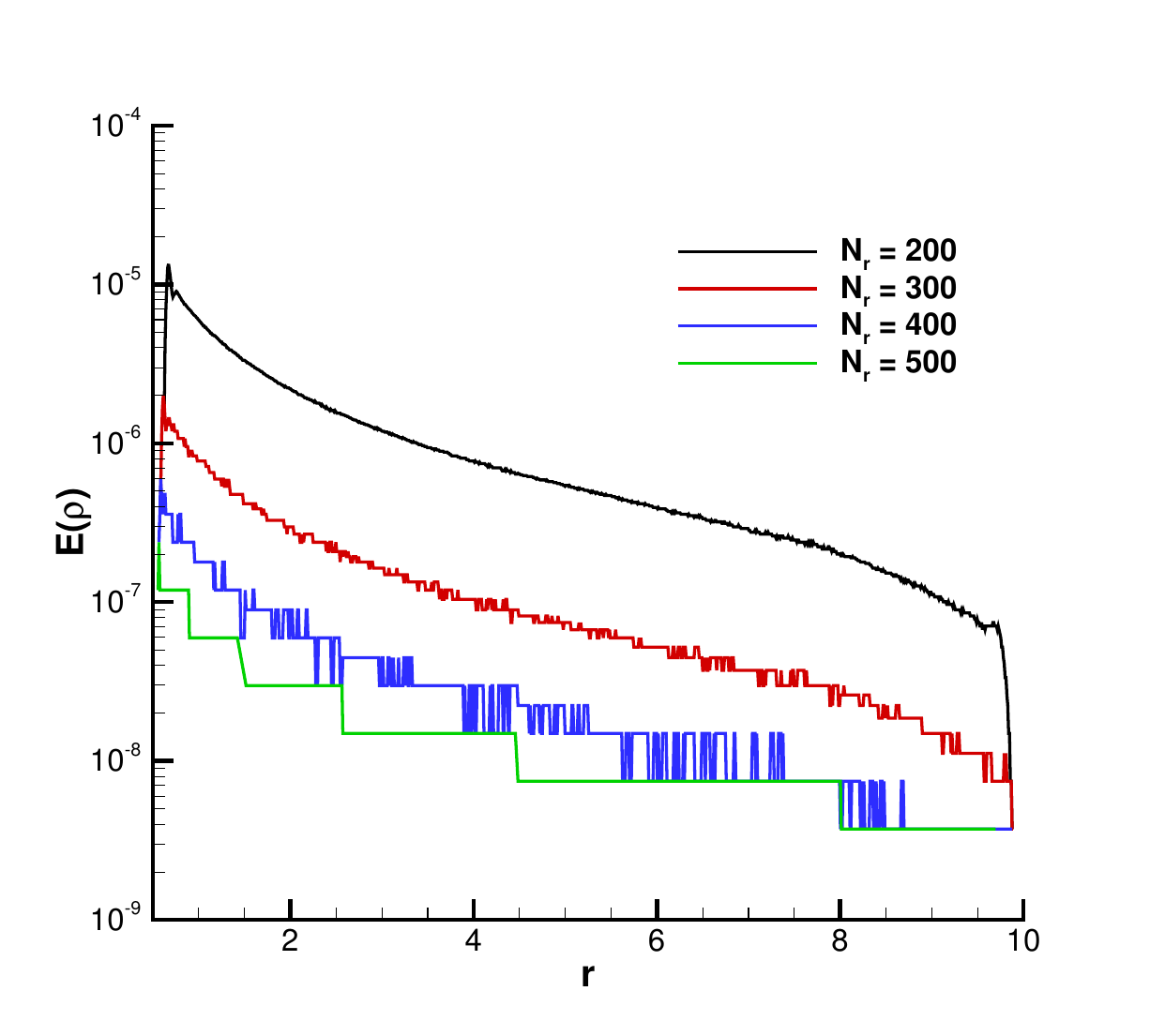}} 
\captionof{figure}{Spherical accretion of matter onto a Schwarzschild black hole using Kerr-Schild coordinates. Absolute error at different resolutions for the CWENO5 scheme.} 
\label{fig:SA-1D}
	\end{center}
\end{figure}

Again in the Cowling approximation,
we can consider the motion of a gas that is accreting onto a non rotating black hole through a spherically symmetric
stationary flow. A detailed presentation of the solution can be found  
in~\cite{Rezzolla_book:2013,Anton06}. 
We just note here that in Kerr--Schild coordinates the computation of the quantities at the critical radius follows the
same relations valid in Schwarzschild coordinates. For example, using as free parameters the critical radius $r_c$ and 
the critical density $\rho_c$, the $r-$component of the four velocity at the critical radius is still given by $|u^r_c|=1/\sqrt{2r_c}$,
just like in Schwarzschild coordinates. What changes, though, is the value of $u^t$, which follows from the normalization
condition $u^\mu u_\mu=-1$ and is given by
\begin{equation}
u^t  = \frac{-z u^r - \sqrt{(u^r)^2 - z + 1}}{z - 1}\hspace{1cm}{\textrm{with}}\,\,\,\,\,u^r<0\,,\,\,\,\,\,z=2/r\,.
\end{equation}
We have solved this test on a 
computational domain  $(r,\theta)\in [0.5;10]\times[0+\epsilon;\pi-\epsilon]$, with $\epsilon=0.05$ and covered by a uniform grid.
The relevant parameters are chosen as
$r_c=8$ and $\rho_c=1/16$ like in \cite{DelZanna2007}, while an ideal gas with adiabatic index $\gamma=5/3$ is assumed. No well--balancing is used for this test, while analytic boundary conditions are used both at the inner and at the outer radii.
Fig.~\ref{fig:SA-1D} reports the absolute errors (with respect to the exact solution) 
computed at $t=1000$ while increasing the number $N_r$ of radial gridpoints for the CWENO5 scheme, thus showing mesh convergence. In addition, Tab.~\ref{tab:SphericalAccretion} shows the high order of convergence in a series of companion runs.

\begin{table}[!b]
	\caption{Numerical convergence results for the spherical accretion
			of matter onto a Schwarzschild black hole at a final time  $t=10$.  The $L^2$
			errors and the corresponding order of convergence  are reported for
			the variable $\rho$.} 
	\vspace{0.5cm}
	\renewcommand{\arraystretch}{1.0}
	\begin{center}
	\begin{tabular}{ccc|ccc}
	\hline
	$N_r$ & ${L^2}$ error $\rho$ & $\mathcal{O}(\rho)$  & 
	$N_r$ & ${L^2}$ error $\rho$ & $\mathcal{O}(\rho)$  \\
	\hline
	\multicolumn{3}{c}{CWENO5}  & \multicolumn{3}{|c}{CWENO7}  \\
	\hline
	$200$   & 4.9639E-05   &       &  $200$  & 2.7232E-07  &     \\
	$300$   & 6.6499E-06   & 4.96  &  $300$  & 1.9132E-08  & 6.55 \\
	$400$   & 1.5811E-06   & 4.99  &  $400$  & 3.0359E-09  & 6.40 \\
	$500$   & 5.1810E-07   & 5.00  &  $500$  & 7.6260E-10  & 6.19\\
	\hline
\end{tabular}
	\end{center}
\label{tab:SphericalAccretion}
\end{table}

\subsection{The equilibrium TOV star }
\label{sec:TOV}
Another fundamental test in numerical general relativity consists of keeping 
an equilibrium star model stationary over long timescales, providing the first case in our sample of tests where the full Einstein-Euler equations are addressed. 
Hence, we have first solved the
standard Tolman--Oppenheimer--Volkoff (TOV) system~\cite{Tolman,Oppenheimer39b,Rezzolla_book:2013,Camenzind2007} 
for a simple polytropic gas obeying $p= K\,\rho^\gamma$.

After choosing the parameters as in~\cite{Font2002}, namely
a central rest mass density 
$\rho_c=1.28\times 10^{-3}$, $K=100$ and $\gamma=2$,
we have integrated the ODEs of the TOV system by a Discontinuous Galerkin solver like in \cite{DumbserZanottiGaburroPeshkov2023}, obtaining
a total mass $M=1.4\,M_{\odot}$ and a radius $R=14.15\,{\textrm {km}}$. 
A small perturbation in the energy density is then added to the initial model, which is evolved 
with the  CWENO7 version of the scheme. 
The logical  computational domain is given by the box $\Omega_{\ell} = [-15;15]^3$,
covered by a uniform grid formed by $100\times100\times100$ points. 
This is also the first test 
where we have used the re-mapping algorithm for the numerical grid described in Sect.~\ref{sec:grid},
producing a physical computational domain that extends up to $\sim 300 M$ in each direction. 
In particular,
the parameters of the polynomial stretching in the grid
function of Eq.~\eqref{coordmap} 
are $a = 2.28$, $b = -68.4$, $c = 685$, $d = -2280$, $\xi_c=10$.

\begin{figure}[!t]
	\begin{center}
		\begin{tabular}{cc} 
			{\includegraphics[angle=0,width=7.3cm,height=7.3cm]{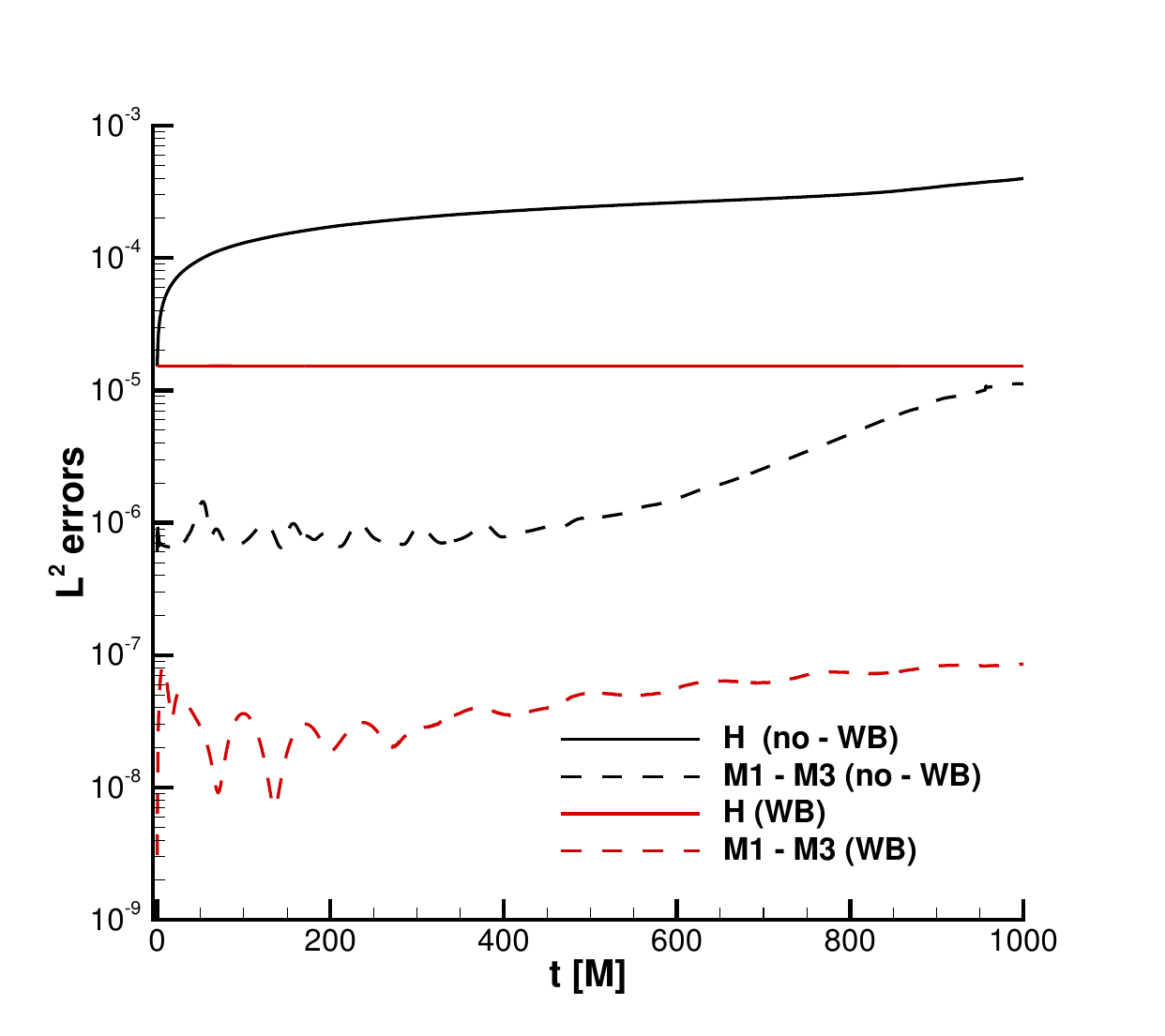}} &
			{\includegraphics[angle=0,width=7.3cm,height=7.3cm]{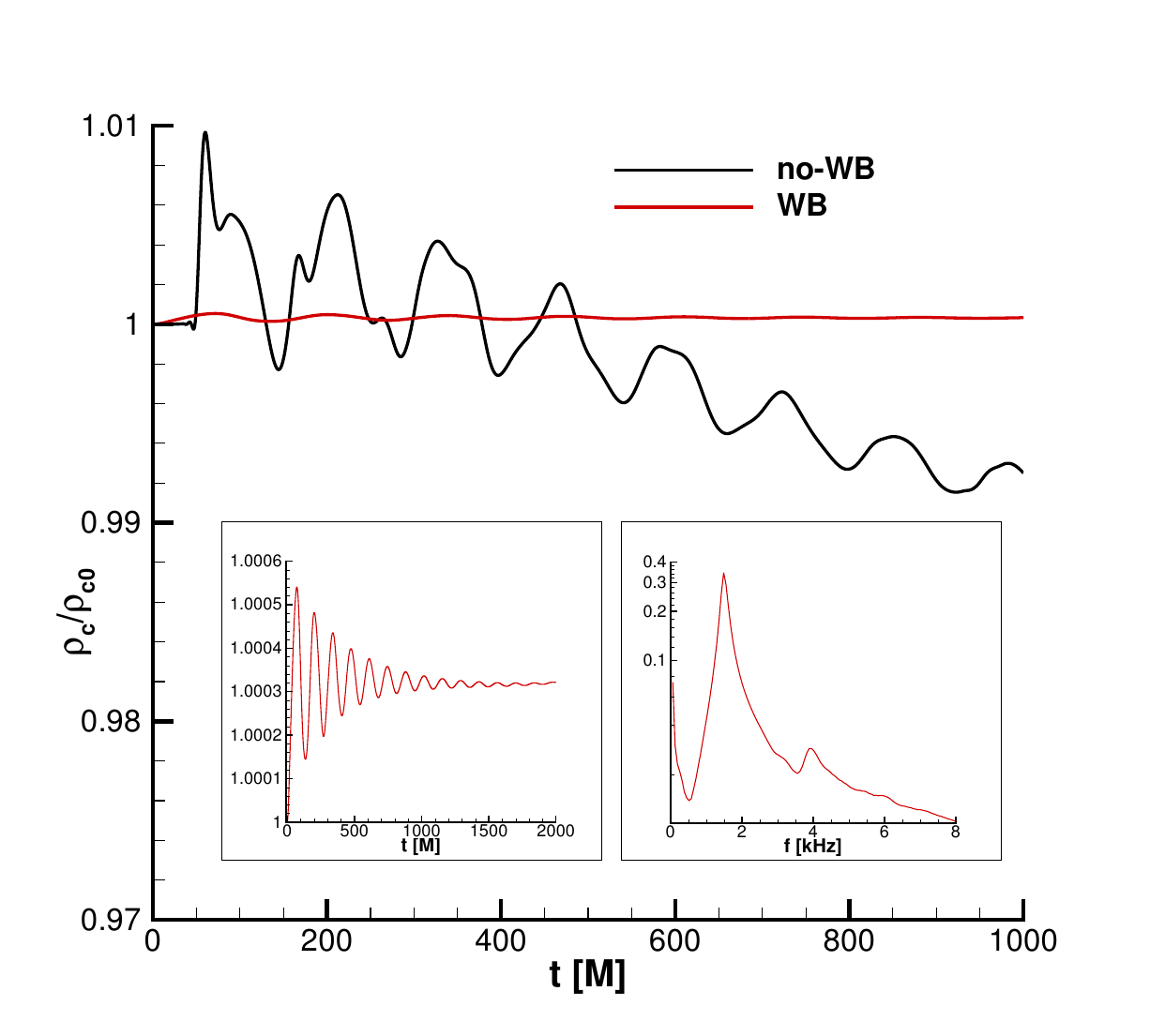}} 
		\end{tabular} 
		\caption{Time evolution of the Einstein constraints (left panel) and of the central density (right panel) for the 3D TOV star solved with the CWENO7 scheme. A comparison among the Well Balanced (WB) scheme and the non Well Balanced (no-WB) is also visible. 
		} 
		\label{fig:TOV-timeseries}
	\end{center}
\end{figure}

We also stress that in this test we take full advantage of the new algorithm proposed by
\cite{DumbserZanottiGaburroPeshkov2023}, which allows to convert from the conserved to the primitive
variables even in a $\rho=0$ atmosphere in the exterior of the star.
Since this is a test for which an exact equilibrium of the Einstein equations is available, the well balancing technique described in Sect.~\ref{sec:wb} can be activated.
To show the advantages of this approach, 
in Fig.~\ref{fig:TOV-timeseries} we compare the results obtained  with and without WB, 
by reporting the evolution of the normalized Einstein constraints (left panel) and of the central density 
(right panel) in the two cases. We plan to perform a systematic study about the benefits of WB
on the extraction of the NS oscillation modes in a future publication.
For the moment, we have just performed a simple Fourier Transform of the central density obtained through the WB evolution, getting the frequencies that are reported in the bottom right inset of Fig.~\ref{fig:TOV-timeseries}. In particular, the first two frequencies are found at $f=1.49\, \rm{kHz},\,3.92 \,\rm{kHz}$, which are in good agreement with those reported in Tab.~II of \cite{Font2002}.
%

\subsection{Single puncture black hole }
\label{sec:Trumphet}

As next test case we consider a single puncture black hole, as originally proposed in \cite{Brandt1997} (the so--called {\emph{puncture (or trumpet)}} solution). 
The initial condition for the conformal metric tensor is simply the identity matrix, i.e. $\tilde{\gamma}_{ij} = \delta_{ij} = \mathbf{I}$, while the conformal factor at the initial time is set to $\Psi=1+M/(2r)$ with $r = \left\| \mathbf{x} - \mathbf{x}_c \right\|$.  
The black hole is centered in the origin $\mathbf{x}_c=(0,0,0)$, has unit mass $M=1$ and  zero spin. 
The initial extrinsic curvature and the initial shift are set to zero, i.e. $K_{ij}=0$, $\beta^i=0$, while the lapse is initialized with $\alpha = \psi^{-2}$. 
The (logical) three dimensional computational domain is given by $\Omega_{\ell} = [-13, 13]^3$ covered by  $132^3$ gridpoints with grid-stretching activated. 
The parameters of the polynomial stretching in the grid
function \eqref{coordmap} are $a = 0.32$, $b = -4.8$, $c = 25.0$, $d = -40.0$, $\xi_c=5$, producing a physical domain of size
$\Omega = [-176, 176]^3$.  
To avoid the singularity in $r=0$, the computational mesh is built in such a way that no grid point coincides with the origin $\mathbf{x}_c$, i.e. the grid points are located in the centers of the equidistant control volumes of the logical grid described above, see also Section \ref{sec:CWENO} for details. 
The \emph{gamma--driver} is turned on, i.e. $s=1$, and the damping parameter in the gamma driver is set to $\eta=2$. We furthermore set $\mu=0$. 
We have solved this problem with the CWENO7 scheme using exponent $r=4$ in Eq.~\eqref{WENOwr},
$\lambda_0 = 10^{10}$, $\lambda_C = 10^6$, and $\lambda_L = \lambda_R = 1$ in Eq.~\eqref{eqn.optimalorderl2}
  and stopping the simulation at $t=1000$. 
	In this test we impose  non-reflecting Sommerfeld boundary conditions at the outer border of the domain.  
  
\begin{figure}[!htbp]
	\begin{center}
		\begin{tabular}{cc}
			\includegraphics[width=0.45\textwidth]{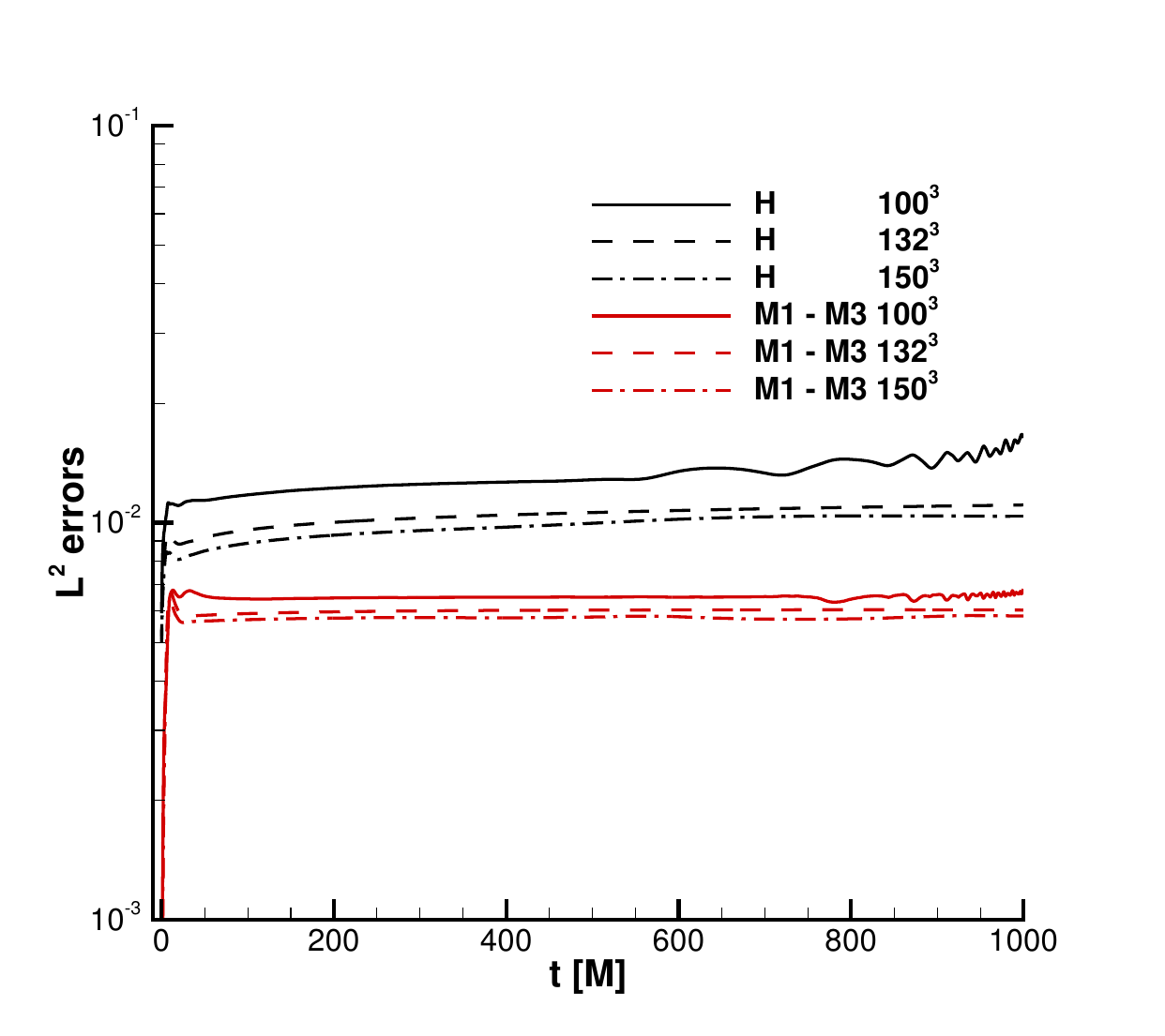} & 
			\includegraphics[width=0.45\textwidth]{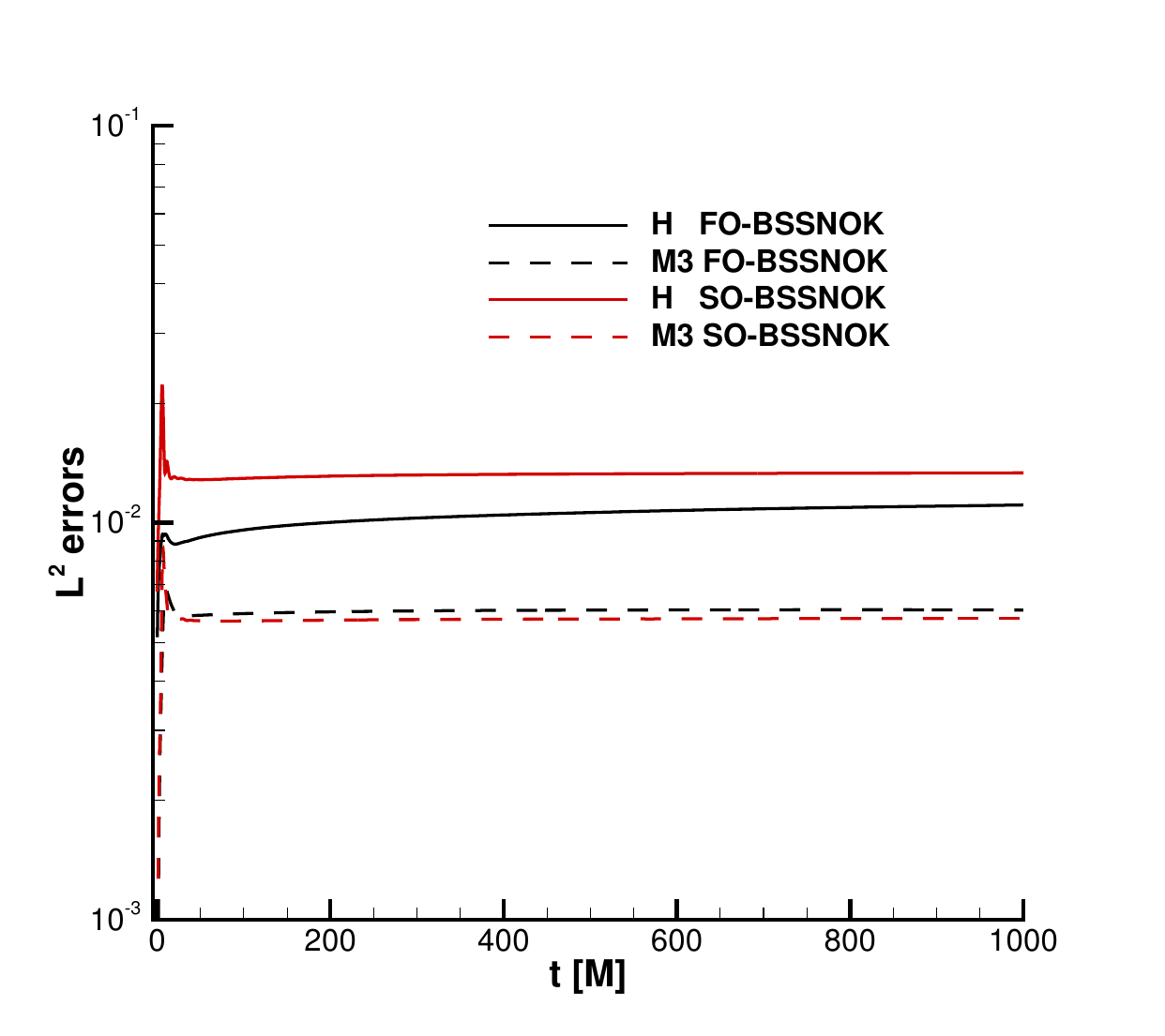}  
		\end{tabular}
		\caption{Single puncture black hole: Time evolution of the normalized Einstein constraints as grid resolution is changed (left panel). Comparison of the normalized Einstein constraints among FO-BSSNOK and SO-BSSNOK (right panel).
		}
		\label{fig.OnePuncture.a}
	\end{center}
\end{figure}

\begin{figure}[!htbp]
	\begin{center}
		\begin{tabular}{cc}
			\includegraphics[width=0.45\textwidth]{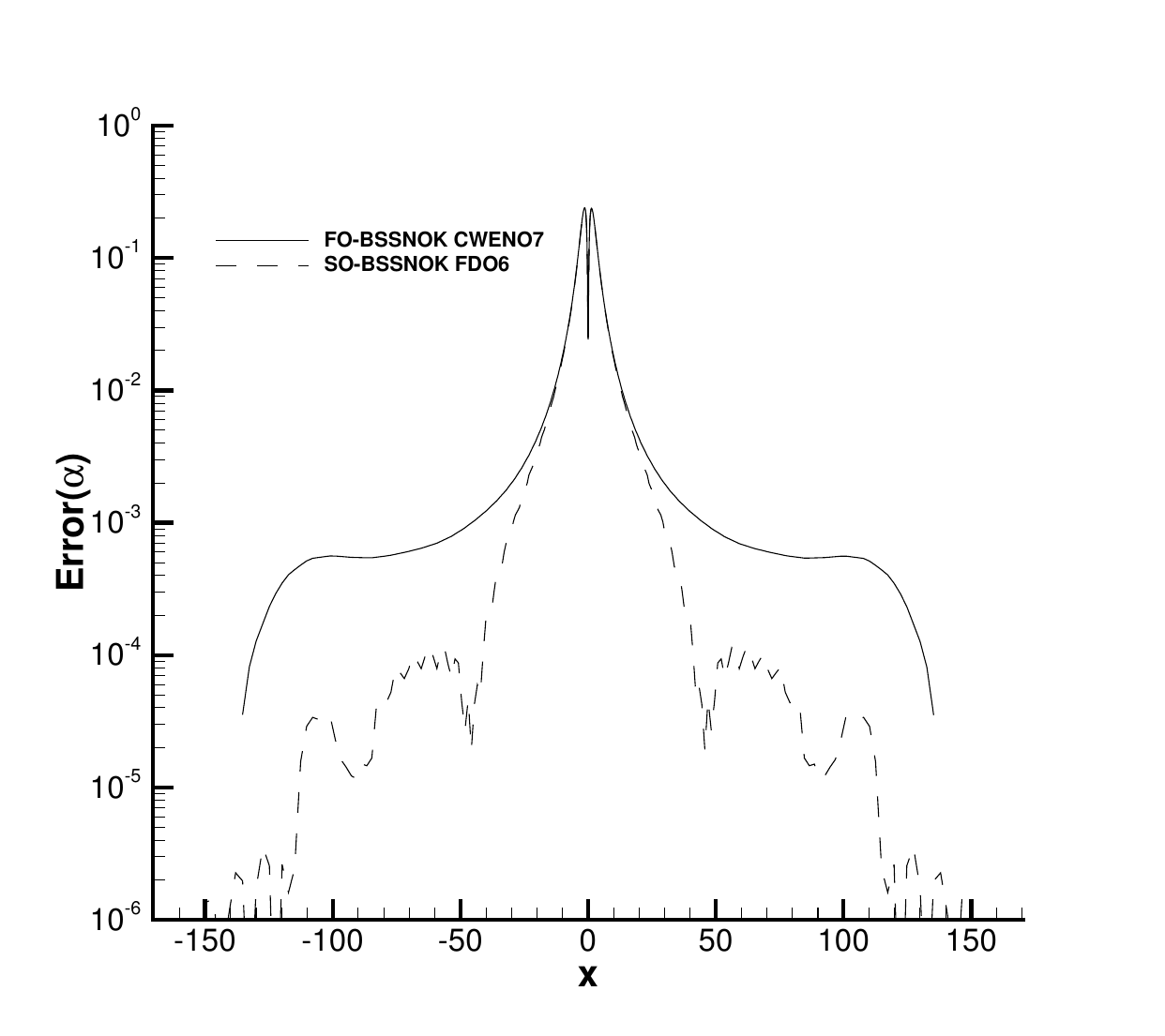} &
			\includegraphics[width=0.45\textwidth]{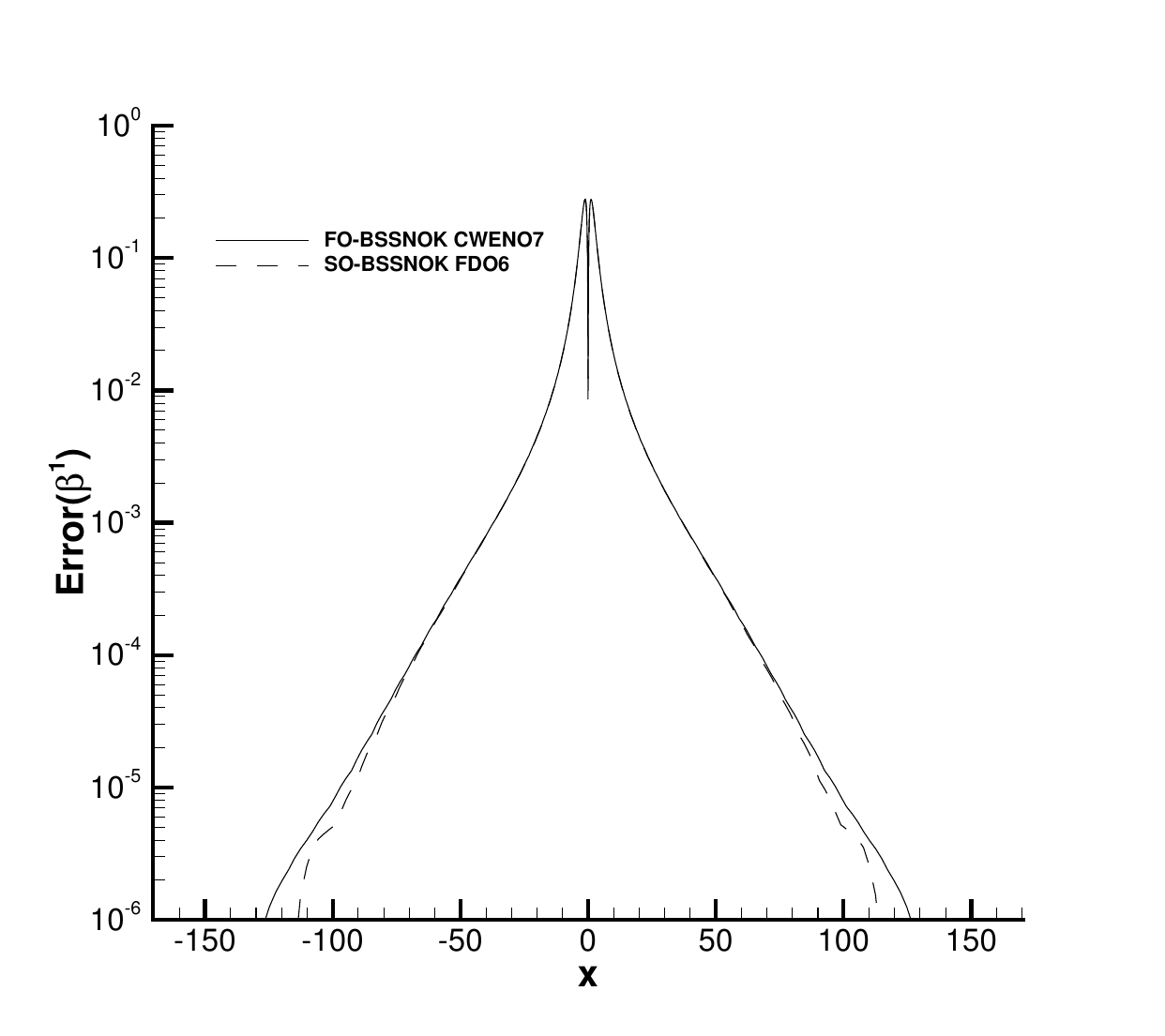} \\
			\includegraphics[width=0.45\textwidth]{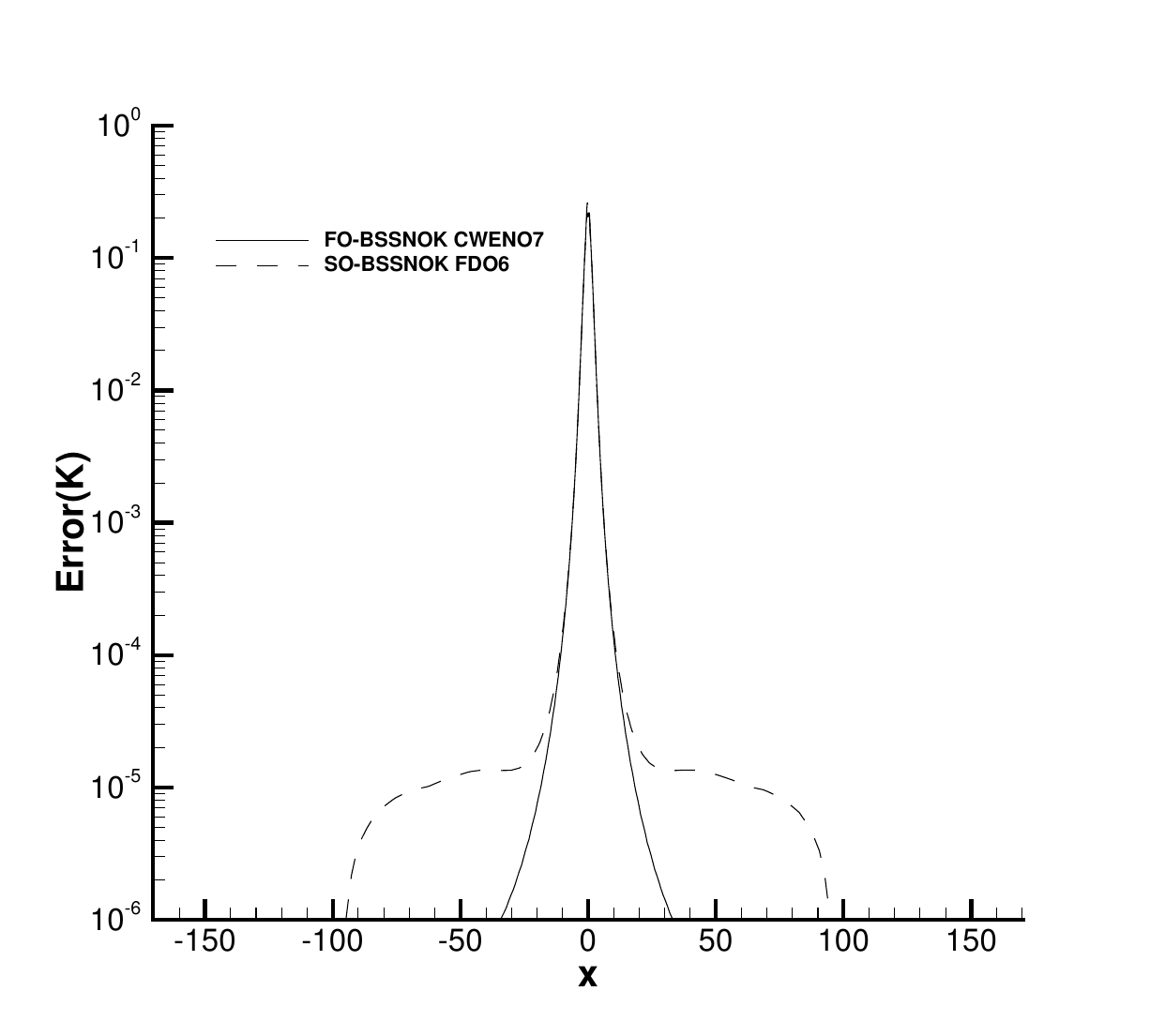} &
			\includegraphics[width=0.45\textwidth]{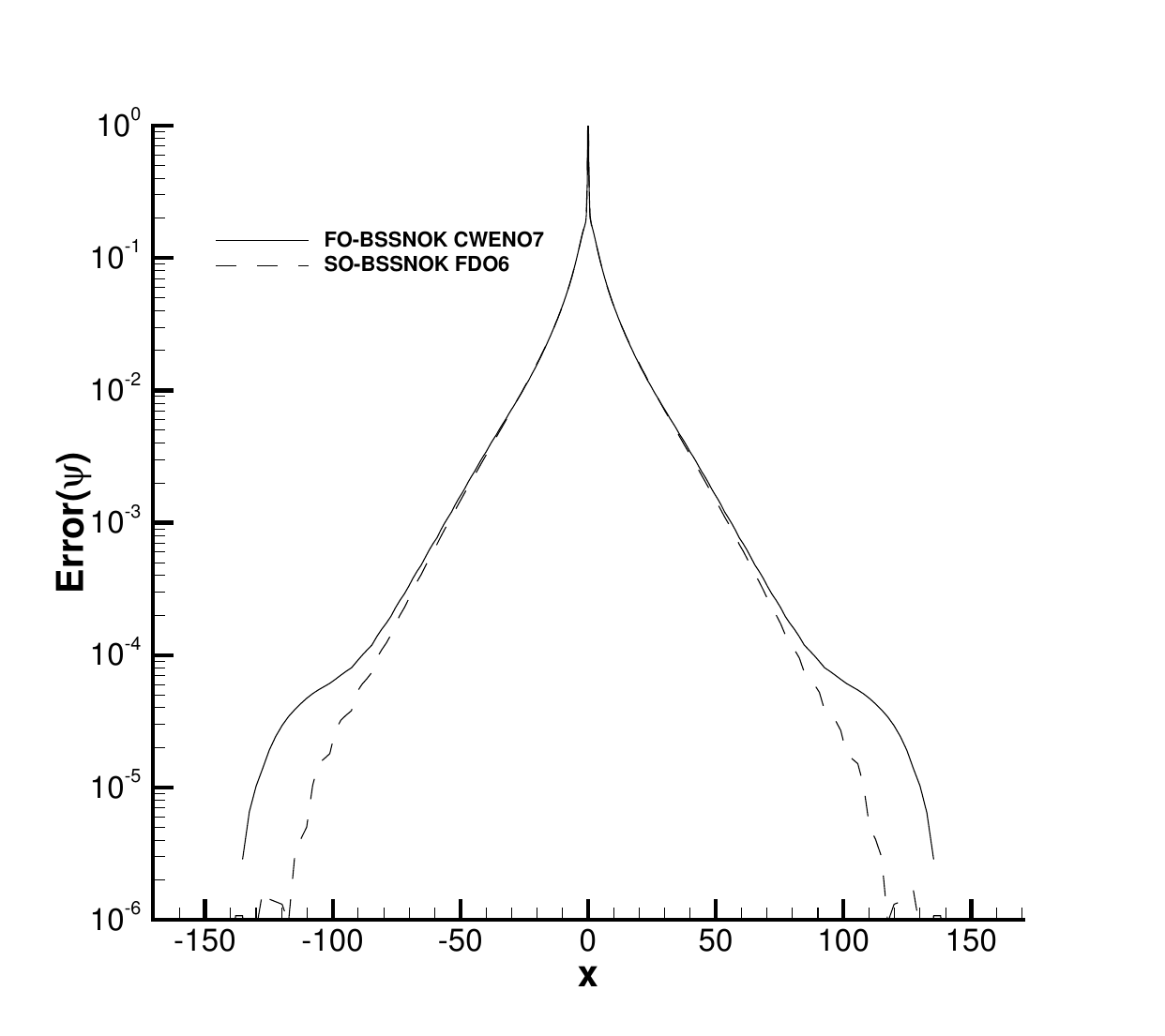} 
		\end{tabular}
\caption{Comparison among the numerical solutions at time $t=1000$ for the single puncture black hole obtained with (i) our new	seventh order CWENO finite difference scheme in the first order FO-BSSNOK system and (ii) a classical sixth order central finite difference scheme applied to the standard second order formulation of BSSNOK (SO-BSSNOK).   
Absolute errors with respect to the initial conditions (computed along			1D cuts)  are shown for a few representative quantities: lapse $\alpha$ (top left), shift vector component $\beta^1$ (top right), trace of the extrinsic curvature $K$ (bottom left) and conformal factor $\psi$ (bottom right). 		}
		\label{fig.OnePuncture.b}
	\end{center}
\end{figure}

The left panel of Fig.~\ref{fig.OnePuncture.a} provides a first glance of the results, by reporting
the time evolution of the Einstein constraints in a series of tests aimed at showing mesh
convergence.
In addition,   
and limited to this test, we have taken the opportunity to perform a close comparison among the results
obtained with our new first--order formulation (FO-BSSNOK) and the much more popular
second--order formulation (SO-BSSNOK).
First of all, in the right panel of
Fig.~\ref{fig.OnePuncture.a} we show the Einstein constraints obtained with the two approaches,
which manifest essentially the same trend.
Furthermore, Fig.~\ref{fig.OnePuncture.b} enters the details of such a comparison, by
showing
the errors of the 1D profiles of  $\alpha$, $\beta^1$, $K$ and $\psi$ using the two schemes, computed at the final time $t=1000$
	with respect to the initial condition.
In particular, we compare	 a seventh order CWENO for FO-BSSNOK
and a sixth order linear central finite difference scheme for SO-BSSNOK.
We note that both methods give pretty much the same error in the inner region close to the puncture.
 
Concerning a quantitative CPU time comparison, the numerical scheme based on the new FO-BSSNOK is more expensive when compared to the classical SO-BSSNOK formulation, the amount being $87\%$ and $82\%$ at the fifth and seventh orders, respectively. This is essentially due to the larger amount of variables that are evolved in the FO formulation.


\subsection{Head on collision of two puncture black holes }
\label{sec:Head}
Here and in the following Section, we use the procedure initially proposed by \cite{Brandt1997,Ansorg:2004ds} to create the initial conditions corresponding to two puncture black holes without excision. 

The initial metric and the lapse are provided by the TWOPUNCTURES initial data code of \cite{Ansorg:2004ds}.
As a first set up, we consider two non--spinning black holes, at rest with respect to each other, and
at an initial distance $d=2\,M$. This is the so--called {\emph{head--on collision}}. 
The \emph{gamma--driver} is necessarily turned on, i.e. $s=1$, again with damping parameter $\eta=2$ and $\mu=0$. We have evolved this configuration with the
CWENO7 scheme, over a logical computational domain  $\Omega_{\ell}=[-15, 15]^3$ covered by  $150^3$ gridpoints, and grid-stretching activated. 
The parameters of the polynomial stretching in the grid
function of Eq.~\eqref{coordmap} 
are $a = 0.32$, $b = -4.8$, $c = 25.0$, $d = -40.0$, $\xi_c=5$, producing a physical domain of size
$\Omega = [-335, 335]^3$. 
Also in this test we impose non-reflecting Sommerfeld boundary conditions at the outer boundaries of the computational domain.

\begin{figure}[!htbp]
	\begin{center}
			\includegraphics[width=0.45\textwidth]{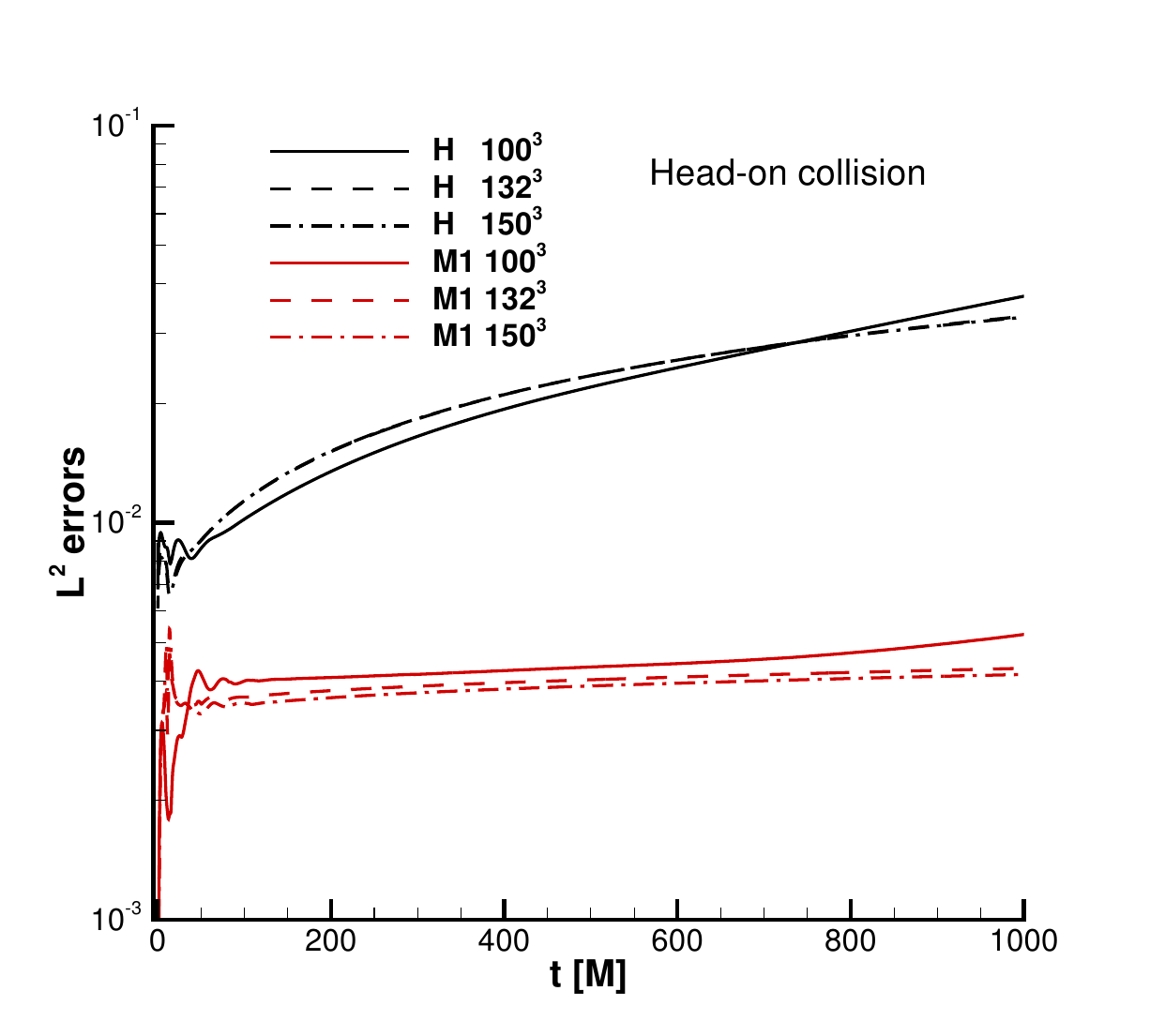} 
		\caption{Time evolution of the normalized Einstein constraints
for the Head on collision of Sect.~\ref{sec:Head}.
		}
		\label{fig.HeadOn}
	\end{center}
\end{figure}

The two black holes merge as expected and at time $t=20$ a single black hole is already formed, that we subsequently evolved until $t=1000$ in order to show long-time stability of our simulation and robustness of both, the underlying mathematical formulation of the equations as well as of the adopted numerical scheme.
In Fig.~\ref{fig.HeadOn} we show the convergence of the normalized Einstein constraints
as resolution is increased.
The Hamiltonian constraint slightly increases in time, but only in a very mild form.

%

\subsection{Inspiralling merger of two black holes}
\label{sec:BBHs}

\begin{figure}[!htbp]
\begin{center}
    \begin{tabular}{cc}
\includegraphics[width=0.45\textwidth]{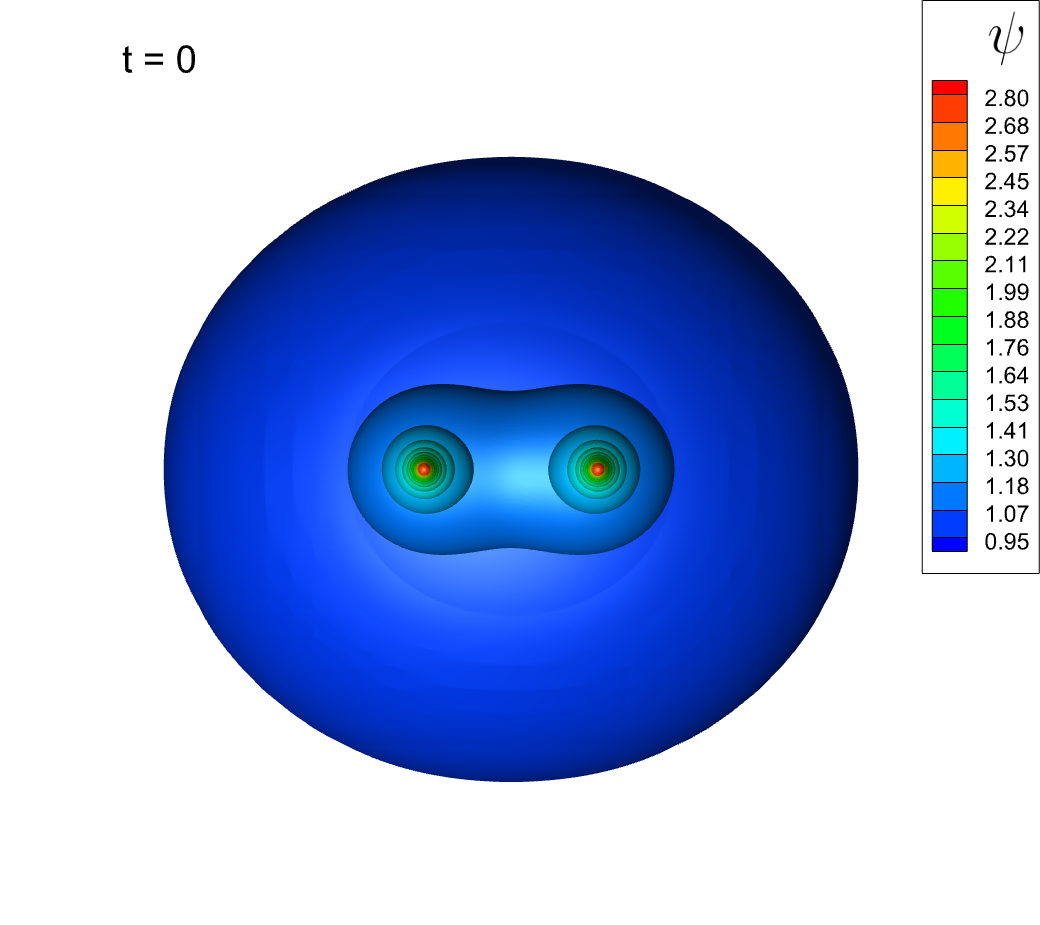} &
\includegraphics[width=0.45\textwidth]{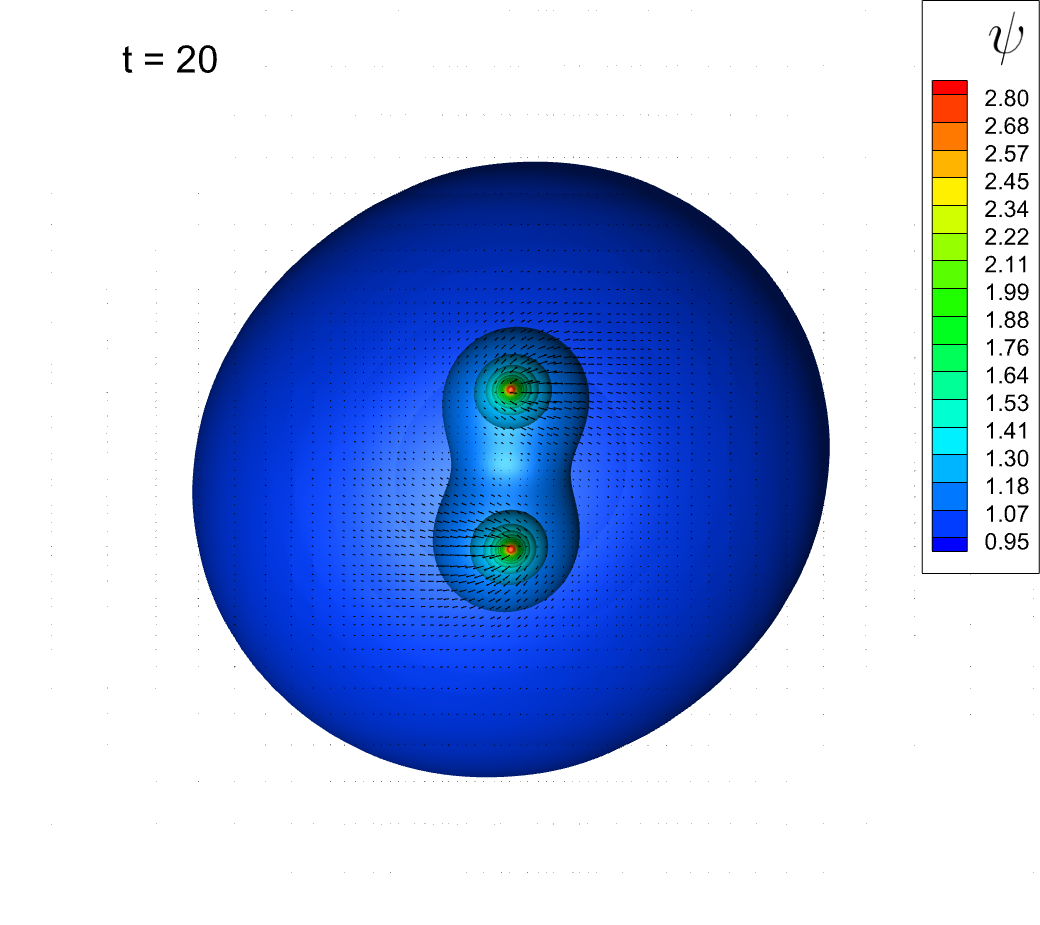} \\
\includegraphics[width=0.45\textwidth]{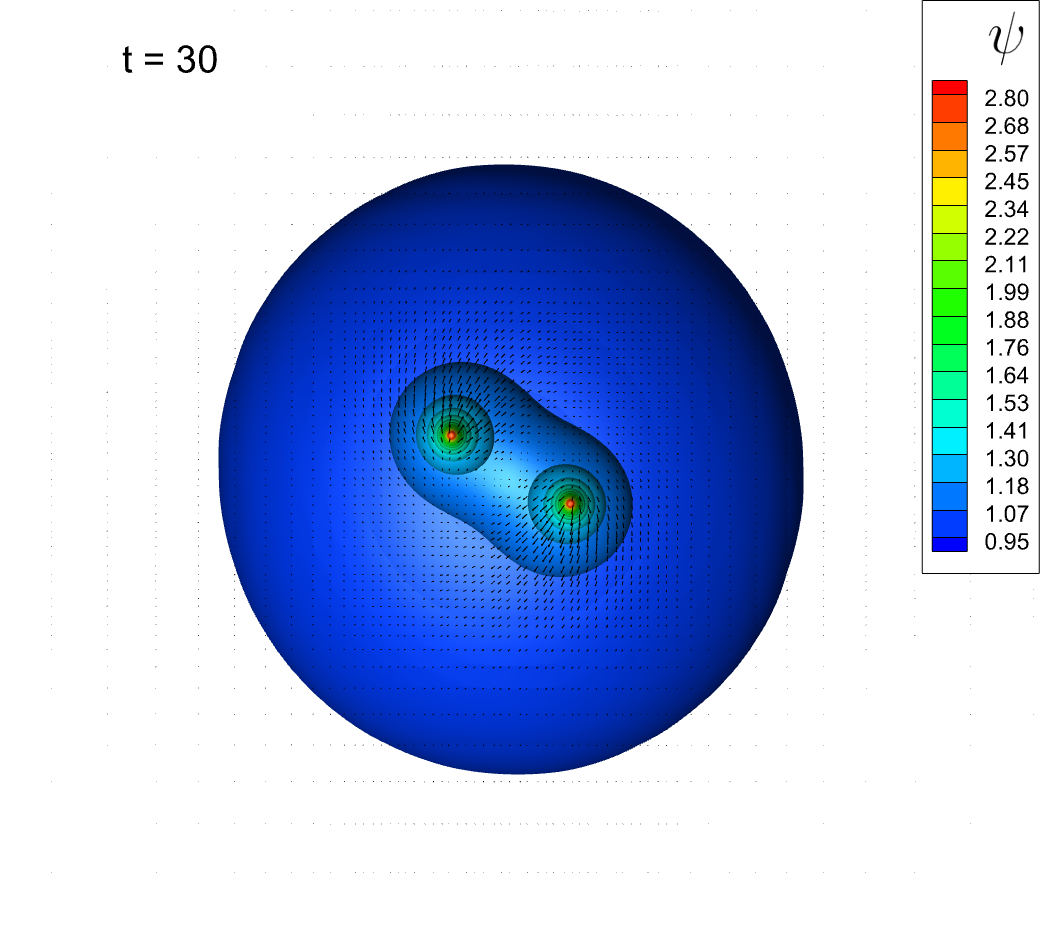} &
\includegraphics[width=0.45\textwidth]{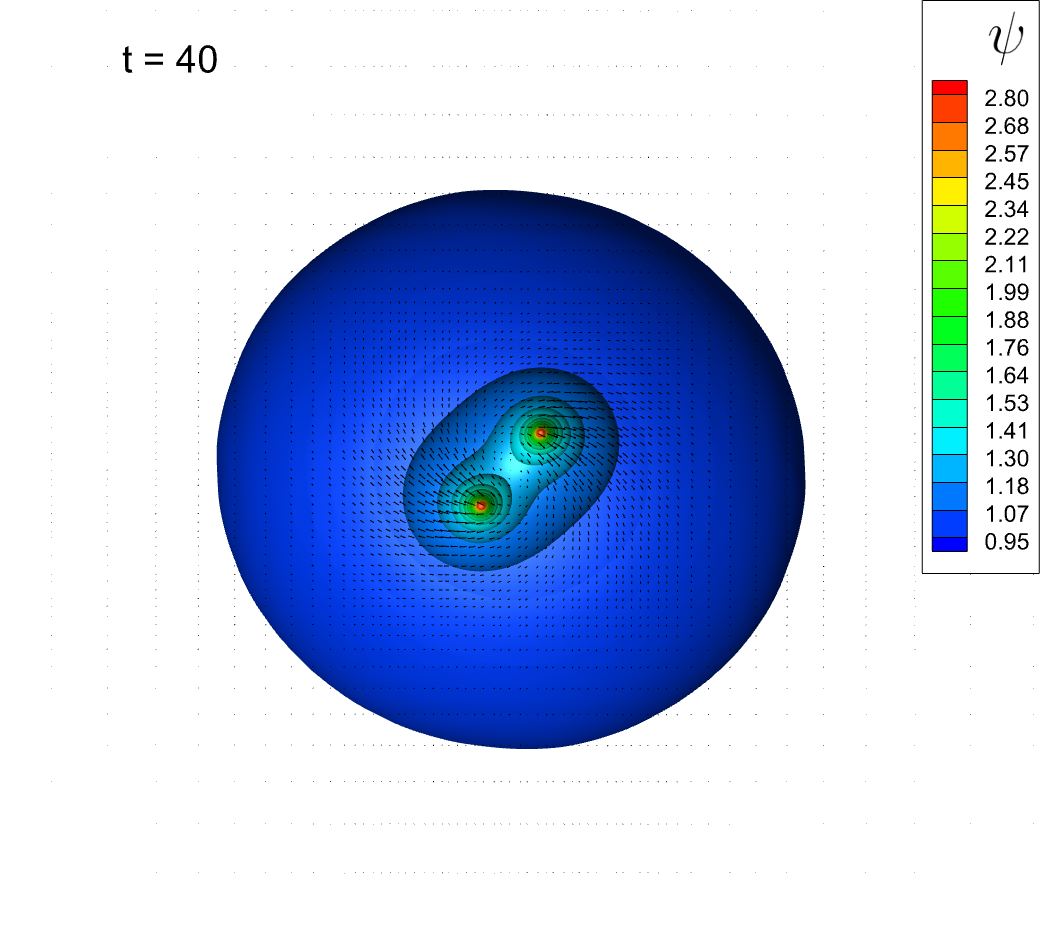} \\
\includegraphics[width=0.45\textwidth]{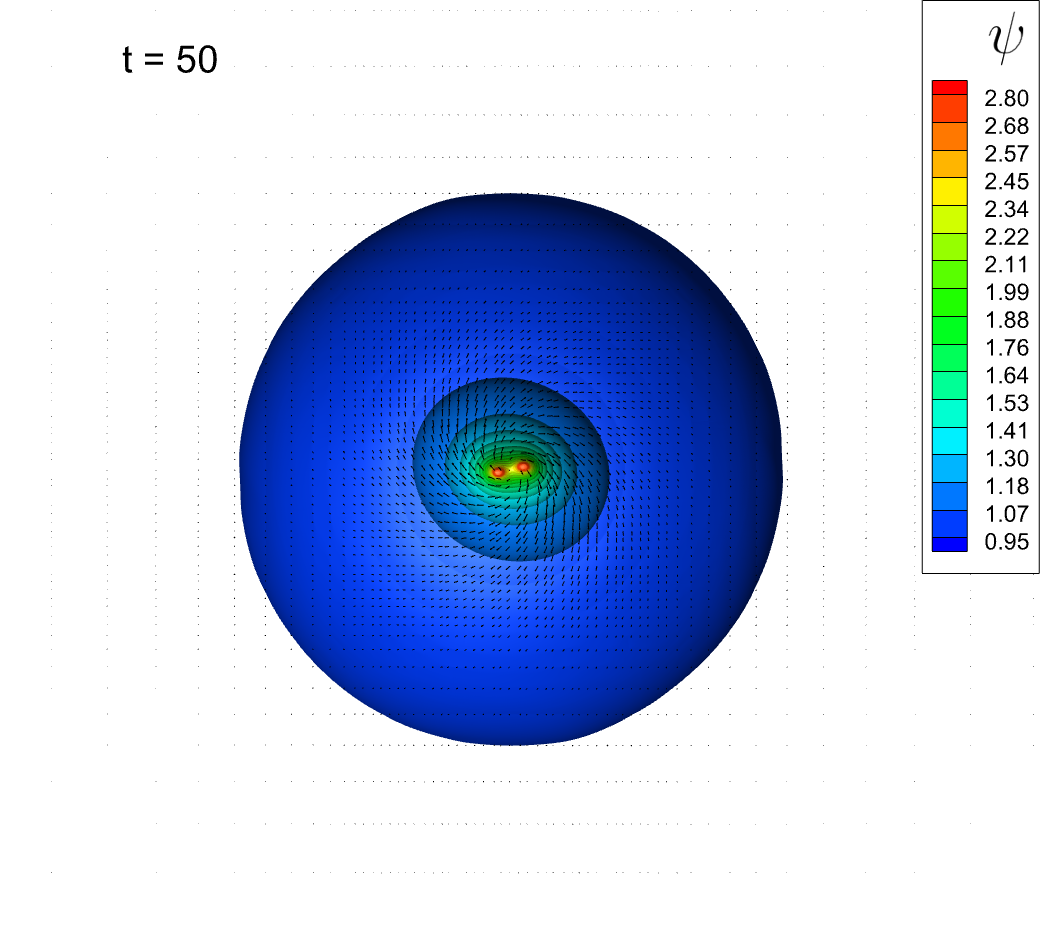} &
\includegraphics[width=0.45\textwidth]{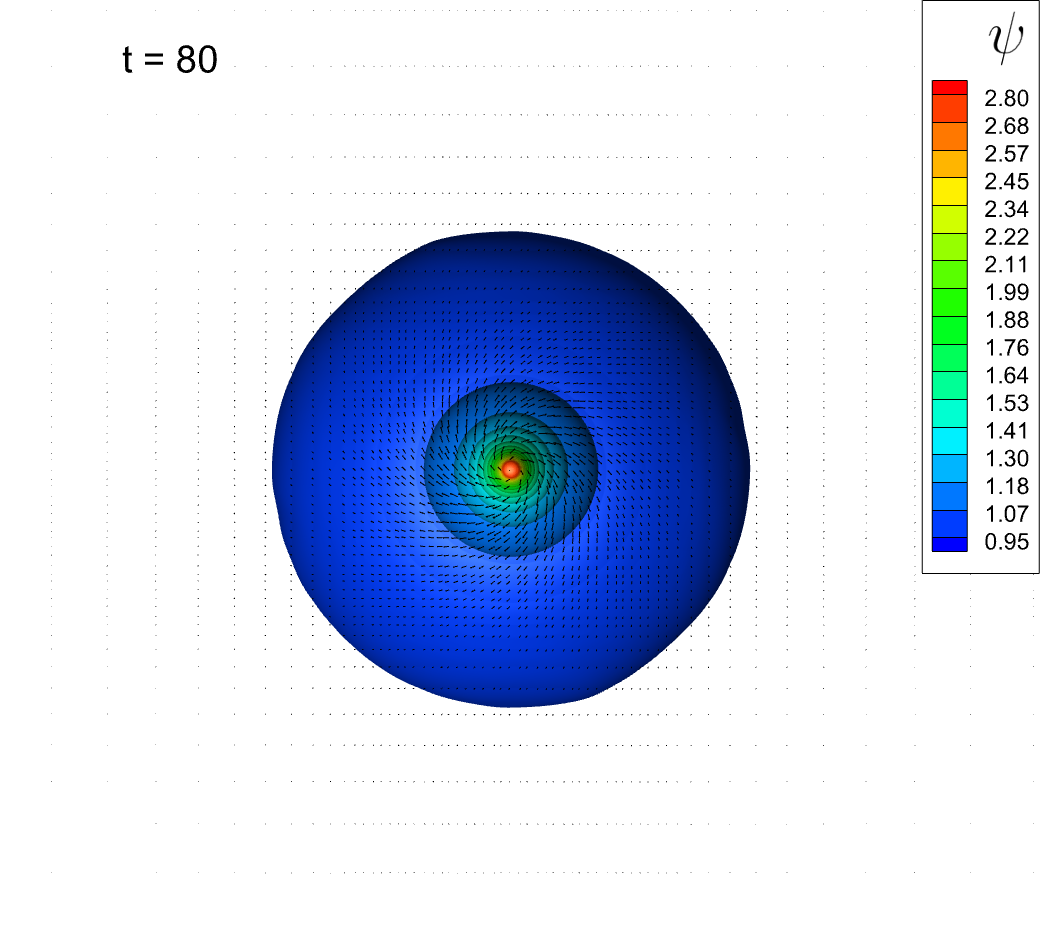} \\
    \end{tabular}
    \caption{Inspiralling merger of two black holes obtained with CWENO7. }
    \label{fig.BinaryMerger}
\end{center}
\end{figure}

\begin{figure}[!htbp]
\begin{center}
    \begin{tabular}{cc}
  \includegraphics[width=0.45\textwidth]{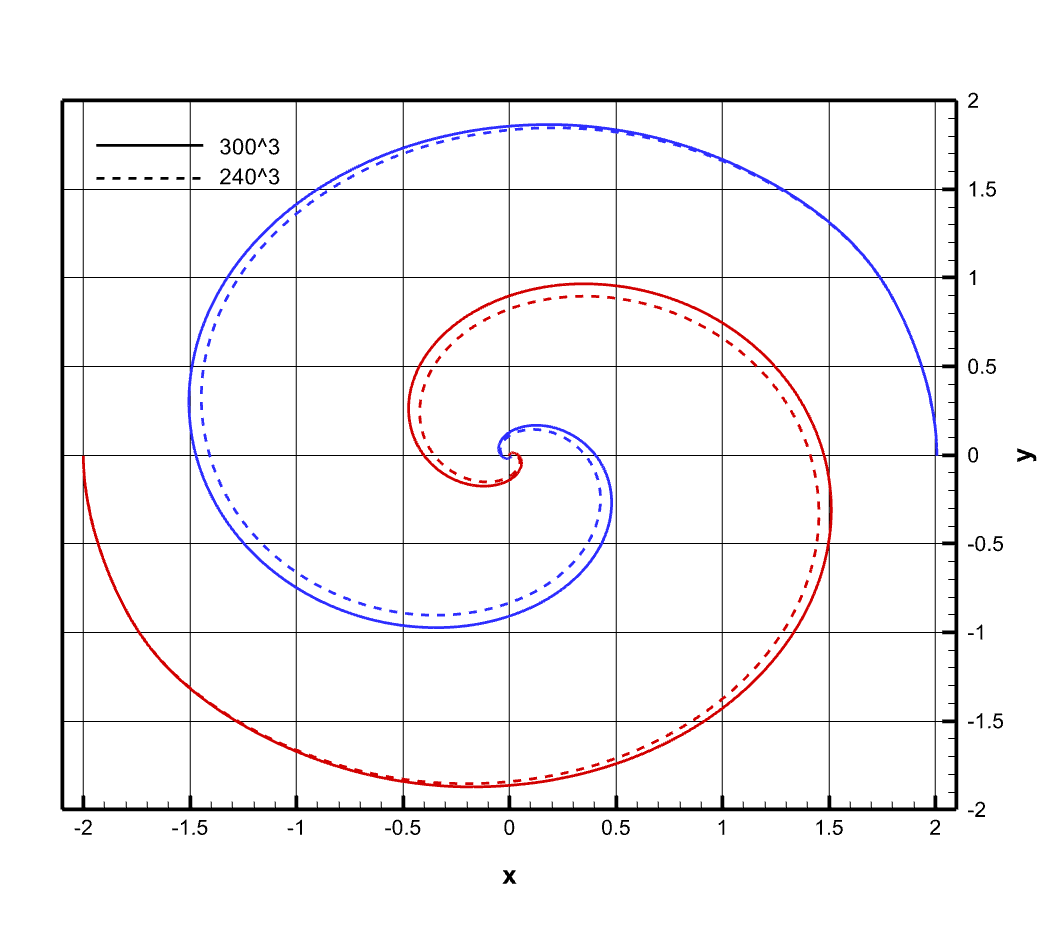} &
  \includegraphics[width=0.45\textwidth]{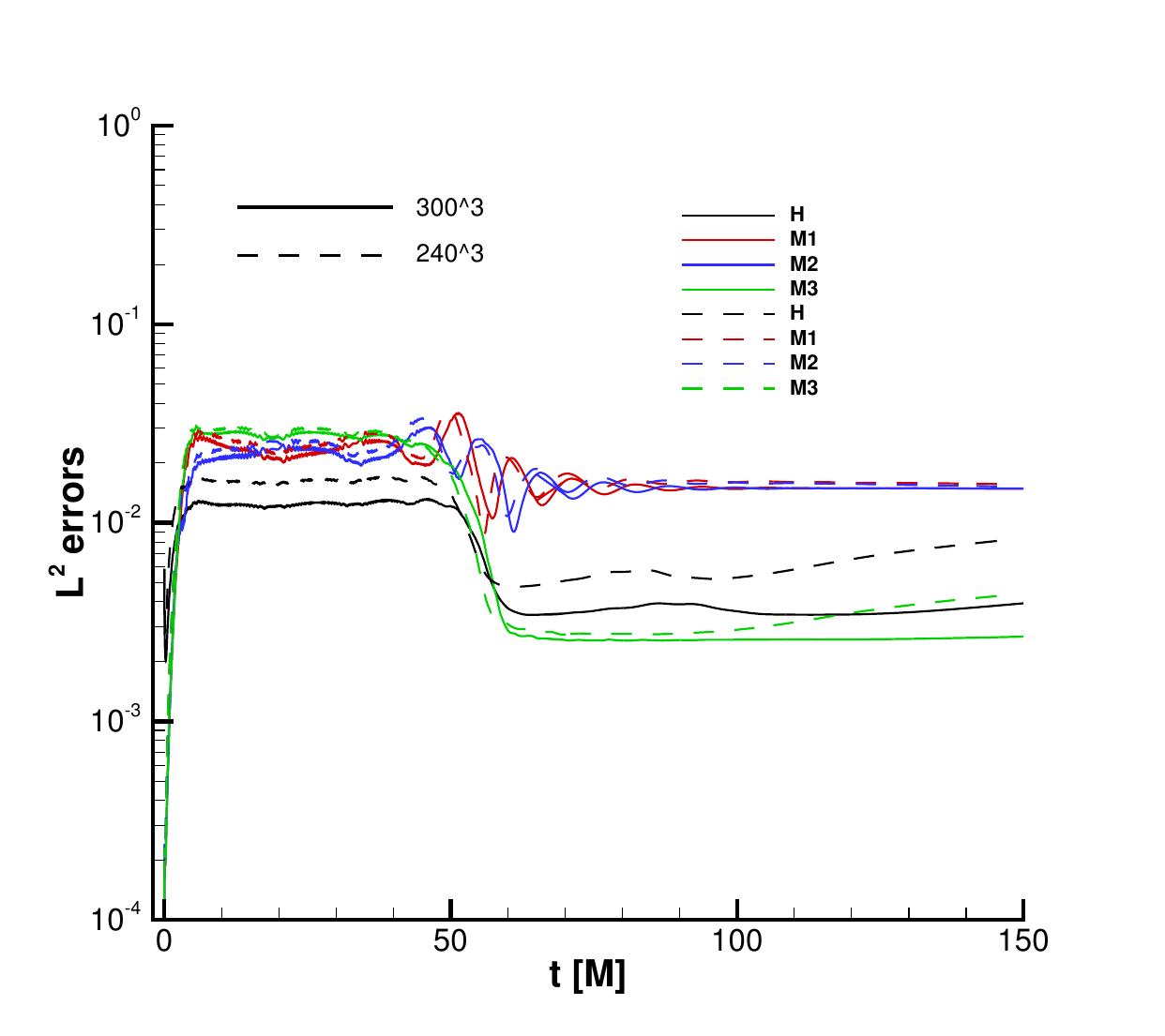} \\
    \end{tabular}
	\caption{
Inspiralling merger of Sect.~\ref{sec:BBHs} obtained with the CWENO7 scheme. Left: orbits followed by the black holes on the $z=0$ plane. Right:	time evolution of the normalized Einstein constraints.
	}
    \label{fig.binaryConstraints}
\end{center}
\end{figure}

Numerical investigations of inspiralling black hole mergers started well before the detection of gravitational waves, 
and a vast literature exists about their behaviour (see, among the other, \cite{Babiuc_2008,Reisswig2009,Hannam2010}).
Using again the TWOPUNCTURES library of \cite{Ansorg:2004ds}, 
we place the two black holes at a distance $d=4.0\,M$ among each other, zero individual spins, opposite linear momenta along the $y$ direction, \ie $p_1=(0,0.19243,0)$, $p_2=(0,-0.19243,0)$, and bare masses of the two black holes $m_1=m_2=0.46477$. This is one of the models considered by \cite{Tichy2004} in which the two black holes perform $\sim 2$ orbits before merging.
The \emph{gamma--driver} is again turned on, i.e. $s=1$, with $\eta=2$ and $\mu=0$. In these conditions we have evolved the system
with the CWENO7 scheme
over a logical computational domain 
 $\Omega_{\ell}=[-6, 6]^3$ covered by  $300^3$ gridpoints, 
and grid-stretching activated. 
The parameters of the polynomial stretching in the grid
function of Eq.~\eqref{coordmap} 
are $a = 14.59$, $b = -131.33$, $c = 395.0$, $d = -394.0$, $\xi_c=3$, producing a physical domain of size
$\Omega = [-400, 400]^3$ and an inner maximum grid resolution of $h=0.02$. 
	We impose again Sommerfeld (non-reflecting)  boundary conditions.
%
\newline
The merger takes place around $t\sim 70$, after which we let the system evolve 
to show the stability of the newly formed black hole.
The merging phase is shown in Fig.~\ref{fig.BinaryMerger},
where we report the contour plots of the scalar $\psi=e^{\phi}$ over the $z=0$ plane.
The left panel of 
Fig.~\ref{fig.binaryConstraints} shows the trajectories of the two punctures at two different grid resolutions, while in the right panel we report the evolution of the Einstein constraints. 

%

\section{Conclusions}
\label{sec:conclusions}
The second--order BSSNOK formulation of the Einstein equations 
is one of the most commonly used formulation in numerical general relativity,
but it also contains an unpleasant feature,
in so far it requires quite different numerical schemes for the solution of the Einstein sector
and of the matter subsystem. More specifically, writing the Einstein equations as a second--order PDE system prevents from applying to them 
the whole class of modern numerical schemes 
developed over the last decades developed  having first--order systems in mind. 
In this paper we thus propose two fundamental advancements in this respect:
\begin{itemize}
	\item We have provided a new first--order BSSNOK
	formulation of the Einstein--Euler equations which is non--conservative for the Einstein equations, while of course 
	conservative for the matter equations. 
	Moreover, the full PDE system is provably strongly hyperbolic.
	\item We have presented a new high order (verified up to the seventh) path-conservative central WENO finite difference scheme that is able to account for the mixed nature
	of the first--order PDE system.
\end{itemize}
Concerning the new first--order formulation, our work has been inspired by \cite{Brown2012}, from which, however, we differ 
in the crucial aspect that we carefully avoided to insert first order constraints in the equations. 
Such terms, we believe, introduce artificial Jordan blocks in the overall PDE system and thus may lead to a loss of strong hyperbolicity.

Concerning the numerical scheme, our work resumes a sequence of works by \cite{Levy1999,Levy2000,Levy2001}, who proposed central WENO
schemes more than twenty years ago for classical gas dynamics.
Much more recently, on the other hand, \cite{Balsara2024c} considered a new class of alternative finite difference WENO schemes for the non--conservative Z4 formulation of the vacuum Einstein equations, and to their achievements we are also highly indebted. 

Our new implementation can successfully solve all the standard tests of numerical relativity~\cite{Alcubierre2004}, including
the long term evolution of a black hole binary system.

We believe that the NR community could potentially benefit from this work, both regarding the new first order BSSNOK formulation of the Einstein-Euler equations, as well as the proposed CWENO finite difference scheme for its monolithic numerical discretization. 

\section{Acknowledgments}

This work was financially supported by i) the Italian Ministry of Education, University 
and Research (MIUR) in the framework of the PRIN 2022 project \textit{High order structure-preserving semi-implicit schemes for hyperbolic equations} and via the  Departments of Excellence  Initiative 2018--2027 attributed to DICAM of the University of Trento (grant L. 232/2016) 
and ii) by the Departments of Excellence  Initiative 2022--2027 attributed to the Department of Mathematics of the University of Roma La Sapienza (grant L. 232/2016). 
MD was also funded by the Fondazione Caritro via the project SOPHOS and by the European Research Council (ERC) under the European Union’s Horizon 2020 research and innovation programme, Grant agreement No. ERC-ADG-2021-101052956-BEYOND.  

The authors of this work  are all members of the INdAM GNCS group in Italy.  

\noindent
We acknowledge the CINECA award under the ISCRA initiative, for the availability of high--performance computing (HPC) resources and technical support.

\noindent 
We would like to thank Ilya Peshkov, Luciano Rezzolla, Federico Guercilena 
and Han Zhang for inspiring discussions. 
The authors would also kindly like to thank the anonymous referee for the numerous  constructive comments and questions, which contributed to improve the overall quality and clarity of this paper substantially.
The authors declare that not a single line of this paper has been written with AI.
The Fortran source code of the computer software which has generated the results shown in this paper can be obtained under CC BY-NC 4.0 license from the authors after written request. 

\appendix 

\section{Eigenvalues and eigenvectors of the new FO-BSSNOK system}
\label{app.eigenstructure}

The eigenstructure of the Einstein--Euler
system can be studied by focusing on the Einstein block only, more specifically by setting to zero all the hydrodynamic variables $(D, S_1, S_2, S_3, E)$, whose eigenvalues and eigenvectors are well known.
In addition, we assume for simplicity the case in which the gamma--driver is switched off ($s=0$) and the \emph{ 1+log gauge} condition is adopted, hence getting a reduced system composed of 34 variables, given by
\begin{eqnarray}
	\tilde{\boldsymbol Q}^T &=& \Big(
	\tilde{A}_{11},  \tilde{A}_{12},  \tilde{A}_{13},
	\tilde{A}_{22},  \tilde{A}_{23},  \tilde{A}_{33}, 	K,
	\tilde{\Gamma}^1, \tilde{\Gamma}^2, \tilde{\Gamma}^3,
	A_1, A_2, A_3,
	D_{111}, D_{112}, D_{113},
	D_{122}, D_{123}, D_{133},
	\nonumber \\
	&&	
	D_{211}, D_{212}, D_{213},
	D_{222}, D_{223}, D_{233},
	D_{311}, D_{312}, D_{313},
	D_{322}, D_{323}, D_{333},
	P_1, P_2, P_3 
	\Big)
	\,.
\end{eqnarray}

The eigenvalues of the reduced system in $x_1$-direction are

\begin{eqnarray}
\lambda_{1,\ldots,5} &=& +\alpha{\psi}^{-2}\sqrt {\tilde\gamma^{11} }-\beta^1
\\
\lambda_{6,\ldots,10} &=& -\alpha{\psi}^{-2}\sqrt {\tilde\gamma^{11} }-\beta^1
\\	
\lambda_{11,\ldots,32} &=& -\beta^1\,,
\\
\lambda_{33} &=&  +\sqrt{2\,\alpha}\,\psi^{-2}\sqrt{\tilde\gamma^{11}}-\beta^1\\
\lambda_{34} &=& - \sqrt{2\, \alpha}\,\psi^{-2}\sqrt{\tilde\gamma^{11}}-\beta^1
\end{eqnarray}
\normalsize
while the corresponding eigenvectors are
\footnotesize
\begin{eqnarray}
	%
	\boldsymbol{r}_{1}^T & = & \Big( -{\frac {\tilde{\gamma}^{33}}{\sqrt {\tilde{\gamma}^{11}}{\psi}^{2}}},0,0,0,0,{\frac {
			\sqrt {\tilde{\gamma}^{11}}}{{\psi}^{2}}},0,0,0,0,0,0,0,-{\frac {\tilde{\gamma}^{33}}{\tilde{\gamma}^{11}}},0,0,0,0,1,0,0,0,0,0,0,0,0,0,0,0,0,0,0,0	
	\Big) \\
	%
	\boldsymbol{r}_{2}^T & = & \Big( -2\,{\frac {\tilde{\gamma}^{23}}{\sqrt {\tilde{\gamma}^{11}}{\psi}^{2}}},0,0,0,{\frac {
			\sqrt {\tilde{\gamma}^{11}}}{{\psi}^{2}}},0,0,0,0,0,0,0,0,-2\,{\frac {\tilde{\gamma}^{23}}{\tilde{\gamma}^{11}}},0,0,0,1,0,0,0,0,0,0,0,0,0,0,0,0,0,0,0,0
	\Big) \\
	%
	\boldsymbol{r}_{3}^T & = & \Big(-{\frac {\tilde{\gamma}^{22}}{\sqrt {\tilde{\gamma}^{11}}{\psi}^{2}}},0,0,{\frac {
			\sqrt {\tilde{\gamma}^{11}}}{{\psi}^{2}}},0,0,0,0,0,0,0,0,0,-{\frac {\tilde{\gamma}^{22}
		}{\tilde{\gamma}^{11}}},0,0,1,0,0,0,0,0,0,0,0,0,0,0,0,0,0,0,0,0
	\Big) \\
	%
	\boldsymbol{r}_{4}^T & = & \Big(-2\,{\frac {\tilde{\gamma}^{13}}{\sqrt {\tilde{\gamma}^{11}}{\psi}^{2}}},0,{\frac {
			\sqrt {\tilde{\gamma}^{11}}}{{\psi}^{2}}},0,0,0,0,0,0,0,0,0,0,-2\,{\frac {\tilde{\gamma}^{13}}{\tilde{\gamma}^{11}}},0,1,0,0,0,0,0,0,0,0,0,0,0,0,0,0,0,0,0,0
	\Big) \\
	%
	\boldsymbol{r}_{5}^T & = & \Big(-2\,{\frac {\tilde{\gamma}^{12}}{\sqrt {\tilde{\gamma}^{11}}{\psi}^{2}}},{\frac {\sqrt 
			{\tilde{\gamma}^{11}}}{{\psi}^{2}}},0,0,0,0,0,0,0,0,0,0,0,-2\,{\frac {\tilde{\gamma}^{12}}{\tilde{\gamma}^{11}}},1,0,0,0,0,0,0,0,0,0,0,0,0,0,0,0,0,0,0,0
	\Big) \\
	%
	\boldsymbol{r}_{6}^T & = & \Big(2\,{\frac {\tilde{\gamma}^{13}}{\sqrt {\tilde{\gamma}^{11}}{\psi}^{2}}},0,-{\frac {
			\sqrt {\tilde{\gamma}^{11}}}{{\psi}^{2}}},0,0,0,0,0,0,0,0,0,0,-2\,{\frac {\tilde{\gamma}^{13}}{\tilde{\gamma}^{11}}},0,1,0,0,0,0,0,0,0,0,0,0,0,0,0,0,0,0,0,0
	\Big) \\
	%
	\boldsymbol{r}_{7}^T & = & \Big(2\,{\frac {\tilde{\gamma}^{12}}{\sqrt {\tilde{\gamma}^{11}}{\psi}^{2}}},-{\frac {\sqrt 
			{\tilde{\gamma}^{11}}}{{\psi}^{2}}},0,0,0,0,0,0,0,0,0,0,0,-2\,{\frac {\tilde{\gamma}^{12}}{\tilde{\gamma}^{11}}},1,0,0,0,0,0,0,0,0,0,0,0,0,0,0,0,0,0,0,0
	\Big) \\
	%
	\boldsymbol{r}_{8}^T & = & \Big({\frac {\tilde{\gamma}^{33}}{\sqrt {\tilde{\gamma}^{11}}{\psi}^{2}}},0,0,0,0,-{\frac {
			\sqrt {\tilde{\gamma}^{11}}}{{\psi}^{2}}},0,0,0,0,0,0,0,-{\frac {\tilde{\gamma}^{33}}{\tilde{\gamma}^{11}}},0,0,0,0,1,0,0,0,0,0,0,0,0,0,0,0,0,0,0,0
	\Big) \\
	%
	\boldsymbol{r}_{9}^T & = & \Big(2\,{\frac {\tilde{\gamma}^{23}}{\sqrt {\tilde{\gamma}^{11}}{\psi}^{2}}},0,0,0,-{\frac {
			\sqrt {\tilde{\gamma}^{11}}}{{\psi}^{2}}},0,0,0,0,0,0,0,0,-2\,{\frac {\tilde{\gamma}^{23}}{\tilde{\gamma}^{11}}},0,0,0,1,0,0,0,0,0,0,0,0,0,0,0,0,0,0,0,0
	\Big) \\
    %
	\boldsymbol{r}_{10}^T & = & \Big({\frac {\tilde{\gamma}^{22}}{\sqrt {\tilde{\gamma}^{11}}{\psi}^{2}}},0,0,-{\frac {
			\sqrt {\tilde{\gamma}^{11}}}{{\psi}^{2}}},0,0,0,0,0,0,0,0,0,-{\frac {\tilde{\gamma}^{22}
		}{\tilde{\gamma}^{11}}},0,0,1,0,0,0,0,0,0,0,0,0,0,0,0,0,0,0,0,0
	\Big) \\
	%
	\boldsymbol{r}_{11}^T & = & \Big(0,0,0,0,0,0,0,0,0,0,0,0,0,0,0,0,0,-{\frac {\tilde{\gamma}^{13}}{\tilde{\gamma}^{11}}},0,0,0,0,0,0,0,0,0,0,0,1,0,0,0
	,0
	\Big) \\
    %
	\boldsymbol{r}_{12}^T & = & \Big(0,0,0,0,0,0,0,2\,\tilde{\gamma}^{11},2\,\tilde{\gamma}^{12},2\,\tilde{\gamma}^{13},0,0,0,0,0,0,0
	,0,0,0,0,0,0,0,0,0,0,0,0,0,0,1,0,0
	\Big) \\
	%
	\boldsymbol{r}_{13}^T & = & \Big(0,0,0,0,0,0,0,2\,\tilde{\gamma}^{13},-2\,{\frac {\tilde{\gamma}_{23}
			-\tilde{\gamma}^{12}\,\tilde{\gamma}^{13}}{\tilde{\gamma}^{11}}},
		-2\,{
		\frac {\Lambda_1}{\tilde{\gamma}_{33}\,\tilde{\gamma}^{11}}},
		0,0,0,0,0,0,0,0,0,0,0,0,0,0,0,0,0,0,0,0,0,0,0,1
	\Big) \\
	%
	\boldsymbol{r}_{14}^T & = & \Big(0,0,0,0,0,0,0,0,{\frac {\tilde{\gamma}_{33}}{\tilde{\gamma}^{11}}},-{\frac {\tilde{\gamma}_{23}}{\tilde{\gamma}^{11}}},-{\frac {\tilde{\gamma}^{12}}{\tilde{\gamma}^{11}}},1,0,0,0,0,0,0,0,0,0,0,0,0
	,0,0,0,0,0,0,0,0,0,0
	\Big) \\
	%
	\boldsymbol{r}_{15}^T & = & \Big({\frac {\tilde{\gamma}_{11}}{\tilde{\gamma}_{33}}},{\frac {\tilde{\gamma}_{12}}{\tilde{\gamma}_{33}}},{\frac 
		{\tilde{\gamma}_{13}}{\tilde{\gamma}_{33}}},{\frac {\tilde{\gamma}_{22}}{\tilde{\gamma}_{33}}},{\frac {\tilde{\gamma}_{23}}{\tilde{\gamma}_{33}}},1,0,0,0,0,0,0,0,0,0,0,0,0,0,0,0,0,0,0,0,0,0,0,0,0,0,0
	,0,0
	\Big) \\
	%
	\boldsymbol{r}_{16}^T & = & \Big(0,0,0,0,0,0,0,2\,\tilde{\gamma}^{12},2\,{\frac {(\tilde{\gamma}^{12})^2+\tilde{\gamma}_{33}}{\tilde{\gamma}^{11}}},-2\,{\frac {\tilde{\gamma}_{23}-\tilde{\gamma}^{12}\,
				\tilde{\gamma}^{13}}{\tilde{\gamma}^{11}}},0,0,0,0,0
	,0,0,0,0,0,0,0,0,0,0,0,0,0,0,0,0,0,1,0	
	\Big) \\
	%
	\boldsymbol{r}_{17}^T & = & \Big(0,0,0,0,0,0,0,0,-{\frac {\tilde{\gamma}_{23}}{\tilde{\gamma}^{11}}},{\frac {\tilde{\gamma}_{22}}{\tilde{\gamma}^{11}}},-
	{\frac{\tilde{\gamma}^{13}}{\tilde{\gamma}^{11}}},0,1,0,0,0,0,0,0,0,0,0,0,0
	,0,0,0,0,0,0,0,0,0,0
	\Big) \\
	%
	\boldsymbol{r}_{18}^T & = & \Big(0,0,0,0,0,0,0,(\tilde{\gamma}^{11})^2,\tilde{\gamma}^{11}\,\tilde{\gamma}^{12},\tilde{\gamma}^{11}\,\tilde{\gamma}^{13},0,0,0,1,0,0,0,0,0,0,0,0,0,0,0,0,0,0,0,0,0,0,0,0
	\Big) \\
	%
	%
	\boldsymbol{r}_{19}^T & = & \Big(0,0,0,0,0,0,0,2\,\tilde{\gamma}^{11}\,\tilde{\gamma}^{13},2\,{\frac {\Lambda_2}{\tilde{\gamma}_{33}\,\tilde{\gamma}^{11}}},-2\,{\frac {\Lambda_3}{\tilde{\gamma}_{33}\,\tilde{\gamma}^{11}}},0,0,0,0,0,1,0,0,0,0,0,0,0,0,0,0,0,0,0,0,0,0,0,0
	\Big) \\
	%
%
	\boldsymbol{r}_{20}^T & = & \Big(0,0,0,0,0,0,0,2\,\tilde{\gamma}^{11}\,\tilde{\gamma}^{12},2\,(\tilde{\gamma}^{12})^2+2\,\tilde{\gamma}_{33},2\,{\frac {\Lambda_2}{\tilde{\gamma}_{33}\,\tilde{\gamma}^{11}}},0
	,0,0,0,1,0,0,0,0,0,0,0,0,0,0,0,0,0,0,0,0,0,0,0
	\Big) \\
	%
	\boldsymbol{r}_{21}^T & = & \Big(0,0,0,0,0,0,0,-{\frac {\tilde{\gamma}_{12}\,\tilde{\gamma}^{11}\,\tilde{\gamma}^{12}-\tilde{\gamma}_{23}\,\tilde{\gamma}^{12}\,\tilde{\gamma}^{13}-\tilde{\gamma}_{33}\,(\tilde{\gamma}^{13})^2+\tilde{\gamma}^{11}}{\tilde{\gamma}_{33}}},-{\frac {\Lambda_5}{\tilde{\gamma}_{33}\,\tilde{\gamma}^{11}}},-{\frac {\Lambda_6}{\tilde{\gamma}_{33}\,\tilde{\gamma}^{11}}},0,0,0,0,0,0
	,{\frac {\tilde{\gamma}_{22}}{\tilde{\gamma}_{33}}},{\frac {\tilde{\gamma}_{23}}{\tilde{\gamma}_{33}}},1, \nonumber\\
	&&0,0,0
	,0,0,0,0,0,0,0,0,0,0,0,0
	\Big) \\
	%
    %
	\boldsymbol{r}_{22}^T & = & \Big(0,0,0,0,0,0,0,{\tilde\gamma^{11}}\,{\tilde\gamma^{12} },({\tilde\gamma^{12} })^{2},{\frac {\Lambda_7}{\tilde\gamma_{33}\,\tilde\gamma^{11}}},0,0,0,0,0,0,0,0,0,1,0,0,0,0,0,0,0,0,0,0,0,0,0,0
	\Big) \\
	%
	\boldsymbol{r}_{23}^T & = & \Big(0,0,0,0,0,0,0,2\,\tilde{\gamma}^{12}\,\tilde{\gamma}^{13}\,,-2\,{\frac {\tilde{\gamma}^{12}\,
			\left(\tilde{\gamma}_{23}- \tilde{\gamma}^{12}\,\tilde{\gamma}^{13} \right) }{\tilde{\gamma}^{11}}},-2\,{\frac {\Lambda_8}{\tilde{\gamma}_{33}\,\tilde{\gamma}^{11}}},0,0,0,0,0,0,0,0,0,0,0,1,\nonumber \\
		&& 0,0,0,0,0,0,0,0,0,0,0
	,0
	\Big) \\
	%
	\boldsymbol{r}_{24}^T & = & \Big(0,0,0,0,0,0,0,2\,(\tilde{\gamma}^{12})^2,-2\,{\frac {\Lambda_9}{\tilde{\gamma}^{11}}},-2\,{\frac {\tilde{\gamma}^{12}\, \left( \tilde{\gamma}_{23}-\tilde{\gamma}^{12}
			\,\tilde{\gamma}^{13}\right) }{\tilde{\gamma}^{11}}},0,0,0,0,0,0,0,0,0,0,1,0,0,0,0,0,0,0,0,0,0,0,0,0
	\Big) \\
	%
%
	\boldsymbol{r}_{25}^T & = & \Big(0,0,0,0,0,0,0,0,0,0,0,0,0,0,0,0,-{\frac {\tilde{\gamma}^{12}}{\tilde{\gamma}^{11}}},0,0
	,0,0,0,1,0,0,0,0,0,0,0,0,0,0,0
	\Big) \\
	%
	\boldsymbol{r}_{26}^T & = & \Big(0,0,0,0,0,0,0,0,0,0,0,0,0,0,0,0,0,-{\frac {\tilde{\gamma}^{12}}{\tilde{\gamma}^{11}}},0
	,0,0,0,0,1,0,0,0,0,0,0,0,0,0,0
	\Big) \\
	\boldsymbol{r}_{27}^T & = & \Big(0,0,0,0,0,0,0,-{\frac {\Lambda_{10}}{\tilde{\gamma}_{33}}},-{\frac {\Lambda_{11}}{\tilde{\gamma}_{33}\,\tilde{\gamma}^{11}}},-{\frac {\Lambda_{12}}{\tilde{\gamma}_{33}\,\tilde{\gamma}^{11}}},0,0,0,0,0,0,-{\frac {\tilde{\gamma}_{12}\,(\tilde{\gamma}_{23})^2-\tilde{\gamma}_{13}\,\tilde{\gamma}_{22}\,\tilde{\gamma}_{23}+\tilde{\gamma}_{12}\,\tilde{\gamma}^{11}}{\tilde{\gamma}_{33}\,\tilde{\gamma}^{11}}},{\frac {\tilde{\gamma}_{23}\,\tilde{\gamma}^{12}}{\tilde{\gamma}_{33}\,\tilde{\gamma}^{11}}}, \nonumber \\
	&& 0,0,0,0
	,0,0,1,0,0,0,0,0,0,0,0,0
	\Big) \\
	\boldsymbol{r}_{28}^T & = & \Big(0,0,0,0,0,0,0,2\,\tilde{\gamma}^{12}\,\tilde{\gamma}^{13},-2\,{\frac {\Lambda_{13}}{\tilde{\gamma}_{33}\,\tilde{\gamma}^{11}}},-2\,{\frac {\Lambda_{14}}{\tilde{\gamma}_{33}\,\tilde{\gamma}^{11}}},0,0,0,0,0,0,0,0,0,0
	,0,0,0,0,0,0,1,0,0,0,0,0,0,0
	\Big) \\
	\boldsymbol{r}_{29}^T & = & \Big(0,0,0,0,0,0,0,\tilde{\gamma}^{11}\,\tilde{\gamma}^{13},{\frac {\Lambda_{15}}{\tilde{\gamma}_{33}\,\tilde{\gamma}^{11}}}
	,-{\frac {\Lambda_{16}}{\tilde{\gamma}_{33}\,\tilde{\gamma}^{11}}},0,0,0,0,0,0,0,0,0,0,0,0,0,0,0,1,0,0,0,0,0,0,0,0
	\Big) \\
	\boldsymbol{r}_{30}^T & = & \Big(0,0,0,0,0,0,0,-2\,{\frac {\Lambda_{17}}{\tilde{\gamma}_{33}}},-2\,{\frac {\Lambda_{14}}{\tilde{\gamma}_{33}\,\tilde{\gamma}^{11}}},-2\,{\frac {\Lambda_{19}}{\tilde{\gamma}_{33}\,\tilde{\gamma}^{11}}},0,0,0,0,0,0,0,0,0,0,0,0,0,0,0,0,0,1,0,0,0,0,0,0
	\Big) \\
	\boldsymbol{r}_{31}^T & = & \Big(0,0,0,0,0,0,0,0,0,0,0,0,0,0,0,0,-{\frac {\tilde{\gamma}^{13}}{\tilde{\gamma}^{11}}},0,0,0,0,0,0,0,0,0,0,0,1,0,0,0
	,0,0
	\Big) \\
	%
	\boldsymbol{r}_{32}^T & = & \Big(0,0,0,0,0,0,0,-{\frac {\Lambda_{20}}{\tilde{\gamma}_{33}}},-{\frac {\Lambda_{21}}{\tilde{\gamma}_{33}\,\tilde{\gamma}^{11}}},-{\frac {\Lambda_{22}}{\tilde{\gamma}_{33}\,\tilde{\gamma}^{11}}},0,0,0,0,0,0,{\frac {\tilde{\gamma}_{22}\,\tilde{\gamma}^{13}}{\tilde{\gamma}_{33}\,\tilde{\gamma}^{11}}},{\frac {\tilde{\gamma}_{23}\,\tilde{\gamma}^{13}}{\tilde{\gamma}_{33}\,\tilde{\gamma}^{11}}},0,0
	,0,0,0,0,0,0,0,0,0,0,1,0,0,0
	\Big) \\
	%
	\boldsymbol{r}_{33}^T & = & \Big(2\,{\frac {\sqrt {2} \left( \tilde{\gamma}^{12}\,\tilde{\gamma}_{12}+\tilde{\gamma}^{13}\,\tilde{\gamma}_{13}+2 \right) }{{\psi}^{2}\sqrt {\alpha\,\tilde{\gamma}^{11}}}},-2\,{\frac {\tilde{\gamma}_{12}\,\sqrt {2}\tilde{\gamma}^{11}}{{\psi}^{2}\sqrt {\alpha\,\tilde{\gamma}^{11}}}},2\,{\frac {\sqrt {2} \left( \tilde{\gamma}_{23}\,\tilde{\gamma}^{12}+\tilde{\gamma}_{33}\,\tilde{\gamma}^{13} \right) }{{\psi}^{2}\sqrt {\alpha\,\tilde{\gamma}^{11}}}}
	,	-2\,{\frac {\sqrt {2}\tilde{\gamma}_{22}\,\tilde{\gamma}^{11}}{{\psi}^{2}\sqrt {{\alpha}\,\tilde{\gamma}^{11}}}},-2\,{\frac {\sqrt {2}\tilde{\gamma}_{23}\,\tilde{\gamma}^{11}}{{
				\psi}^{2}\sqrt {\alpha\,\tilde{\gamma}^{11}}}},\nonumber\\
&&	-2\,{\frac {\sqrt {2}\tilde{\gamma}_{33}\,\tilde{\gamma}^{11}}{{\psi}^{2}\sqrt {\alpha\,\tilde{\gamma}^{11}}}},6\,{
		\frac {\sqrt {2}\tilde{\gamma}^{11}}{{\psi}^{2}\sqrt {\alpha\,\tilde{\gamma}^{11}}}
	},8\,\tilde{\gamma}^{11},8\,\tilde{\gamma}^{12},8\,\tilde{\gamma}^{13},12\,{\alpha}^{-1},0,0
	,2\,{\frac {\tilde{\gamma}^{12}\,\tilde{\gamma}_{12}+\tilde{\gamma}^{13}\,\tilde{\gamma}_{13}+2}{\tilde{\gamma}^{11}}
	},-2\,\tilde{\gamma}_{12},-2\,\tilde{\gamma}_{13},-2\,\tilde{\gamma}_{22},\nonumber\\
	&&-2\,\tilde{\gamma}_{23},-2\,\tilde{\gamma}_{33},0,0,0,0,0,0,0,0,0,0,0,0,1,0,0
	\Big) \\
	%
	\boldsymbol{r}_{34}^T & = & \Big(-2\,{\frac {\sqrt {2} \left( \tilde{\gamma}^{12}\,\tilde{\gamma}_{12}+\tilde{\gamma}^{13}\,\tilde{\gamma}_{13}+2 \right) }{{\psi}^{2}\sqrt {\alpha\,\tilde{\gamma}^{11}}}},2\,{
		\frac {\tilde{\gamma}_{12}\,\sqrt {2}\tilde{\gamma}^{11}}{{\psi}^{2}\sqrt {\alpha\,\tilde{\gamma}^{11}}}},-2\,{\frac {\sqrt {2} \left( \tilde{\gamma}_{23}\,\tilde{\gamma}^{12}+\tilde{\gamma}_{33}\,\tilde{\gamma}^{13} \right) }{{\psi}^{2}\sqrt {\alpha\,\tilde{\gamma}^{11}}}}
	,2\,{\frac {\sqrt {2}\tilde{\gamma}_{22}\,\tilde{\gamma}^{11}}{{\psi}^{2}\sqrt {{\alpha}\,\tilde{\gamma}^{11}}}},2\,{\frac {\sqrt {2}\tilde{\gamma}_{23}\,\tilde{\gamma}^{11}}{{\psi
			}^{2}\sqrt {\alpha\,\tilde{\gamma}^{11}}}},\nonumber\\
		&&2\,{\frac {\sqrt {2}\tilde{\gamma}_{33}\,
			\tilde{\gamma}^{11}}{{\psi}^{2}\sqrt {\alpha\,\tilde{\gamma}^{11}}}},-6\,{\frac {\sqrt {2}\tilde{\gamma}^{11}}{{\psi}^{2}\sqrt {\alpha\,\tilde{\gamma}^{11}}}},8\,\tilde{\gamma}^{11},8\,\tilde{\gamma}^{12},8\,\tilde{\gamma}^{13},12\,(\alpha)^{-1},0,0,2\,{
		\frac {\tilde{\gamma}^{12}\,\tilde{\gamma}_{12}+\tilde{\gamma}^{13}\,\tilde{\gamma}_{13}+2}{\tilde{\gamma}^{11}}},-2
	\,\tilde{\gamma}_{12},-2\,\tilde{\gamma}_{13},-2\,\tilde{\gamma}_{22},\nonumber\\
	&&-2\,\tilde{\gamma}_{23},-2\,\tilde{\gamma}_{33},0,0
	,0,0,0,0,0,0,0,0,0,0,1,0,0
	\Big) 
\end{eqnarray}
\normalsize
where we have defined
\footnotesize
\begin{eqnarray}
	\Lambda_1 &=& 2\,\tilde{\gamma}_{12}\,\tilde{\gamma}_{13}\,(\tilde{\gamma}_{23})^3-2\,(\tilde{\gamma}_{13})^2\tilde{\gamma}_{22}\,(\tilde{\gamma}_{23})^2+2\,\tilde{\gamma}_{12}\,\tilde{\gamma}_{13}\,\tilde{\gamma}_{23}\,\tilde{\gamma}^{11}-(\tilde{\gamma}_{13})^2\tilde{\gamma}_{22}\,\tilde{\gamma}^{11}+\tilde{\gamma}_{13}\,\tilde{\gamma}_{22}\,\tilde{\gamma}_{23}\,\tilde{\gamma}^{12}-\tilde{\gamma}_{13}\,(\tilde{\gamma}_{23})^2\tilde{\gamma}^{13}+\nonumber\\
	&&\tilde{\gamma}_{23}\,\tilde{\gamma}^{12}\,\tilde{\gamma}^{13}-(\tilde{\gamma}_{23})^2-\tilde{\gamma}^{11}\\
	\Lambda_2&=&(\tilde{\gamma}_{13})^2(\tilde{\gamma}_{23})^3\tilde{\gamma}^{11}+\tilde{\gamma}_{13}\,(\tilde{\gamma}_{23})^4\tilde{\gamma}^{12}+\tilde{\gamma}_{13}
	\,(\tilde{\gamma}_{23})^3\tilde{\gamma}_{33}\,\tilde{\gamma}^{13}-2\,\tilde{\gamma}_{13}\,(\tilde{\gamma}_{23})^2\tilde{\gamma}^{11}\,\tilde{\gamma}^{12}-2\,(\tilde{\gamma}_{23})^3(\tilde{\gamma}^{12})^2-2\,{\tilde{\gamma}_{23}
	}^{2}\tilde{\gamma}_{33}\,\tilde{\gamma}^{12}\,\tilde{\gamma}^{13}\nonumber\\
	&&-\tilde{\gamma}_{13}\,(\tilde{\gamma}^{11})^2\tilde{\gamma}^{12}-\tilde{\gamma}_{23}\,\tilde{\gamma}^{11}\,(\tilde{\gamma}^{12})^2-\tilde{\gamma}_{23}\,\tilde{\gamma}_{33}
	\,\tilde{\gamma}^{11}\\
	\Lambda_3 &=&2\,\tilde{\gamma}_{12}\,\tilde{\gamma}_{13}\,(\tilde{\gamma}_{23})^3\tilde{\gamma}^{11}-(\tilde{\gamma}_{13})^2\tilde{\gamma}_{22}\,(\tilde{\gamma}_{23})^{2}\tilde{\gamma}^{11}+\tilde{\gamma}_{13}\,\tilde{\gamma}_{22}\,(\tilde{\gamma}_{23})^3\tilde{\gamma}^{12}+\tilde{\gamma}_{13}\,\tilde{\gamma}_{22}\,(\tilde{\gamma}_{23})^2\tilde{\gamma}_{33}\,\tilde{\gamma}^{13}+2\,\tilde{\gamma}_{12}\,\tilde{\gamma}_{13}\,\tilde{\gamma}_{23}\,(\tilde{\gamma}^{11})^2\nonumber\\
	&&
	-(\tilde{\gamma}_{13})^2\tilde{\gamma}_{22}\,(\tilde{\gamma}^{11})^{2}
	-\tilde{\gamma}_{22}\,(\tilde{\gamma}_{23})^2(\tilde{\gamma}^{12})^2-\tilde{\gamma}_{22}\,\tilde{\gamma}_{23}
	\,\tilde{\gamma}_{33}\,\tilde{\gamma}^{12}\,\tilde{\gamma}^{13}+(\tilde{\gamma}_{23})^3\tilde{\gamma}^{12}\,\tilde{\gamma}^{13}+(\tilde{\gamma}_{23})^2\tilde{\gamma}_{33}\,(\tilde{\gamma}^{13})^2+\tilde{\gamma}_{23}\,\tilde{\gamma}^{11}
	\,\tilde{\gamma}^{12}\,\tilde{\gamma}^{13}\nonumber\\
	&&-(\tilde{\gamma}_{23})^2\tilde{\gamma}^{11}-(\tilde{\gamma}^{11})^2\\
	%
	%
	\Lambda_5
	&=&3\,\tilde{\gamma}_{12}\,(\tilde{\gamma}_{13})^2(\tilde{\gamma}_{23})^2\tilde{\gamma}^{11}+3
	\,(\tilde{\gamma}_{13})^2\tilde{\gamma}_{22}\,(\tilde{\gamma}_{23})^2\tilde{\gamma}^{12}+3\,{\tilde{\gamma}_{13}}^{
		2}\tilde{\gamma}_{22}\,\tilde{\gamma}_{23}\,\tilde{\gamma}_{33}\,\tilde{\gamma}^{13}+\tilde{\gamma}_{12}\,(\tilde{\gamma}_{13})^{
		2}(\tilde{\gamma}^{11})^2-\tilde{\gamma}_{13}\,\tilde{\gamma}_{22}\,\tilde{\gamma}_{23}\,(\tilde{\gamma}^{12})^2\nonumber\\
		&&-\tilde{\gamma}_{13}\,\tilde{\gamma}_{22}\,\tilde{\gamma}_{33}\,\tilde{\gamma}^{12}\,\tilde{\gamma}^{13}+
		2\,\tilde{\gamma}_{13}\,(\tilde{\gamma}_{23})^{2}\tilde{\gamma}^{12}\,\tilde{\gamma}^{13}+2\,\tilde{\gamma}_{13}\,\tilde{\gamma}_{23}\,\tilde{\gamma}_{33}\,(\tilde{\gamma}^{13})^2+\tilde{\gamma}_{12}\,\tilde{\gamma}^{11}\,(\tilde{\gamma}^{12})^2-\tilde{\gamma}_{23}
	\,(\tilde{\gamma}^{12})^2\tilde{\gamma}^{13}-\tilde{\gamma}_{33}\,\tilde{\gamma}^{12}\,(\tilde{\gamma}^{13})^2\nonumber\\
	&&-\tilde{\gamma}^{11}\,\tilde{\gamma}^{12}\\
	\Lambda_6
	&=&3\,{\tilde{\gamma}_{12}}
	^{2}\tilde{\gamma}_{13}\,(\tilde{\gamma}_{23})^2\tilde{\gamma}^{11}+3\,(\tilde{\gamma}_{13})^2{\tilde{\gamma}_{22}}
	^{2}\tilde{\gamma}_{23}\,\tilde{\gamma}^{12}+3\,(\tilde{\gamma}_{13})^2(\tilde{\gamma}_{22})^2\tilde{\gamma}_{33}\,
	\tilde{\gamma}^{13}+(\tilde{\gamma}_{12})^2\tilde{\gamma}_{13}\,(\tilde{\gamma}^{11})^2
	-\tilde{\gamma}_{12}\,\tilde{\gamma}_{22}\,\tilde{\gamma}_{23}\,(\tilde{\gamma}^{12})^2\nonumber\\
	&&-\tilde{\gamma}_{12}\,\tilde{\gamma}_{22}\,\tilde{\gamma}_{33}\,\tilde{\gamma}^{12}\,\tilde{\gamma}^{13}
	+5\,\tilde{\gamma}_{13}\,\tilde{\gamma}_{22}\,\tilde{\gamma}_{23}\,\tilde{\gamma}^{12}\,
	\tilde{\gamma}^{13}+5\,\tilde{\gamma}_{13}\,\tilde{\gamma}_{22}\,\tilde{\gamma}_{33}\,(\tilde{\gamma}^{13})^2+\tilde{\gamma}_{12}\,\tilde{\gamma}^{11}\,\tilde{\gamma}^{12}\,\tilde{\gamma}^{13}-\tilde{\gamma}_{13}\,\tilde{\gamma}^{11}\,(\tilde{\gamma}^{13})^{2}-\tilde{\gamma}^{11}\,\tilde{\gamma}^{13}\\
	\Lambda_7&=&  \Lambda_2+\tilde{\gamma}_{23}\,\tilde{\gamma}_{33}
	\,\tilde{\gamma}^{11}\\
	%
	%
%
	\Lambda_8&=& 3\,
	\tilde{\gamma}_{12}\,(\tilde{\gamma}_{13})^2(\tilde{\gamma}_{23})^4-3\,(\tilde{\gamma}_{13})^3\tilde{\gamma}_{22}
	\,(\tilde{\gamma}_{23})^3+4\,\tilde{\gamma}_{12}\,(\tilde{\gamma}_{13})^2(\tilde{\gamma}_{23})^2\tilde{\gamma}^{11}+3\,\tilde{\gamma}_{12}\,\tilde{\gamma}_{13}\,(\tilde{\gamma}_{23})^3\tilde{\gamma}^{12}-3\,{\tilde{\gamma}_{13}}^{3}\tilde{\gamma}_{22}\,\tilde{\gamma}_{23}\,\tilde{\gamma}^{11}\nonumber\\
	&&-2\,(\tilde{\gamma}_{13})^2\tilde{\gamma}_{22}\,(\tilde{\gamma}_{23})^{2}\tilde{\gamma}^{12}
	-2\,(\tilde{\gamma}_{13})^2(\tilde{\gamma}_{23})^3\tilde{\gamma}^{13}+\tilde{\gamma}_{12}\,(\tilde{\gamma}_{13})^2(\tilde{\gamma}^{11})^2+2\,\tilde{\gamma}_{12}\,\tilde{\gamma}_{13}\,\tilde{\gamma}_{23}\,\tilde{\gamma}^{11}\,\tilde{\gamma}^{12}
	-2\,(\tilde{\gamma}_{13})^2\tilde{\gamma}_{23}\,\tilde{\gamma}^{11}\,\tilde{\gamma}^{13}\nonumber\\
	&&+\tilde{\gamma}_{13}\,\tilde{\gamma}_{22}\,\tilde{\gamma}_{23}\,(\tilde{\gamma}^{12})^2
	-2
	\,\tilde{\gamma}_{13}\,(\tilde{\gamma}_{23})^2\tilde{\gamma}^{12}\,\tilde{\gamma}^{13}+\tilde{\gamma}_{23}\,(\tilde{\gamma}^{12})^{2}\tilde{\gamma}^{13}-(\tilde{\gamma}_{23})^2\tilde{\gamma}^{12}-\tilde{\gamma}^{11}\,\tilde{\gamma}^{12}\\
	\Lambda_9&=&
	3\,\tilde{\gamma}_{12}\,(\tilde{\gamma}_{13})^{2}(\tilde{\gamma}_{23})^2\tilde{\gamma}_{33}-3\,(\tilde{\gamma}_{13})^3(\tilde{\gamma}_{23})^3+3\,(\tilde{\gamma}_{13})^{2}(\tilde{\gamma}_{23})^2\tilde{\gamma}^{12}+\tilde{\gamma}_{12}\,(\tilde{\gamma}_{33})^2-\tilde{\gamma}_{13}\,\tilde{\gamma}_{23}\,\tilde{\gamma}_{33}-(\tilde{\gamma}^{12})^3\\
	\Lambda_{10}&=&
	3\,\tilde{\gamma}_{12}\,(\tilde{\gamma}_{13})^2(\tilde{\gamma}_{23})^2-3
	\,\tilde{\gamma}_{23}\,\tilde{\gamma}_{22}\,(\tilde{\gamma}_{13})^3+\tilde{\gamma}_{12}\,(\tilde{\gamma}_{13})^2\tilde{\gamma}^{11}+(\tilde{\gamma}_{13})^2\tilde{\gamma}_{22}\,\tilde{\gamma}^{12}-2\,(\tilde{\gamma}_{13})^2\tilde{\gamma}_{23}\,\tilde{\gamma}^{13}+\tilde{\gamma}_{12}\,(\tilde{\gamma}^{12})^2+\tilde{\gamma}_{13}\,\tilde{\gamma}^{12}\,\tilde{\gamma}^{13}+\tilde{\gamma}^{12}\\
	\Lambda_{11}&=&
	4\,\tilde{\gamma}_{12}\,{\tilde{\gamma}_{13}}^{3}(\tilde{\gamma}_{23})^3-4\,(\tilde{\gamma}_{13})^4\tilde{\gamma}_{22}\,(\tilde{\gamma}_{23})^2+2\,\tilde{\gamma}_{12}\,(\tilde{\gamma}_{13})^3\tilde{\gamma}_{23}\,\tilde{\gamma}^{11}+6\,\tilde{\gamma}_{12}\,(\tilde{\gamma}_{13})^{2}(\tilde{\gamma}_{23})^2\tilde{\gamma}^{12}-(\tilde{\gamma}_{13})^4\tilde{\gamma}_{22}\,\tilde{\gamma}^{11}-5
	\,(\tilde{\gamma}_{13})^3\tilde{\gamma}_{22}\,\tilde{\gamma}_{23}\,\tilde{\gamma}^{12}\nonumber\\
	&&-3\,(\tilde{\gamma}_{13})^3(\tilde{\gamma}_{23})^{2}\tilde{\gamma}^{13}+\tilde{\gamma}_{12}\,(\tilde{\gamma}_{13})^2\tilde{\gamma}^{11}\,\tilde{\gamma}^{12}-(\tilde{\gamma}_{13})^3\tilde{\gamma}^{11}\,\tilde{\gamma}^{13}+(\tilde{\gamma}_{13})^2\tilde{\gamma}_{22}\,
	(\tilde{\gamma}^{12})^2-5\,(\tilde{\gamma}_{13})^2\tilde{\gamma}_{23}\,\tilde{\gamma}^{12}\,\tilde{\gamma}^{13}+\tilde{\gamma}_{12}\,(\tilde{\gamma}^{12})^3\nonumber\\
	&&+\tilde{\gamma}_{13}\,(\tilde{\gamma}^{12})^2\tilde{\gamma}^{13}-(\tilde{\gamma}^{12})^{2}\\
	\Lambda_{12}&=&
6\,(\tilde{\gamma}_{12})^2(\tilde{\gamma}_{13})^{2}(\tilde{\gamma}_{23})^3-8\,\tilde{\gamma}_{12}\,(\tilde{\gamma}_{13})^3\tilde{\gamma}_{22}\,(\tilde{\gamma}_{23})^{2}+2\,(\tilde{\gamma}_{13})^4(\tilde{\gamma}_{22})^2\tilde{\gamma}_{23}+4\,(\tilde{\gamma}_{12})^{2}(\tilde{\gamma}_{13})^2\tilde{\gamma}_{23}\,\tilde{\gamma}^{11}+4\,(\tilde{\gamma}_{12})^2\tilde{\gamma}_{13}
\,(\tilde{\gamma}_{23})^2\tilde{\gamma}^{12}\nonumber\\
&&-2\,\tilde{\gamma}_{12}\,(\tilde{\gamma}_{13})^3\tilde{\gamma}_{22}\,\tilde{\gamma}^{11}-4\,\tilde{\gamma}_{12}\,(\tilde{\gamma}_{13})^2(\tilde{\gamma}_{23})^2\tilde{\gamma}^{13}-2\,(\tilde{\gamma}_{13})^{3}(\tilde{\gamma}_{22})^2\tilde{\gamma}^{12}+2\,(\tilde{\gamma}_{13})^3\tilde{\gamma}_{22}\,\tilde{\gamma}_{23}\,\tilde{\gamma}^{13}+(\tilde{\gamma}_{12})^2\tilde{\gamma}_{13}\,\tilde{\gamma}^{11}\,\tilde{\gamma}^{12}\nonumber\\
&&-
\tilde{\gamma}_{12}\,(\tilde{\gamma}_{13})^2\tilde{\gamma}^{11}\,\tilde{\gamma}^{13}
+\tilde{\gamma}_{12}\,\tilde{\gamma}_{13}
\,\tilde{\gamma}_{22}\,(\tilde{\gamma}^{12})^2-4\,(\tilde{\gamma}_{13})^2\tilde{\gamma}_{22}\,\tilde{\gamma}^{12}
\,\tilde{\gamma}^{13}+(\tilde{\gamma}_{13})^2\tilde{\gamma}_{23}\,(\tilde{\gamma}^{13})^2-2\,\tilde{\gamma}_{12}\,
\tilde{\gamma}_{13}\,(\tilde{\gamma}_{23})^2+\tilde{\gamma}_{12}\,(\tilde{\gamma}^{12})^2\tilde{\gamma}^{13}\nonumber\\
&&+2\,(\tilde{\gamma}_{13})^{2}\tilde{\gamma}_{22}\,\tilde{\gamma}_{23}
-2\,\tilde{\gamma}_{13}\,\tilde{\gamma}^{12}\,(\tilde{\gamma}^{13})^{2}-\tilde{\gamma}_{12}\,\tilde{\gamma}_{13}\,\tilde{\gamma}^{11}-\tilde{\gamma}_{13}\,\tilde{\gamma}_{22}\,\tilde{\gamma}^{12}+\tilde{\gamma}_{13}\,\tilde{\gamma}_{23}\,\tilde{\gamma}^{13}-\tilde{\gamma}^{12}\,\tilde{\gamma}^{13}	\\
\Lambda_{13}&=&
\tilde{\gamma}_{13}\,\tilde{\gamma}^{11}\,(\tilde{\gamma}^{12})^2+\tilde{\gamma}_{23}\,(\tilde{\gamma}^{12})^3+\tilde{\gamma}_{13}\,\tilde{\gamma}_{33}\,\tilde{\gamma}^{11}+\tilde{\gamma}_{23}\,\tilde{\gamma}_{33}\,\tilde{\gamma}^{12}\\
\Lambda_{14}&=&
3\,\tilde{\gamma}_{12}\,(\tilde{\gamma}_{13})^2(\tilde{\gamma}_{23})^4-3
\,(\tilde{\gamma}_{13})^3\tilde{\gamma}_{22}\,(\tilde{\gamma}_{23})^3+4\,\tilde{\gamma}_{12}\,{\tilde{\gamma}_{13}}^{2}(\tilde{\gamma}_{23})^2\tilde{\gamma}^{11}+3\,\tilde{\gamma}_{12}\,\tilde{\gamma}_{13}\,(\tilde{\gamma}_{23})^3\tilde{\gamma}^{12}-3\,(\tilde{\gamma}_{13})^3\tilde{\gamma}_{22}\,\tilde{\gamma}_{23}\,\tilde{\gamma}^{11}\nonumber\\
&&-2\,(\tilde{\gamma}_{13})^{2}\tilde{\gamma}_{22}\,(\tilde{\gamma}_{23})^2\tilde{\gamma}^{12}
-2\,(\tilde{\gamma}_{13})^2(\tilde{\gamma}_{23})^3\tilde{\gamma}^{13}+\tilde{\gamma}_{12}\,(\tilde{\gamma}_{13})^2(\tilde{\gamma}^{11})^2+2\,\tilde{\gamma}_{12}\,\tilde{\gamma}_{13}\,\tilde{\gamma}_{23}\,\tilde{\gamma}^{11}\,\tilde{\gamma}^{12}-2\,{\tilde{\gamma}_{13}}^{2}\tilde{\gamma}_{23}\,\tilde{\gamma}^{11}\,\tilde{\gamma}^{13}\nonumber\\
&&
+\tilde{\gamma}_{13}\,\tilde{\gamma}_{22}\,\tilde{\gamma}_{23}\,
(\tilde{\gamma}^{12})^2
-2\,\tilde{\gamma}_{13}\,(\tilde{\gamma}_{23})^2\tilde{\gamma}^{12}\,\tilde{\gamma}^{13}+\tilde{\gamma}_{23}\,(\tilde{\gamma}^{12})^2\tilde{\gamma}^{13}-\tilde{\gamma}_{13}\,\tilde{\gamma}_{23}\,\tilde{\gamma}^{11}-
(\tilde{\gamma}_{23})^2\tilde{\gamma}^{12}\\
\Lambda_{15}&=&
(\tilde{\gamma}_{13})^2(\tilde{\gamma}_{23})^3\tilde{\gamma}^{11}+\tilde{\gamma}_{13}\,(\tilde{\gamma}_{23})^4\tilde{\gamma}^{12}+\tilde{\gamma}_{13}\,(\tilde{\gamma}_{23})^{3}\tilde{\gamma}_{33}\,\tilde{\gamma}^{13}-2\,\tilde{\gamma}_{13}\,(\tilde{\gamma}_{23})^2\tilde{\gamma}^{11}\,\tilde{\gamma}^{12}-2\,(\tilde{\gamma}_{23})^3(\tilde{\gamma}^{12})^2-2\,(\tilde{\gamma}_{23})^{2
}\tilde{\gamma}_{33}\,\tilde{\gamma}^{12}\,\tilde{\gamma}^{13}\nonumber\\
&&-\tilde{\gamma}_{13}\,(\tilde{\gamma}^{11})^2\tilde{\gamma}^{12}-\tilde{\gamma}_{23}\,\tilde{\gamma}^{11}\,(\tilde{\gamma}^{12})^2\\
\Lambda_{16}&=&\Lambda_3
+(\tilde{\gamma}_{23})^2\tilde{\gamma}^{11}+(\tilde{\gamma}^{11})^2\\
%
%
\Lambda_{17}&=&
2\,\tilde{\gamma}_{12}\,\tilde{\gamma}_{13}\,(\tilde{\gamma}_{23})^3-2
\,(\tilde{\gamma}_{13})^2\tilde{\gamma}_{22}\,(\tilde{\gamma}_{23})^2+2\,\tilde{\gamma}_{12}\,\tilde{\gamma}_{13}\,\tilde{\gamma}_{23}\,\tilde{\gamma}^{11}-(\tilde{\gamma}_{13})^2\tilde{\gamma}_{22}\,\tilde{\gamma}^{11}+\tilde{\gamma}_{13}\,\tilde{\gamma}_{22}\,\tilde{\gamma}_{23}\,\tilde{\gamma}^{12}-\tilde{\gamma}_{13}\,(\tilde{\gamma}_{23})^2\tilde{\gamma}^{13}+\tilde{\gamma}_{23}\,\tilde{\gamma}^{12}\,\tilde{\gamma}^{13}\\
%
%
\Lambda_{19}&=&
3\,(\tilde{\gamma}_{12})^2\tilde{\gamma}_{13}\,(\tilde{\gamma}_{23})^{4}-3\,\tilde{\gamma}_{12}\,(\tilde{\gamma}_{13})^2\tilde{\gamma}_{22}\,(\tilde{\gamma}_{23})^3+3\,{\tilde{\gamma}_{12}}^{2}\tilde{\gamma}_{13}\,(\tilde{\gamma}_{23})^2\tilde{\gamma}^{11}-3\,\tilde{\gamma}_{12}\,(\tilde{\gamma}_{13})^{2}\tilde{\gamma}_{22}\,\tilde{\gamma}_{23}\,\tilde{\gamma}^{11}+(\tilde{\gamma}_{13})^3(\tilde{\gamma}_{22})^{2}\tilde{\gamma}^{11}\nonumber\\
&&+(\tilde{\gamma}_{13})^2(\tilde{\gamma}_{22})^2\tilde{\gamma}_{23}\,\tilde{\gamma}^{12}-2
\,(\tilde{\gamma}_{13})^2\tilde{\gamma}_{22}\,(\tilde{\gamma}_{23})^2\tilde{\gamma}^{13}+2\,\tilde{\gamma}_{13}\,\tilde{\gamma}_{22}\,\tilde{\gamma}_{23}\,\tilde{\gamma}^{12}\,\tilde{\gamma}^{13}-\tilde{\gamma}_{13}\,(\tilde{\gamma}_{23})^2
(\tilde{\gamma}^{13})^2-\tilde{\gamma}_{12}\,(\tilde{\gamma}_{23})^3+\tilde{\gamma}_{13}\,\tilde{\gamma}_{22}\,(\tilde{\gamma}_{23})^{2}\nonumber\\
&&+\tilde{\gamma}_{23}\,\tilde{\gamma}^{12}\,(\tilde{\gamma}^{13})^2-\tilde{\gamma}_{12}\,\tilde{\gamma}_{23}\,\tilde{\gamma}^{11}+\tilde{\gamma}_{13}\,\tilde{\gamma}_{22}\,\tilde{\gamma}^{11}\\
\Lambda_{20}&=&
3\,(\tilde{\gamma}_{23})^2\tilde{\gamma}_{13}\,(\tilde{\gamma}_{12})^2-3
\,(\tilde{\gamma}_{22})^2(\tilde{\gamma}_{13})^3+(\tilde{\gamma}_{12})^2\tilde{\gamma}_{13}\,\tilde{\gamma}^{11}+
\tilde{\gamma}_{12}\,\tilde{\gamma}_{13}\,\tilde{\gamma}_{22}\,\tilde{\gamma}^{12}-5\,(\tilde{\gamma}_{13})^2\tilde{\gamma}_{22}\,\tilde{\gamma}^{13}+\tilde{\gamma}_{12}\,\tilde{\gamma}^{12}\,\tilde{\gamma}^{13}-\tilde{\gamma}_{13}\,(\tilde{\gamma}^{13})^{2}+\tilde{\gamma}^{13}\\
\Lambda_{21}&=&
6\,(\tilde{\gamma}_{12})^2{\tilde{\gamma}_{13}}^{2}(\tilde{\gamma}_{23})^3-8\,\tilde{\gamma}_{12}\,(\tilde{\gamma}_{13})^3\tilde{\gamma}_{22}\,(\tilde{\gamma}_{23})^{2}+2\,(\tilde{\gamma}_{13})^4(\tilde{\gamma}_{22})^2\tilde{\gamma}_{23}+4\,(\tilde{\gamma}_{12})^2(\tilde{\gamma}_{13})^{2}\tilde{\gamma}_{23}\,\tilde{\gamma}^{11}+4\,(\tilde{\gamma}_{12})^2\tilde{\gamma}_{13}\,(\tilde{\gamma}_{23})^{2}\tilde{\gamma}^{12}\nonumber\\
&&-2\,\tilde{\gamma}_{12}\,(\tilde{\gamma}_{13})^3\tilde{\gamma}_{22}\,\tilde{\gamma}^{11}
-4\,\tilde{\gamma}_{12}\,(\tilde{\gamma}_{13})^2(\tilde{\gamma}_{23})^2\tilde{\gamma}^{13}-2\,{\tilde{\gamma}_{13}}^{3}(\tilde{\gamma}_{22})^2\tilde{\gamma}^{12}+2\,(\tilde{\gamma}_{13})^3\tilde{\gamma}_{22}\,\tilde{\gamma}_{23}\,
\tilde{\gamma}^{13}+(\tilde{\gamma}_{12})^2\tilde{\gamma}_{13}\,\tilde{\gamma}^{11}\,\tilde{\gamma}^{12}\nonumber\\
&&-\tilde{\gamma}_{12}
\,(\tilde{\gamma}_{13})^2\tilde{\gamma}^{11}\,\tilde{\gamma}^{13}+\tilde{\gamma}_{12}\,\tilde{\gamma}_{13}\,\tilde{\gamma}_{22}\,(\tilde{\gamma}^{12})^2-4\,(\tilde{\gamma}_{13})^2\tilde{\gamma}_{22}\,\tilde{\gamma}^{12}\,\tilde{\gamma}^{13}+(\tilde{\gamma}_{13})^2\tilde{\gamma}_{23}\,(\tilde{\gamma}^{13})^2\nonumber\\
&&-2\,\tilde{\gamma}_{12}\,\tilde{\gamma}_{13}\,
(\tilde{\gamma}_{23})^2+\tilde{\gamma}_{12}\,(\tilde{\gamma}^{12})^2\tilde{\gamma}^{13}+2\,{\tilde{\gamma}_{13}}^{2}\tilde{\gamma}_{22}\,\tilde{\gamma}_{23}-2\,\tilde{\gamma}_{13}\,\tilde{\gamma}^{12}\,(\tilde{\gamma}^{13})^2-\tilde{\gamma}_{12}\,\tilde{\gamma}_{13}\,\tilde{\gamma}^{11}-\tilde{\gamma}_{13}\,\tilde{\gamma}_{22}\,\tilde{\gamma}^{12}+\tilde{\gamma}_{13}\,\tilde{\gamma}_{23}\,\tilde{\gamma}^{13}-\tilde{\gamma}^{12}\,\tilde{\gamma}^{13}\\
\Lambda_{22}&=&
4\,(\tilde{\gamma}_{12})^3\tilde{\gamma}_{13}\,(\tilde{\gamma}_{23})^3-6\,(\tilde{\gamma}_{12})^{2}(\tilde{\gamma}_{13})^2\tilde{\gamma}_{22}\,(\tilde{\gamma}_{23})^2+2\,{\tilde{\gamma}_{13}}^{4}(\tilde{\gamma}_{22})^3+2\,(\tilde{\gamma}_{12})^3\tilde{\gamma}_{13}\,\tilde{\gamma}_{23}\,\tilde{\gamma}^{11}
-
(\tilde{\gamma}_{12})^2(\tilde{\gamma}_{13})^2\tilde{\gamma}_{22}\,\tilde{\gamma}^{11}\nonumber\\
&&+\tilde{\gamma}_{12}\,(\tilde{\gamma}_{13})^{2}(\tilde{\gamma}_{22})^2\tilde{\gamma}^{12}
+(\tilde{\gamma}_{13})^3(\tilde{\gamma}_{22})^2\tilde{\gamma}^{13}+2\,\tilde{\gamma}_{12}\,\tilde{\gamma}_{13}\,\tilde{\gamma}_{22}\,\tilde{\gamma}^{12}\,\tilde{\gamma}^{13}-3\,(\tilde{\gamma}_{13})^{2}\tilde{\gamma}_{22}\,(\tilde{\gamma}^{13})^2+\tilde{\gamma}_{12}\,\tilde{\gamma}^{12}\,(\tilde{\gamma}^{13})^{2}\nonumber\\
&&-\tilde{\gamma}_{13}\,(\tilde{\gamma}^{13})^{3}-(\tilde{\gamma}^{13})^2
\end{eqnarray}

\bibliographystyle{apsrev4-1} 
\bibliography{referencesZ4}

\begin{thebibliography}{121}%
\makeatletter
\providecommand \@ifxundefined [1]{%
 \@ifx{#1\undefined}
}%
\providecommand \@ifnum [1]{%
 \ifnum #1\expandafter \@firstoftwo
 \else \expandafter \@secondoftwo
 \fi
}%
\providecommand \@ifx [1]{%
 \ifx #1\expandafter \@firstoftwo
 \else \expandafter \@secondoftwo
 \fi
}%
\providecommand \natexlab [1]{#1}%
\providecommand \enquote  [1]{``#1''}%
\providecommand \bibnamefont  [1]{#1}%
\providecommand \bibfnamefont [1]{#1}%
\providecommand \citenamefont [1]{#1}%
\providecommand \href@noop [0]{\@secondoftwo}%
\providecommand \href [0]{\begingroup \@sanitize@url \@href}%
\providecommand \@href[1]{\@@startlink{#1}\@@href}%
\providecommand \@@href[1]{\endgroup#1\@@endlink}%
\providecommand \@sanitize@url [0]{\catcode `\\12\catcode `\$12\catcode
  `\&12\catcode `\#12\catcode `\^12\catcode `\_12\catcode `\%12\relax}%
\providecommand \@@startlink[1]{}%
\providecommand \@@endlink[0]{}%
\providecommand \url  [0]{\begingroup\@sanitize@url \@url }%
\providecommand \@url [1]{\endgroup\@href {#1}{\urlprefix }}%
\providecommand \urlprefix  [0]{URL }%
\providecommand \Eprint [0]{\href }%
\providecommand \doibase [0]{http://dx.doi.org/}%
\providecommand \selectlanguage [0]{\@gobble}%
\providecommand \bibinfo  [0]{\@secondoftwo}%
\providecommand \bibfield  [0]{\@secondoftwo}%
\providecommand \translation [1]{[#1]}%
\providecommand \BibitemOpen [0]{}%
\providecommand \bibitemStop [0]{}%
\providecommand \bibitemNoStop [0]{.\EOS\space}%
\providecommand \EOS [0]{\spacefactor3000\relax}%
\providecommand \BibitemShut  [1]{\csname bibitem#1\endcsname}%
\let\auto@bib@innerbib\@empty
\bibitem [{\citenamefont {Abbott}\ \emph {et~al.}(2016)\citenamefont {Abbott},
  \citenamefont {Abbott}, \citenamefont {Abbott}, \citenamefont {Abernathy},
  \citenamefont {Acernese}, \citenamefont {Ackley}, \citenamefont {Adams},
  \citenamefont {Adams},\ and\ \citenamefont {Addesso}}]{Abbott2016}%
  \BibitemOpen
  \bibfield  {author} {\bibinfo {author} {\bibfnamefont {B.~P.}\ \bibnamefont
  {Abbott}}, \bibinfo {author} {\bibfnamefont {R.}~\bibnamefont {Abbott}},
  \bibinfo {author} {\bibfnamefont {T.~D.}\ \bibnamefont {Abbott}}, \bibinfo
  {author} {\bibfnamefont {M.~R.}\ \bibnamefont {Abernathy}}, \bibinfo {author}
  {\bibfnamefont {F.}~\bibnamefont {Acernese}}, \bibinfo {author}
  {\bibfnamefont {K.}~\bibnamefont {Ackley}}, \bibinfo {author} {\bibfnamefont
  {C.}~\bibnamefont {Adams}}, \bibinfo {author} {\bibfnamefont
  {T.}~\bibnamefont {Adams}}, \ and\ \bibinfo {author} {\bibfnamefont
  {P.~e.~a.}\ \bibnamefont {Addesso}} (\bibinfo {collaboration} {LIGO
  Scientific Collaboration and Virgo Collaboration}),\ }\href {\doibase
  10.1103/PhysRevLett.116.061102} {\bibfield  {journal} {\bibinfo  {journal}
  {Phys. Rev. Lett.}\ }\textbf {\bibinfo {volume} {116}},\ \bibinfo {pages}
  {061102} (\bibinfo {year} {2016})}\BibitemShut {NoStop}%
\bibitem [{\citenamefont {{Bailes}}\ \emph {et~al.}(2021)\citenamefont
  {{Bailes}}, \citenamefont {{Berger}}, \citenamefont {{Brady}}, \citenamefont
  {{Branchesi}}, \citenamefont {{Danzmann}}, \citenamefont {{Evans}},
  \citenamefont {{Holley-Bockelmann}}, \citenamefont {{Iyer}}, \citenamefont
  {{Kajita}}, \citenamefont {{Katsanevas}}, \citenamefont {{Kramer}},
  \citenamefont {{Lazzarini}}, \citenamefont {{Lehner}}, \citenamefont
  {{Losurdo}}, \citenamefont {{L{\"u}ck}}, \citenamefont {{McClelland}},
  \citenamefont {{McLaughlin}}, \citenamefont {{Punturo}}, \citenamefont
  {{Ransom}}, \citenamefont {{Raychaudhury}}, \citenamefont {{Reitze}},
  \citenamefont {{Ricci}}, \citenamefont {{Rowan}}, \citenamefont {{Saito}},
  \citenamefont {{Sanders}}, \citenamefont {{Sathyaprakash}}, \citenamefont
  {{Schutz}}, \citenamefont {{Sesana}}, \citenamefont {{Shinkai}},
  \citenamefont {{Siemens}}, \citenamefont {{Shoemaker}}, \citenamefont
  {{Thorpe}}, \citenamefont {{van den Brand}},\ and\ \citenamefont
  {{Vitale}}}]{Bailes2021}%
  \BibitemOpen
  \bibfield  {author} {\bibinfo {author} {\bibfnamefont {M.}~\bibnamefont
  {{Bailes}}}, \bibinfo {author} {\bibfnamefont {B.~K.}\ \bibnamefont
  {{Berger}}}, \bibinfo {author} {\bibfnamefont {P.~R.}\ \bibnamefont
  {{Brady}}}, \bibinfo {author} {\bibfnamefont {M.}~\bibnamefont
  {{Branchesi}}}, \bibinfo {author} {\bibfnamefont {K.}~\bibnamefont
  {{Danzmann}}}, \bibinfo {author} {\bibfnamefont {M.}~\bibnamefont {{Evans}}},
  \bibinfo {author} {\bibfnamefont {K.}~\bibnamefont {{Holley-Bockelmann}}},
  \bibinfo {author} {\bibfnamefont {B.~R.}\ \bibnamefont {{Iyer}}}, \bibinfo
  {author} {\bibfnamefont {T.}~\bibnamefont {{Kajita}}}, \bibinfo {author}
  {\bibfnamefont {S.}~\bibnamefont {{Katsanevas}}}, \bibinfo {author}
  {\bibfnamefont {M.}~\bibnamefont {{Kramer}}}, \bibinfo {author}
  {\bibfnamefont {A.}~\bibnamefont {{Lazzarini}}}, \bibinfo {author}
  {\bibfnamefont {L.}~\bibnamefont {{Lehner}}}, \bibinfo {author}
  {\bibfnamefont {G.}~\bibnamefont {{Losurdo}}}, \bibinfo {author}
  {\bibfnamefont {H.}~\bibnamefont {{L{\"u}ck}}}, \bibinfo {author}
  {\bibfnamefont {D.~E.}\ \bibnamefont {{McClelland}}}, \bibinfo {author}
  {\bibfnamefont {M.~A.}\ \bibnamefont {{McLaughlin}}}, \bibinfo {author}
  {\bibfnamefont {M.}~\bibnamefont {{Punturo}}}, \bibinfo {author}
  {\bibfnamefont {S.}~\bibnamefont {{Ransom}}}, \bibinfo {author}
  {\bibfnamefont {S.}~\bibnamefont {{Raychaudhury}}}, \bibinfo {author}
  {\bibfnamefont {D.~H.}\ \bibnamefont {{Reitze}}}, \bibinfo {author}
  {\bibfnamefont {F.}~\bibnamefont {{Ricci}}}, \bibinfo {author} {\bibfnamefont
  {S.}~\bibnamefont {{Rowan}}}, \bibinfo {author} {\bibfnamefont
  {Y.}~\bibnamefont {{Saito}}}, \bibinfo {author} {\bibfnamefont {G.~H.}\
  \bibnamefont {{Sanders}}}, \bibinfo {author} {\bibfnamefont {B.~S.}\
  \bibnamefont {{Sathyaprakash}}}, \bibinfo {author} {\bibfnamefont {B.~F.}\
  \bibnamefont {{Schutz}}}, \bibinfo {author} {\bibfnamefont {A.}~\bibnamefont
  {{Sesana}}}, \bibinfo {author} {\bibfnamefont {H.}~\bibnamefont {{Shinkai}}},
  \bibinfo {author} {\bibfnamefont {X.}~\bibnamefont {{Siemens}}}, \bibinfo
  {author} {\bibfnamefont {D.~H.}\ \bibnamefont {{Shoemaker}}}, \bibinfo
  {author} {\bibfnamefont {J.}~\bibnamefont {{Thorpe}}}, \bibinfo {author}
  {\bibfnamefont {J.~F.~J.}\ \bibnamefont {{van den Brand}}}, \ and\ \bibinfo
  {author} {\bibfnamefont {S.}~\bibnamefont {{Vitale}}},\ }\href {\doibase
  10.1038/s42254-021-00303-8} {\bibfield  {journal} {\bibinfo  {journal}
  {Nature Reviews Physics}\ }\textbf {\bibinfo {volume} {3}},\ \bibinfo {pages}
  {344} (\bibinfo {year} {2021})}\BibitemShut {NoStop}%
\bibitem [{\citenamefont {Jim\'enez-Forteza}\ \emph {et~al.}(2017)\citenamefont
  {Jim\'enez-Forteza}, \citenamefont {Keitel}, \citenamefont {Husa},
  \citenamefont {Hannam}, \citenamefont {Khan},\ and\ \citenamefont
  {P\"urrer}}]{Jimenez2017}%
  \BibitemOpen
  \bibfield  {author} {\bibinfo {author} {\bibfnamefont {X.}~\bibnamefont
  {Jim\'enez-Forteza}}, \bibinfo {author} {\bibfnamefont {D.}~\bibnamefont
  {Keitel}}, \bibinfo {author} {\bibfnamefont {S.}~\bibnamefont {Husa}},
  \bibinfo {author} {\bibfnamefont {M.}~\bibnamefont {Hannam}}, \bibinfo
  {author} {\bibfnamefont {S.}~\bibnamefont {Khan}}, \ and\ \bibinfo {author}
  {\bibfnamefont {M.}~\bibnamefont {P\"urrer}},\ }\href {\doibase
  10.1103/PhysRevD.95.064024} {\bibfield  {journal} {\bibinfo  {journal} {Phys.
  Rev. D}\ }\textbf {\bibinfo {volume} {95}},\ \bibinfo {pages} {064024}
  (\bibinfo {year} {2017})}\BibitemShut {NoStop}%
\bibitem [{\citenamefont {Abbott}\ \emph {et~al.}(2019)\citenamefont {Abbott},
  \citenamefont {Abbott}, \citenamefont {Abbott}, \citenamefont {Abraham},
  \citenamefont {Acernese}, \citenamefont {Ackley},\ and\ \citenamefont
  {Adams}}]{Abbott2019}%
  \BibitemOpen
  \bibfield  {author} {\bibinfo {author} {\bibfnamefont {B.~P.}\ \bibnamefont
  {Abbott}}, \bibinfo {author} {\bibfnamefont {R.}~\bibnamefont {Abbott}},
  \bibinfo {author} {\bibfnamefont {T.~D.}\ \bibnamefont {Abbott}}, \bibinfo
  {author} {\bibfnamefont {S.}~\bibnamefont {Abraham}}, \bibinfo {author}
  {\bibfnamefont {F.}~\bibnamefont {Acernese}}, \bibinfo {author}
  {\bibfnamefont {K.}~\bibnamefont {Ackley}}, \ and\ \bibinfo {author}
  {\bibfnamefont {C.~e.~a.}\ \bibnamefont {Adams}} (\bibinfo {collaboration}
  {LIGO Scientific Collaboration and Virgo Collaboration}),\ }\href {\doibase
  10.1103/PhysRevX.9.031040} {\bibfield  {journal} {\bibinfo  {journal} {Phys.
  Rev. X}\ }\textbf {\bibinfo {volume} {9}},\ \bibinfo {pages} {031040}
  (\bibinfo {year} {2019})}\BibitemShut {NoStop}%
\bibitem [{\citenamefont {{Event Horizon Telescope
  Collaboration}}(2022{\natexlab{a}})}]{EventHorizon2022a}%
  \BibitemOpen
  \bibfield  {author} {\bibinfo {author} {\bibnamefont {{Event Horizon
  Telescope Collaboration}}},\ }\href {\doibase 10.3847/2041-8213/ac6674}
  {\bibfield  {journal} {\bibinfo  {journal} {Astrophysical Journal Letters}\
  }\textbf {\bibinfo {volume} {930}},\ \bibinfo {eid} {L12} (\bibinfo {year}
  {2022}{\natexlab{a}})}\BibitemShut {NoStop}%
\bibitem [{\citenamefont {{Event Horizon Telescope
  Collaboration}}(2022{\natexlab{b}})}]{EventHorizon2022b}%
  \BibitemOpen
  \bibfield  {author} {\bibinfo {author} {\bibnamefont {{Event Horizon
  Telescope Collaboration}}},\ }\href {\doibase 10.3847/2041-8213/ac6429}
  {\bibfield  {journal} {\bibinfo  {journal} {Astrophysical Journal Letters}\
  }\textbf {\bibinfo {volume} {930}},\ \bibinfo {eid} {L14} (\bibinfo {year}
  {2022}{\natexlab{b}})}\BibitemShut {NoStop}%
\bibitem [{\citenamefont {{Aranguren}}\ \emph {et~al.}(2023)\citenamefont
  {{Aranguren}}, \citenamefont {{Font}}, \citenamefont {{Sanchis-Gual}},\ and\
  \citenamefont {{Vera}}}]{Aranguren2023}%
  \BibitemOpen
  \bibfield  {author} {\bibinfo {author} {\bibfnamefont {E.}~\bibnamefont
  {{Aranguren}}}, \bibinfo {author} {\bibfnamefont {J.~A.}\ \bibnamefont
  {{Font}}}, \bibinfo {author} {\bibfnamefont {N.}~\bibnamefont
  {{Sanchis-Gual}}}, \ and\ \bibinfo {author} {\bibfnamefont {R.}~\bibnamefont
  {{Vera}}},\ }\href {\doibase 10.1103/PhysRevD.108.104065} {\bibfield
  {journal} {\bibinfo  {journal} {Physical Review D}\ }\textbf {\bibinfo
  {volume} {108}},\ \bibinfo {eid} {104065} (\bibinfo {year}
  {2023})}\BibitemShut {NoStop}%
\bibitem [{\citenamefont {{Camilletti}}\ \emph {et~al.}(2022)\citenamefont
  {{Camilletti}}, \citenamefont {{Chiesa}}, \citenamefont {{Ricigliano}},
  \citenamefont {{Perego}}, \citenamefont {{Lippold}}, \citenamefont
  {{Padamata}}, \citenamefont {{Bernuzzi}}, \citenamefont {{Radice}},
  \citenamefont {{Logoteta}},\ and\ \citenamefont
  {{Guercilena}}}]{Camilletti2022}%
  \BibitemOpen
  \bibfield  {author} {\bibinfo {author} {\bibfnamefont {A.}~\bibnamefont
  {{Camilletti}}}, \bibinfo {author} {\bibfnamefont {L.}~\bibnamefont
  {{Chiesa}}}, \bibinfo {author} {\bibfnamefont {G.}~\bibnamefont
  {{Ricigliano}}}, \bibinfo {author} {\bibfnamefont {A.}~\bibnamefont
  {{Perego}}}, \bibinfo {author} {\bibfnamefont {L.~C.}\ \bibnamefont
  {{Lippold}}}, \bibinfo {author} {\bibfnamefont {S.}~\bibnamefont
  {{Padamata}}}, \bibinfo {author} {\bibfnamefont {S.}~\bibnamefont
  {{Bernuzzi}}}, \bibinfo {author} {\bibfnamefont {D.}~\bibnamefont
  {{Radice}}}, \bibinfo {author} {\bibfnamefont {D.}~\bibnamefont
  {{Logoteta}}}, \ and\ \bibinfo {author} {\bibfnamefont {F.~M.}\ \bibnamefont
  {{Guercilena}}},\ }\href {\doibase 10.1093/mnras/stac2333} {\bibfield
  {journal} {\bibinfo  {journal} {Monthly Notices of the Royal Astronomical
  Society}\ }\textbf {\bibinfo {volume} {516}},\ \bibinfo {pages} {4760}
  (\bibinfo {year} {2022})}\BibitemShut {NoStop}%
\bibitem [{\citenamefont {{Palenzuela}}\ \emph {et~al.}(2022)\citenamefont
  {{Palenzuela}}, \citenamefont {{Liebling}},\ and\ \citenamefont
  {{Minano}}}]{Palenzuela2022}%
  \BibitemOpen
  \bibfield  {author} {\bibinfo {author} {\bibfnamefont {C.}~\bibnamefont
  {{Palenzuela}}}, \bibinfo {author} {\bibfnamefont {S.}~\bibnamefont
  {{Liebling}}}, \ and\ \bibinfo {author} {\bibfnamefont {B.}~\bibnamefont
  {{Minano}}},\ }\href {\doibase 10.1103/PhysRevD.105.103020} {\bibfield
  {journal} {\bibinfo  {journal} {Physical Review D}\ }\textbf {\bibinfo
  {volume} {105}},\ \bibinfo {eid} {103020} (\bibinfo {year}
  {2022})}\BibitemShut {NoStop}%
\bibitem [{\citenamefont {{Camilletti}}\ \emph {et~al.}(2024)\citenamefont
  {{Camilletti}}, \citenamefont {{Perego}}, \citenamefont {{Guercilena}},
  \citenamefont {{Bernuzzi}},\ and\ \citenamefont {{Radice}}}]{Camilletti2024}%
  \BibitemOpen
  \bibfield  {author} {\bibinfo {author} {\bibfnamefont {A.}~\bibnamefont
  {{Camilletti}}}, \bibinfo {author} {\bibfnamefont {A.}~\bibnamefont
  {{Perego}}}, \bibinfo {author} {\bibfnamefont {F.~M.}\ \bibnamefont
  {{Guercilena}}}, \bibinfo {author} {\bibfnamefont {S.}~\bibnamefont
  {{Bernuzzi}}}, \ and\ \bibinfo {author} {\bibfnamefont {D.}~\bibnamefont
  {{Radice}}},\ }\href {\doibase 10.1103/PhysRevD.109.063023} {\bibfield
  {journal} {\bibinfo  {journal} {Physical Review D}\ }\textbf {\bibinfo
  {volume} {109}},\ \bibinfo {eid} {063023} (\bibinfo {year}
  {2024})}\BibitemShut {NoStop}%
\bibitem [{\citenamefont {{Breschi}}\ \emph {et~al.}(2024)\citenamefont
  {{Breschi}}, \citenamefont {{Bernuzzi}}, \citenamefont {{Chakravarti}},
  \citenamefont {{Camilletti}}, \citenamefont {{Prakash}},\ and\ \citenamefont
  {{Perego}}}]{Breschi2024}%
  \BibitemOpen
  \bibfield  {author} {\bibinfo {author} {\bibfnamefont {M.}~\bibnamefont
  {{Breschi}}}, \bibinfo {author} {\bibfnamefont {S.}~\bibnamefont
  {{Bernuzzi}}}, \bibinfo {author} {\bibfnamefont {K.}~\bibnamefont
  {{Chakravarti}}}, \bibinfo {author} {\bibfnamefont {A.}~\bibnamefont
  {{Camilletti}}}, \bibinfo {author} {\bibfnamefont {A.}~\bibnamefont
  {{Prakash}}}, \ and\ \bibinfo {author} {\bibfnamefont {A.}~\bibnamefont
  {{Perego}}},\ }\href {\doibase 10.1103/PhysRevD.109.064009} {\bibfield
  {journal} {\bibinfo  {journal} {Physical Review D}\ }\textbf {\bibinfo
  {volume} {109}},\ \bibinfo {eid} {064009} (\bibinfo {year}
  {2024})}\BibitemShut {NoStop}%
\bibitem [{\citenamefont {{Topolski}}\ \emph {et~al.}(2024)\citenamefont
  {{Topolski}}, \citenamefont {{Tootle}},\ and\ \citenamefont
  {{Rezzolla}}}]{Topolski2024}%
  \BibitemOpen
  \bibfield  {author} {\bibinfo {author} {\bibfnamefont {K.}~\bibnamefont
  {{Topolski}}}, \bibinfo {author} {\bibfnamefont {S.~D.}\ \bibnamefont
  {{Tootle}}}, \ and\ \bibinfo {author} {\bibfnamefont {L.}~\bibnamefont
  {{Rezzolla}}},\ }\href {\doibase 10.3847/1538-4357/ad0152} {\bibfield
  {journal} {\bibinfo  {journal} {\apj}\ }\textbf {\bibinfo {volume} {960}},\
  \bibinfo {eid} {86} (\bibinfo {year} {2024})}\BibitemShut {NoStop}%
\bibitem [{\citenamefont {{Lindblom}}\ \emph {et~al.}(2006)\citenamefont
  {{Lindblom}}, \citenamefont {{Scheel}}, \citenamefont {{Kidder}},
  \citenamefont {{Owen}},\ and\ \citenamefont {{Rinne}}}]{Lindblom2006}%
  \BibitemOpen
  \bibfield  {author} {\bibinfo {author} {\bibfnamefont {L.}~\bibnamefont
  {{Lindblom}}}, \bibinfo {author} {\bibfnamefont {M.~A.}\ \bibnamefont
  {{Scheel}}}, \bibinfo {author} {\bibfnamefont {L.~E.}\ \bibnamefont
  {{Kidder}}}, \bibinfo {author} {\bibfnamefont {R.}~\bibnamefont {{Owen}}}, \
  and\ \bibinfo {author} {\bibfnamefont {O.}~\bibnamefont {{Rinne}}},\
  }\href@noop {} {\bibfield  {journal} {\bibinfo  {journal} {Classical and
  Quantum Gravity}\ }\textbf {\bibinfo {volume} {23}},\ \bibinfo {pages} {S447}
  (\bibinfo {year} {2006})}\BibitemShut {NoStop}%
\bibitem [{\citenamefont {{Pretorius}}(2005)}]{Pretorius2005a}%
  \BibitemOpen
  \bibfield  {author} {\bibinfo {author} {\bibfnamefont {F.}~\bibnamefont
  {{Pretorius}}},\ }\href@noop {} {\bibfield  {journal} {\bibinfo  {journal}
  {\prl}\ }\textbf {\bibinfo {volume} {95}},\ \bibinfo {eid} {121101} (\bibinfo
  {year} {2005})}\BibitemShut {NoStop}%
\bibitem [{\citenamefont {Baumgarte}\ and\ \citenamefont
  {Shapiro}(1998)}]{Baumgarte99}%
  \BibitemOpen
  \bibfield  {author} {\bibinfo {author} {\bibfnamefont {T.~W.}\ \bibnamefont
  {Baumgarte}}\ and\ \bibinfo {author} {\bibfnamefont {S.~L.}\ \bibnamefont
  {Shapiro}},\ }\href@noop {} {\bibfield  {journal} {\bibinfo  {journal}
  {Physical Review D}\ }\textbf {\bibinfo {volume} {59}},\ \bibinfo {pages}
  {024007} (\bibinfo {year} {1998})}\BibitemShut {NoStop}%
\bibitem [{\citenamefont {Alcubierre}(2008)}]{Alcubierre:2008}%
  \BibitemOpen
  \bibfield  {author} {\bibinfo {author} {\bibfnamefont {M.}~\bibnamefont
  {Alcubierre}},\ }\href@noop {} {\emph {\bibinfo {title} {{Introduction to 3+1
  numerical relativity}}}},\ Vol.\ \bibinfo {volume} {140}\ (\bibinfo
  {publisher} {Oxford University Press},\ \bibinfo {year} {2008})\BibitemShut
  {NoStop}%
\bibitem [{\citenamefont {Baumgarte}\ and\ \citenamefont
  {Shapiro}(2010)}]{Baumgarte2010}%
  \BibitemOpen
  \bibfield  {author} {\bibinfo {author} {\bibfnamefont {T.~W.}\ \bibnamefont
  {Baumgarte}}\ and\ \bibinfo {author} {\bibfnamefont {S.~L.}\ \bibnamefont
  {Shapiro}},\ }\href@noop {} {\emph {\bibinfo {title} {{Numerical relativity:
  solving Einstein's equations on the computer}}}}\ (\bibinfo  {publisher}
  {Cambridge University Press},\ \bibinfo {year} {2010})\BibitemShut {NoStop}%
\bibitem [{\citenamefont {Bona}\ and\ \citenamefont
  {Palenzuela-Luque}(2005)}]{Bona-and-Palenzuela-Luque-2005:numrel-book}%
  \BibitemOpen
  \bibfield  {author} {\bibinfo {author} {\bibfnamefont {C.}~\bibnamefont
  {Bona}}\ and\ \bibinfo {author} {\bibfnamefont {C.}~\bibnamefont
  {Palenzuela-Luque}},\ }\href {\doibase 10.1007/b135928} {\emph {\bibinfo
  {title} {Elements of Numerical Relativity}}}\ (\bibinfo  {publisher}
  {Springer-Verlag},\ \bibinfo {address} {Berlin},\ \bibinfo {year}
  {2005})\BibitemShut {NoStop}%
\bibitem [{\citenamefont {{Brown}}\ \emph {et~al.}(2012)\citenamefont
  {{Brown}}, \citenamefont {{Diener}}, \citenamefont {{Field}}, \citenamefont
  {{Hesthaven}}, \citenamefont {{Herrmann}}, \citenamefont {{Mrou{\'e}}},
  \citenamefont {{Sarbach}}, \citenamefont {{Schnetter}}, \citenamefont
  {{Tiglio}},\ and\ \citenamefont {{Wagman}}}]{Brown2012}%
  \BibitemOpen
  \bibfield  {author} {\bibinfo {author} {\bibfnamefont {J.~D.}\ \bibnamefont
  {{Brown}}}, \bibinfo {author} {\bibfnamefont {P.}~\bibnamefont {{Diener}}},
  \bibinfo {author} {\bibfnamefont {S.~E.}\ \bibnamefont {{Field}}}, \bibinfo
  {author} {\bibfnamefont {J.~S.}\ \bibnamefont {{Hesthaven}}}, \bibinfo
  {author} {\bibfnamefont {F.}~\bibnamefont {{Herrmann}}}, \bibinfo {author}
  {\bibfnamefont {A.~H.}\ \bibnamefont {{Mrou{\'e}}}}, \bibinfo {author}
  {\bibfnamefont {O.}~\bibnamefont {{Sarbach}}}, \bibinfo {author}
  {\bibfnamefont {E.}~\bibnamefont {{Schnetter}}}, \bibinfo {author}
  {\bibfnamefont {M.}~\bibnamefont {{Tiglio}}}, \ and\ \bibinfo {author}
  {\bibfnamefont {M.}~\bibnamefont {{Wagman}}},\ }\href {\doibase
  10.1103/PhysRevD.85.084004} {\bibfield  {journal} {\bibinfo  {journal}
  {\prd}\ }\textbf {\bibinfo {volume} {85}},\ \bibinfo {eid} {084004} (\bibinfo
  {year} {2012})}\BibitemShut {NoStop}%
\bibitem [{\citenamefont {Dumbser}\ \emph
  {et~al.}(2018{\natexlab{a}})\citenamefont {Dumbser}, \citenamefont
  {Guercilena}, \citenamefont {K{\"o}ppel}, \citenamefont {Rezzolla},\ and\
  \citenamefont {Zanotti}}]{Dumbser2017strongly}%
  \BibitemOpen
  \bibfield  {author} {\bibinfo {author} {\bibfnamefont {M.}~\bibnamefont
  {Dumbser}}, \bibinfo {author} {\bibfnamefont {F.}~\bibnamefont {Guercilena}},
  \bibinfo {author} {\bibfnamefont {S.}~\bibnamefont {K{\"o}ppel}}, \bibinfo
  {author} {\bibfnamefont {L.}~\bibnamefont {Rezzolla}}, \ and\ \bibinfo
  {author} {\bibfnamefont {O.}~\bibnamefont {Zanotti}},\ }\href@noop {}
  {\bibfield  {journal} {\bibinfo  {journal} {Physical Review D}\ }\textbf
  {\bibinfo {volume} {97}},\ \bibinfo {pages} {084053} (\bibinfo {year}
  {2018}{\natexlab{a}})}\BibitemShut {NoStop}%
\bibitem [{\citenamefont {Dumbser}\ \emph {et~al.}(2020)\citenamefont
  {Dumbser}, \citenamefont {Fambri}, \citenamefont {Gaburro},\ and\
  \citenamefont {Reinarz}}]{Dumbser2020GLM}%
  \BibitemOpen
  \bibfield  {author} {\bibinfo {author} {\bibfnamefont {M.}~\bibnamefont
  {Dumbser}}, \bibinfo {author} {\bibfnamefont {F.}~\bibnamefont {Fambri}},
  \bibinfo {author} {\bibfnamefont {E.}~\bibnamefont {Gaburro}}, \ and\
  \bibinfo {author} {\bibfnamefont {A.}~\bibnamefont {Reinarz}},\ }\href@noop
  {} {\bibfield  {journal} {\bibinfo  {journal} {Journal of Computational
  Physics}\ }\textbf {\bibinfo {volume} {404}},\ \bibinfo {pages} {109088}
  (\bibinfo {year} {2020})}\BibitemShut {NoStop}%
\bibitem [{\citenamefont {{Teukolsky}}(2016)}]{Teukolsky2015}%
  \BibitemOpen
  \bibfield  {author} {\bibinfo {author} {\bibfnamefont {S.~A.}\ \bibnamefont
  {{Teukolsky}}},\ }\href {\doibase 10.1016/j.jcp.2016.02.031} {\bibfield
  {journal} {\bibinfo  {journal} {Journal of Computational Physics}\ }\textbf
  {\bibinfo {volume} {312}},\ \bibinfo {pages} {333} (\bibinfo {year}
  {2016})}\BibitemShut {NoStop}%
\bibitem [{\citenamefont {Miller}\ and\ \citenamefont
  {Schnetter}(2016)}]{Miller2017}%
  \BibitemOpen
  \bibfield  {author} {\bibinfo {author} {\bibfnamefont {J.~M.}\ \bibnamefont
  {Miller}}\ and\ \bibinfo {author} {\bibfnamefont {E.}~\bibnamefont
  {Schnetter}},\ }\href {\doibase 10.1088/1361-6382/34/1/015003} {\bibfield
  {journal} {\bibinfo  {journal} {Classical and Quantum Gravity}\ }\textbf
  {\bibinfo {volume} {34}},\ \bibinfo {pages} {015003} (\bibinfo {year}
  {2016})}\BibitemShut {NoStop}%
\bibitem [{\citenamefont {{Kidder}}\ \emph {et~al.}(2017)\citenamefont
  {{Kidder}}, \citenamefont {{Field}}, \citenamefont {{Foucart}}, \citenamefont
  {{Schnetter}}, \citenamefont {{Teukolsky}}, \citenamefont {{Bohn}},
  \citenamefont {{Deppe}}, \citenamefont {{Diener}}, \citenamefont
  {{H{\'e}bert}}, \citenamefont {{Lippuner}}, \citenamefont {{Miller}},
  \citenamefont {{Ott}}, \citenamefont {{Scheel}},\ and\ \citenamefont
  {{Vincent}}}]{Kidder2017}%
  \BibitemOpen
  \bibfield  {author} {\bibinfo {author} {\bibfnamefont {L.~E.}\ \bibnamefont
  {{Kidder}}}, \bibinfo {author} {\bibfnamefont {S.~E.}\ \bibnamefont
  {{Field}}}, \bibinfo {author} {\bibfnamefont {F.}~\bibnamefont {{Foucart}}},
  \bibinfo {author} {\bibfnamefont {E.}~\bibnamefont {{Schnetter}}}, \bibinfo
  {author} {\bibfnamefont {S.~A.}\ \bibnamefont {{Teukolsky}}}, \bibinfo
  {author} {\bibfnamefont {A.}~\bibnamefont {{Bohn}}}, \bibinfo {author}
  {\bibfnamefont {N.}~\bibnamefont {{Deppe}}}, \bibinfo {author} {\bibfnamefont
  {P.}~\bibnamefont {{Diener}}}, \bibinfo {author} {\bibfnamefont
  {F.}~\bibnamefont {{H{\'e}bert}}}, \bibinfo {author} {\bibfnamefont
  {J.}~\bibnamefont {{Lippuner}}}, \bibinfo {author} {\bibfnamefont
  {J.}~\bibnamefont {{Miller}}}, \bibinfo {author} {\bibfnamefont {C.~D.}\
  \bibnamefont {{Ott}}}, \bibinfo {author} {\bibfnamefont {M.~A.}\ \bibnamefont
  {{Scheel}}}, \ and\ \bibinfo {author} {\bibfnamefont {T.}~\bibnamefont
  {{Vincent}}},\ }\href {\doibase 10.1016/j.jcp.2016.12.059} {\bibfield
  {journal} {\bibinfo  {journal} {Journal of Computational Physics}\ }\textbf
  {\bibinfo {volume} {335}},\ \bibinfo {pages} {84} (\bibinfo {year}
  {2017})}\BibitemShut {NoStop}%
\bibitem [{\citenamefont {{H{\'e}bert}}\ \emph {et~al.}(2018)\citenamefont
  {{H{\'e}bert}}, \citenamefont {{Kidder}},\ and\ \citenamefont
  {{Teukolsky}}}]{Hebert2018}%
  \BibitemOpen
  \bibfield  {author} {\bibinfo {author} {\bibfnamefont {F.}~\bibnamefont
  {{H{\'e}bert}}}, \bibinfo {author} {\bibfnamefont {L.~E.}\ \bibnamefont
  {{Kidder}}}, \ and\ \bibinfo {author} {\bibfnamefont {S.~A.}\ \bibnamefont
  {{Teukolsky}}},\ }\href {\doibase 10.1103/PhysRevD.98.044041} {\bibfield
  {journal} {\bibinfo  {journal} {Phys. Rev. D}\ }\textbf {\bibinfo {volume}
  {98}},\ \bibinfo {eid} {044041} (\bibinfo {year} {2018})}\BibitemShut
  {NoStop}%
\bibitem [{\citenamefont {{Tichy}}\ \emph {et~al.}(2021)\citenamefont
  {{Tichy}}, \citenamefont {{Adhikari}},\ and\ \citenamefont
  {{Ji}}}]{Tichy2021}%
  \BibitemOpen
  \bibfield  {author} {\bibinfo {author} {\bibfnamefont {W.}~\bibnamefont
  {{Tichy}}}, \bibinfo {author} {\bibfnamefont {A.}~\bibnamefont {{Adhikari}}},
  \ and\ \bibinfo {author} {\bibfnamefont {L.}~\bibnamefont {{Ji}}},\ }in\
  \href@noop {} {\emph {\bibinfo {booktitle} {APS April Meeting Abstracts}}},\
  \bibinfo {series} {APS Meeting Abstracts}, Vol.\ \bibinfo {volume} {2021}\
  (\bibinfo {year} {2021})\ p.\ \bibinfo {pages} {S08.003}\BibitemShut
  {NoStop}%
\bibitem [{\citenamefont {{Tichy}}\ \emph {et~al.}(2023)\citenamefont
  {{Tichy}}, \citenamefont {{Ji}}, \citenamefont {{Adhikari}}, \citenamefont
  {{Rashti}},\ and\ \citenamefont {{Pirog}}}]{Tichy2023}%
  \BibitemOpen
  \bibfield  {author} {\bibinfo {author} {\bibfnamefont {W.}~\bibnamefont
  {{Tichy}}}, \bibinfo {author} {\bibfnamefont {L.}~\bibnamefont {{Ji}}},
  \bibinfo {author} {\bibfnamefont {A.}~\bibnamefont {{Adhikari}}}, \bibinfo
  {author} {\bibfnamefont {A.}~\bibnamefont {{Rashti}}}, \ and\ \bibinfo
  {author} {\bibfnamefont {M.}~\bibnamefont {{Pirog}}},\ }\href {\doibase
  10.1088/1361-6382/acaae7} {\bibfield  {journal} {\bibinfo  {journal}
  {Classical and Quantum Gravity}\ }\textbf {\bibinfo {volume} {40}},\ \bibinfo
  {eid} {025004} (\bibinfo {year} {2023})}\BibitemShut {NoStop}%
\bibitem [{\citenamefont {Dumbser}\ \emph {et~al.}(2024)\citenamefont
  {Dumbser}, \citenamefont {Zanotti}, \citenamefont {Gaburro},\ and\
  \citenamefont {Peshkov}}]{DumbserZanottiGaburroPeshkov2023}%
  \BibitemOpen
  \bibfield  {author} {\bibinfo {author} {\bibfnamefont {M.}~\bibnamefont
  {Dumbser}}, \bibinfo {author} {\bibfnamefont {O.}~\bibnamefont {Zanotti}},
  \bibinfo {author} {\bibfnamefont {E.}~\bibnamefont {Gaburro}}, \ and\
  \bibinfo {author} {\bibfnamefont {I.}~\bibnamefont {Peshkov}},\ }\href
  {\doibase 10.1016/j.jcp.2024.112875} {\bibfield  {journal} {\bibinfo
  {journal} {J. Comput. Phys.}\ }\textbf {\bibinfo {volume} {504}},\ \bibinfo
  {pages} {112875} (\bibinfo {year} {2024})}\BibitemShut {NoStop}%
\bibitem [{\citenamefont {{Dumbser}}\ \emph {et~al.}(2024)\citenamefont
  {{Dumbser}}, \citenamefont {{Zanotti}},\ and\ \citenamefont
  {{Peshkov}}}]{DumbserZanottiPeshkov2024}%
  \BibitemOpen
  \bibfield  {author} {\bibinfo {author} {\bibfnamefont {M.}~\bibnamefont
  {{Dumbser}}}, \bibinfo {author} {\bibfnamefont {O.}~\bibnamefont
  {{Zanotti}}}, \ and\ \bibinfo {author} {\bibfnamefont {I.}~\bibnamefont
  {{Peshkov}}},\ }\href@noop {} {\bibfield  {journal} {\bibinfo  {journal}
  {Phys. Rev. D}\ }\textbf {\bibinfo {volume} {110}},\ \bibinfo {eid} {084015}
  (\bibinfo {year} {2024})}\BibitemShut {NoStop}%
\bibitem [{\citenamefont {{Deppe}}\ \emph {et~al.}(2024)\citenamefont
  {{Deppe}}, \citenamefont {{Foucart}}, \citenamefont {{Bonilla}},
  \citenamefont {{Boyle}}, \citenamefont {{Corso}}, \citenamefont {{Duez}},
  \citenamefont {{Giesler}}, \citenamefont {{H{\'e}bert}}, \citenamefont
  {{Kidder}}, \citenamefont {{Kim}}, \citenamefont {{Kumar}}, \citenamefont
  {{Legred}}, \citenamefont {{Lovelace}}, \citenamefont {{Most}}, \citenamefont
  {{Moxon}}, \citenamefont {{Nelli}}, \citenamefont {{Pfeiffer}}, \citenamefont
  {{Scheel}}, \citenamefont {{Teukolsky}}, \citenamefont {{Throwe}},\ and\
  \citenamefont {{Vu}}}]{Deppe2024}%
  \BibitemOpen
  \bibfield  {author} {\bibinfo {author} {\bibfnamefont {N.}~\bibnamefont
  {{Deppe}}}, \bibinfo {author} {\bibfnamefont {F.}~\bibnamefont {{Foucart}}},
  \bibinfo {author} {\bibfnamefont {M.~S.}\ \bibnamefont {{Bonilla}}}, \bibinfo
  {author} {\bibfnamefont {M.}~\bibnamefont {{Boyle}}}, \bibinfo {author}
  {\bibfnamefont {N.~J.}\ \bibnamefont {{Corso}}}, \bibinfo {author}
  {\bibfnamefont {M.~D.}\ \bibnamefont {{Duez}}}, \bibinfo {author}
  {\bibfnamefont {M.}~\bibnamefont {{Giesler}}}, \bibinfo {author}
  {\bibfnamefont {F.}~\bibnamefont {{H{\'e}bert}}}, \bibinfo {author}
  {\bibfnamefont {L.~E.}\ \bibnamefont {{Kidder}}}, \bibinfo {author}
  {\bibfnamefont {Y.}~\bibnamefont {{Kim}}}, \bibinfo {author} {\bibfnamefont
  {P.}~\bibnamefont {{Kumar}}}, \bibinfo {author} {\bibfnamefont
  {I.}~\bibnamefont {{Legred}}}, \bibinfo {author} {\bibfnamefont
  {G.}~\bibnamefont {{Lovelace}}}, \bibinfo {author} {\bibfnamefont {E.~R.}\
  \bibnamefont {{Most}}}, \bibinfo {author} {\bibfnamefont {J.}~\bibnamefont
  {{Moxon}}}, \bibinfo {author} {\bibfnamefont {K.~C.}\ \bibnamefont
  {{Nelli}}}, \bibinfo {author} {\bibfnamefont {H.~P.}\ \bibnamefont
  {{Pfeiffer}}}, \bibinfo {author} {\bibfnamefont {M.~A.}\ \bibnamefont
  {{Scheel}}}, \bibinfo {author} {\bibfnamefont {S.~A.}\ \bibnamefont
  {{Teukolsky}}}, \bibinfo {author} {\bibfnamefont {W.}~\bibnamefont
  {{Throwe}}}, \ and\ \bibinfo {author} {\bibfnamefont {N.~L.}\ \bibnamefont
  {{Vu}}},\ }\href@noop {} {\bibfield  {journal} {\bibinfo  {journal}
  {Classical and Quantum Gravity}\ }\textbf {\bibinfo {volume} {41}},\ \bibinfo
  {eid} {245002} (\bibinfo {year} {2024})}\BibitemShut {NoStop}%
\bibitem [{\citenamefont {Boyle}\ \emph
  {et~al.}(2007{\natexlab{a}})\citenamefont {Boyle}, \citenamefont {Lindblom},
  \citenamefont {Pfeiffer}, \citenamefont {Scheel},\ and\ \citenamefont
  {Kidder}}]{Boyle2007}%
  \BibitemOpen
  \bibfield  {author} {\bibinfo {author} {\bibfnamefont {M.}~\bibnamefont
  {Boyle}}, \bibinfo {author} {\bibfnamefont {L.}~\bibnamefont {Lindblom}},
  \bibinfo {author} {\bibfnamefont {H.~P.}\ \bibnamefont {Pfeiffer}}, \bibinfo
  {author} {\bibfnamefont {M.~A.}\ \bibnamefont {Scheel}}, \ and\ \bibinfo
  {author} {\bibfnamefont {L.~E.}\ \bibnamefont {Kidder}},\ }\href {\doibase
  10.1103/PhysRevD.75.024006} {\bibfield  {journal} {\bibinfo  {journal} {Phys.
  Rev. D}\ }\textbf {\bibinfo {volume} {75}},\ \bibinfo {pages} {024006}
  (\bibinfo {year} {2007}{\natexlab{a}})}\BibitemShut {NoStop}%
\bibitem [{\citenamefont {Boyle}\ \emph
  {et~al.}(2007{\natexlab{b}})\citenamefont {Boyle}, \citenamefont {Brown},
  \citenamefont {Kidder}, \citenamefont {Mrou\'e}, \citenamefont {Pfeiffer},
  \citenamefont {Scheel}, \citenamefont {Cook},\ and\ \citenamefont
  {Teukolsky}}]{Boyle2007b}%
  \BibitemOpen
  \bibfield  {author} {\bibinfo {author} {\bibfnamefont {M.}~\bibnamefont
  {Boyle}}, \bibinfo {author} {\bibfnamefont {D.~A.}\ \bibnamefont {Brown}},
  \bibinfo {author} {\bibfnamefont {L.~E.}\ \bibnamefont {Kidder}}, \bibinfo
  {author} {\bibfnamefont {A.~H.}\ \bibnamefont {Mrou\'e}}, \bibinfo {author}
  {\bibfnamefont {H.~P.}\ \bibnamefont {Pfeiffer}}, \bibinfo {author}
  {\bibfnamefont {M.~A.}\ \bibnamefont {Scheel}}, \bibinfo {author}
  {\bibfnamefont {G.~B.}\ \bibnamefont {Cook}}, \ and\ \bibinfo {author}
  {\bibfnamefont {S.~A.}\ \bibnamefont {Teukolsky}},\ }\href {\doibase
  10.1103/PhysRevD.76.124038} {\bibfield  {journal} {\bibinfo  {journal} {Phys.
  Rev. D}\ }\textbf {\bibinfo {volume} {76}},\ \bibinfo {pages} {124038}
  (\bibinfo {year} {2007}{\natexlab{b}})}\BibitemShut {NoStop}%
\bibitem [{\citenamefont {Duez}\ \emph {et~al.}(2008)\citenamefont {Duez},
  \citenamefont {Foucart}, \citenamefont {Kidder}, \citenamefont {Pfeiffer},
  \citenamefont {Scheel},\ and\ \citenamefont {Teukolsky}}]{Duez2008}%
  \BibitemOpen
  \bibfield  {author} {\bibinfo {author} {\bibfnamefont {M.~D.}\ \bibnamefont
  {Duez}}, \bibinfo {author} {\bibfnamefont {F.}~\bibnamefont {Foucart}},
  \bibinfo {author} {\bibfnamefont {L.~E.}\ \bibnamefont {Kidder}}, \bibinfo
  {author} {\bibfnamefont {H.~P.}\ \bibnamefont {Pfeiffer}}, \bibinfo {author}
  {\bibfnamefont {M.~A.}\ \bibnamefont {Scheel}}, \ and\ \bibinfo {author}
  {\bibfnamefont {S.~A.}\ \bibnamefont {Teukolsky}},\ }\href {\doibase
  10.1103/PhysRevD.78.104015} {\bibfield  {journal} {\bibinfo  {journal} {Phys.
  Rev. D}\ }\textbf {\bibinfo {volume} {78}},\ \bibinfo {pages} {104015}
  (\bibinfo {year} {2008})}\BibitemShut {NoStop}%
\bibitem [{\citenamefont {{Scheel}}\ \emph {et~al.}(2009)\citenamefont
  {{Scheel}}, \citenamefont {{Boyle}}, \citenamefont {{Chu}}, \citenamefont
  {{Kidder}}, \citenamefont {{Matthews}},\ and\ \citenamefont
  {{Pfeiffer}}}]{Scheel2009}%
  \BibitemOpen
  \bibfield  {author} {\bibinfo {author} {\bibfnamefont {M.~A.}\ \bibnamefont
  {{Scheel}}}, \bibinfo {author} {\bibfnamefont {M.}~\bibnamefont {{Boyle}}},
  \bibinfo {author} {\bibfnamefont {T.}~\bibnamefont {{Chu}}}, \bibinfo
  {author} {\bibfnamefont {L.~E.}\ \bibnamefont {{Kidder}}}, \bibinfo {author}
  {\bibfnamefont {K.~D.}\ \bibnamefont {{Matthews}}}, \ and\ \bibinfo {author}
  {\bibfnamefont {H.~P.}\ \bibnamefont {{Pfeiffer}}},\ }\href {\doibase
  10.1103/PhysRevD.79.024003} {\bibfield  {journal} {\bibinfo  {journal}
  {\prd}\ }\textbf {\bibinfo {volume} {79}},\ \bibinfo {eid} {024003} (\bibinfo
  {year} {2009})}\BibitemShut {NoStop}%
\bibitem [{\citenamefont {{Grandcl{\'e}ment}}\ and\ \citenamefont
  {{Novak}}(2009)}]{Grandclement2009}%
  \BibitemOpen
  \bibfield  {author} {\bibinfo {author} {\bibfnamefont {P.}~\bibnamefont
  {{Grandcl{\'e}ment}}}\ and\ \bibinfo {author} {\bibfnamefont
  {J.}~\bibnamefont {{Novak}}},\ }\href {\doibase 10.12942/lrr-2009-1}
  {\bibfield  {journal} {\bibinfo  {journal} {Living Reviews in Relativity}\
  }\textbf {\bibinfo {volume} {12}},\ \bibinfo {eid} {1} (\bibinfo {year}
  {2009})}\BibitemShut {NoStop}%
\bibitem [{\citenamefont {Buchman}\ \emph {et~al.}(2012)\citenamefont
  {Buchman}, \citenamefont {Pfeiffer}, \citenamefont {Scheel},\ and\
  \citenamefont {Szil\'agyi}}]{Buchman2012}%
  \BibitemOpen
  \bibfield  {author} {\bibinfo {author} {\bibfnamefont {L.~T.}\ \bibnamefont
  {Buchman}}, \bibinfo {author} {\bibfnamefont {H.~P.}\ \bibnamefont
  {Pfeiffer}}, \bibinfo {author} {\bibfnamefont {M.~A.}\ \bibnamefont
  {Scheel}}, \ and\ \bibinfo {author} {\bibfnamefont {B.}~\bibnamefont
  {Szil\'agyi}},\ }\href {\doibase 10.1103/PhysRevD.86.084033} {\bibfield
  {journal} {\bibinfo  {journal} {Phys. Rev. D}\ }\textbf {\bibinfo {volume}
  {86}},\ \bibinfo {pages} {084033} (\bibinfo {year} {2012})}\BibitemShut
  {NoStop}%
\bibitem [{\citenamefont {{Szil{\'a}gyi}}(2014)}]{Szilagyi2014}%
  \BibitemOpen
  \bibfield  {author} {\bibinfo {author} {\bibfnamefont {B.}~\bibnamefont
  {{Szil{\'a}gyi}}},\ }\href {\doibase 10.1142/S0218271814300146} {\bibfield
  {journal} {\bibinfo  {journal} {International Journal of Modern Physics D}\
  }\textbf {\bibinfo {volume} {23}},\ \bibinfo {eid} {1430014} (\bibinfo {year}
  {2014})}\BibitemShut {NoStop}%
\bibitem [{\citenamefont {Zanotti}\ and\ \citenamefont
  {Dumbser}(2015)}]{Zanotti2015}%
  \BibitemOpen
  \bibfield  {author} {\bibinfo {author} {\bibfnamefont {O.}~\bibnamefont
  {Zanotti}}\ and\ \bibinfo {author} {\bibfnamefont {M.}~\bibnamefont
  {Dumbser}},\ }\href
  {http://www.scopus.com/inward/record.url?eid=2-s2.0-84920650018&partnerID=40&md5=3b1f2cc6325d5e96abf5b76c36a6b886}
  {\bibfield  {journal} {\bibinfo  {journal} {Computer Physics Communications}\
  }\textbf {\bibinfo {volume} {188}},\ \bibinfo {pages} {110} (\bibinfo {year}
  {2015})}\BibitemShut {NoStop}%
\bibitem [{\citenamefont {{Zanotti}}\ \emph {et~al.}(2015)\citenamefont
  {{Zanotti}}, \citenamefont {{Fambri}},\ and\ \citenamefont
  {{Dumbser}}}]{Zanotti2015d}%
  \BibitemOpen
  \bibfield  {author} {\bibinfo {author} {\bibfnamefont {O.}~\bibnamefont
  {{Zanotti}}}, \bibinfo {author} {\bibfnamefont {F.}~\bibnamefont {{Fambri}}},
  \ and\ \bibinfo {author} {\bibfnamefont {M.}~\bibnamefont {{Dumbser}}},\
  }\href {\doibase 10.1093/mnras/stv1510} {\bibfield  {journal} {\bibinfo
  {journal} {Mon. Not. R. Astron. Soc.}\ }\textbf {\bibinfo {volume} {452}},\
  \bibinfo {pages} {3010} (\bibinfo {year} {2015})}\BibitemShut {NoStop}%
\bibitem [{\citenamefont {{Zanotti}}\ and\ \citenamefont
  {{Dumbser}}(2016)}]{Zanotti2016}%
  \BibitemOpen
  \bibfield  {author} {\bibinfo {author} {\bibfnamefont {O.}~\bibnamefont
  {{Zanotti}}}\ and\ \bibinfo {author} {\bibfnamefont {M.}~\bibnamefont
  {{Dumbser}}},\ }\href {\doibase 10.1186/s40668-015-0014-x} {\bibfield
  {journal} {\bibinfo  {journal} {Computational Astrophysics and Cosmology}\
  }\textbf {\bibinfo {volume} {3}},\ \bibinfo {eid} {1} (\bibinfo {year}
  {2016})}\BibitemShut {NoStop}%
\bibitem [{\citenamefont {Fambri}\ \emph {et~al.}(2018)\citenamefont {Fambri},
  \citenamefont {Dumbser}, \citenamefont {K{\"o}ppel}, \citenamefont
  {Rezzolla},\ and\ \citenamefont {Zanotti}}]{ADERGRMHD}%
  \BibitemOpen
  \bibfield  {author} {\bibinfo {author} {\bibfnamefont {F.}~\bibnamefont
  {Fambri}}, \bibinfo {author} {\bibfnamefont {M.}~\bibnamefont {Dumbser}},
  \bibinfo {author} {\bibfnamefont {S.}~\bibnamefont {K{\"o}ppel}}, \bibinfo
  {author} {\bibfnamefont {L.}~\bibnamefont {Rezzolla}}, \ and\ \bibinfo
  {author} {\bibfnamefont {O.}~\bibnamefont {Zanotti}},\ }\href@noop {}
  {\bibfield  {journal} {\bibinfo  {journal} {Monthly Notices of the Royal
  Astronomical Society}\ }\textbf {\bibinfo {volume} {477}},\ \bibinfo {pages}
  {4543} (\bibinfo {year} {2018})}\BibitemShut {NoStop}%
\bibitem [{\citenamefont {Dumbser}\ \emph
  {et~al.}(2018{\natexlab{b}})\citenamefont {Dumbser}, \citenamefont
  {Guercilena}, \citenamefont {K{\"o}ppel}, \citenamefont {Rezzolla},\ and\
  \citenamefont {Zanotti}}]{Dumbser2018conformal}%
  \BibitemOpen
  \bibfield  {author} {\bibinfo {author} {\bibfnamefont {M.}~\bibnamefont
  {Dumbser}}, \bibinfo {author} {\bibfnamefont {F.}~\bibnamefont {Guercilena}},
  \bibinfo {author} {\bibfnamefont {S.}~\bibnamefont {K{\"o}ppel}}, \bibinfo
  {author} {\bibfnamefont {L.}~\bibnamefont {Rezzolla}}, \ and\ \bibinfo
  {author} {\bibfnamefont {O.}~\bibnamefont {Zanotti}},\ }\href@noop {}
  {\bibfield  {journal} {\bibinfo  {journal} {Physical Review D}\ }\textbf
  {\bibinfo {volume} {97}},\ \bibinfo {pages} {084053} (\bibinfo {year}
  {2018}{\natexlab{b}})}\BibitemShut {NoStop}%
\bibitem [{\citenamefont {Gaburro}\ and\ \citenamefont
  {Dumbser}(2021)}]{Gaburro2021PNPMLimiter}%
  \BibitemOpen
  \bibfield  {author} {\bibinfo {author} {\bibfnamefont {E.}~\bibnamefont
  {Gaburro}}\ and\ \bibinfo {author} {\bibfnamefont {M.}~\bibnamefont
  {Dumbser}},\ }\href@noop {} {\bibfield  {journal} {\bibinfo  {journal}
  {Journal of Scientific Computing}\ }\textbf {\bibinfo {volume} {86}},\
  \bibinfo {pages} {1} (\bibinfo {year} {2021})}\BibitemShut {NoStop}%
\bibitem [{\citenamefont {Löffler}\ \emph {et~al.}(2012)\citenamefont
  {Löffler}, \citenamefont {Faber}, \citenamefont {Bentivegna}, \citenamefont
  {Bode}, \citenamefont {Diener}, \citenamefont {Haas}, \citenamefont {Hinder},
  \citenamefont {Mundim}, \citenamefont {Ott}, \citenamefont {Schnetter},
  \citenamefont {Allen}, \citenamefont {Campanelli},\ and\ \citenamefont
  {Laguna}}]{Loffler2012}%
  \BibitemOpen
  \bibfield  {author} {\bibinfo {author} {\bibfnamefont {F.}~\bibnamefont
  {Löffler}}, \bibinfo {author} {\bibfnamefont {J.}~\bibnamefont {Faber}},
  \bibinfo {author} {\bibfnamefont {E.}~\bibnamefont {Bentivegna}}, \bibinfo
  {author} {\bibfnamefont {T.}~\bibnamefont {Bode}}, \bibinfo {author}
  {\bibfnamefont {P.}~\bibnamefont {Diener}}, \bibinfo {author} {\bibfnamefont
  {R.}~\bibnamefont {Haas}}, \bibinfo {author} {\bibfnamefont {I.}~\bibnamefont
  {Hinder}}, \bibinfo {author} {\bibfnamefont {B.~C.}\ \bibnamefont {Mundim}},
  \bibinfo {author} {\bibfnamefont {C.~D.}\ \bibnamefont {Ott}}, \bibinfo
  {author} {\bibfnamefont {E.}~\bibnamefont {Schnetter}}, \bibinfo {author}
  {\bibfnamefont {G.}~\bibnamefont {Allen}}, \bibinfo {author} {\bibfnamefont
  {M.}~\bibnamefont {Campanelli}}, \ and\ \bibinfo {author} {\bibfnamefont
  {P.}~\bibnamefont {Laguna}},\ }\href {\doibase
  10.1088/0264-9381/29/11/115001} {\bibfield  {journal} {\bibinfo  {journal}
  {Classical and Quantum Gravity}\ }\textbf {\bibinfo {volume} {29}},\ \bibinfo
  {pages} {115001} (\bibinfo {year} {2012})}\BibitemShut {NoStop}%
\bibitem [{\citenamefont {Zlochower}\ \emph {et~al.}(2005)\citenamefont
  {Zlochower}, \citenamefont {Baker}, \citenamefont {Campanelli},\ and\
  \citenamefont {Lousto}}]{Zlochower2005}%
  \BibitemOpen
  \bibfield  {author} {\bibinfo {author} {\bibfnamefont {Y.}~\bibnamefont
  {Zlochower}}, \bibinfo {author} {\bibfnamefont {J.~G.}\ \bibnamefont
  {Baker}}, \bibinfo {author} {\bibfnamefont {M.}~\bibnamefont {Campanelli}}, \
  and\ \bibinfo {author} {\bibfnamefont {C.~O.}\ \bibnamefont {Lousto}},\
  }\href {\doibase 10.1103/PhysRevD.72.024021} {\bibfield  {journal} {\bibinfo
  {journal} {Phys. Rev. D}\ }\textbf {\bibinfo {volume} {72}},\ \bibinfo
  {pages} {024021} (\bibinfo {year} {2005})}\BibitemShut {NoStop}%
\bibitem [{\citenamefont {{Lousto}}\ and\ \citenamefont
  {{Healy}}(2023)}]{Lousto2023}%
  \BibitemOpen
  \bibfield  {author} {\bibinfo {author} {\bibfnamefont {C.~O.}\ \bibnamefont
  {{Lousto}}}\ and\ \bibinfo {author} {\bibfnamefont {J.}~\bibnamefont
  {{Healy}}},\ }\href {\doibase 10.1088/1361-6382/acc7ef} {\bibfield  {journal}
  {\bibinfo  {journal} {Classical and Quantum Gravity}\ }\textbf {\bibinfo
  {volume} {40}},\ \bibinfo {eid} {09LT01} (\bibinfo {year}
  {2023})}\BibitemShut {NoStop}%
\bibitem [{\citenamefont {Br\"ugmann}\ \emph {et~al.}(2008)\citenamefont
  {Br\"ugmann}, \citenamefont {Gonz\'alez}, \citenamefont {Hannam},
  \citenamefont {Husa}, \citenamefont {Sperhake},\ and\ \citenamefont
  {Tichy}}]{Bruegmann2008}%
  \BibitemOpen
  \bibfield  {author} {\bibinfo {author} {\bibfnamefont {B.}~\bibnamefont
  {Br\"ugmann}}, \bibinfo {author} {\bibfnamefont {J.~A.}\ \bibnamefont
  {Gonz\'alez}}, \bibinfo {author} {\bibfnamefont {M.}~\bibnamefont {Hannam}},
  \bibinfo {author} {\bibfnamefont {S.}~\bibnamefont {Husa}}, \bibinfo {author}
  {\bibfnamefont {U.}~\bibnamefont {Sperhake}}, \ and\ \bibinfo {author}
  {\bibfnamefont {W.}~\bibnamefont {Tichy}},\ }\href {\doibase
  10.1103/PhysRevD.77.024027} {\bibfield  {journal} {\bibinfo  {journal} {Phys.
  Rev. D}\ }\textbf {\bibinfo {volume} {77}},\ \bibinfo {pages} {024027}
  (\bibinfo {year} {2008})}\BibitemShut {NoStop}%
\bibitem [{\citenamefont {Husa}\ \emph {et~al.}(2008)\citenamefont {Husa},
  \citenamefont {González}, \citenamefont {Hannam}, \citenamefont
  {Brügmann},\ and\ \citenamefont {Sperhake}}]{Husa_2008}%
  \BibitemOpen
  \bibfield  {author} {\bibinfo {author} {\bibfnamefont {S.}~\bibnamefont
  {Husa}}, \bibinfo {author} {\bibfnamefont {J.~A.}\ \bibnamefont {González}},
  \bibinfo {author} {\bibfnamefont {M.}~\bibnamefont {Hannam}}, \bibinfo
  {author} {\bibfnamefont {B.}~\bibnamefont {Brügmann}}, \ and\ \bibinfo
  {author} {\bibfnamefont {U.}~\bibnamefont {Sperhake}},\ }\href {\doibase
  10.1088/0264-9381/25/10/105006} {\bibfield  {journal} {\bibinfo  {journal}
  {Classical and Quantum Gravity}\ }\textbf {\bibinfo {volume} {25}},\ \bibinfo
  {pages} {105006} (\bibinfo {year} {2008})}\BibitemShut {NoStop}%
\bibitem [{\citenamefont {Thierfelder}\ \emph {et~al.}(2011)\citenamefont
  {Thierfelder}, \citenamefont {Bernuzzi},\ and\ \citenamefont
  {Br\"ugmann}}]{Thierfelder2011}%
  \BibitemOpen
  \bibfield  {author} {\bibinfo {author} {\bibfnamefont {M.}~\bibnamefont
  {Thierfelder}}, \bibinfo {author} {\bibfnamefont {S.}~\bibnamefont
  {Bernuzzi}}, \ and\ \bibinfo {author} {\bibfnamefont {B.}~\bibnamefont
  {Br\"ugmann}},\ }\href {\doibase 10.1103/PhysRevD.84.044012} {\bibfield
  {journal} {\bibinfo  {journal} {Phys. Rev. D}\ }\textbf {\bibinfo {volume}
  {84}},\ \bibinfo {pages} {044012} (\bibinfo {year} {2011})}\BibitemShut
  {NoStop}%
\bibitem [{\citenamefont {Clough}\ \emph {et~al.}(2015)\citenamefont {Clough},
  \citenamefont {Figueras}, \citenamefont {Finkel}, \citenamefont {Kunesch},
  \citenamefont {Lim},\ and\ \citenamefont {Tunyasuvunakool}}]{Clough_2015}%
  \BibitemOpen
  \bibfield  {author} {\bibinfo {author} {\bibfnamefont {K.}~\bibnamefont
  {Clough}}, \bibinfo {author} {\bibfnamefont {P.}~\bibnamefont {Figueras}},
  \bibinfo {author} {\bibfnamefont {H.}~\bibnamefont {Finkel}}, \bibinfo
  {author} {\bibfnamefont {M.}~\bibnamefont {Kunesch}}, \bibinfo {author}
  {\bibfnamefont {E.~A.}\ \bibnamefont {Lim}}, \ and\ \bibinfo {author}
  {\bibfnamefont {S.}~\bibnamefont {Tunyasuvunakool}},\ }\href {\doibase
  10.1088/0264-9381/32/24/245011} {\bibfield  {journal} {\bibinfo  {journal}
  {Classical and Quantum Gravity}\ }\textbf {\bibinfo {volume} {32}},\ \bibinfo
  {pages} {245011} (\bibinfo {year} {2015})}\BibitemShut {NoStop}%
\bibitem [{\citenamefont {Andrade}(2021)}]{Andrade2021}%
  \BibitemOpen
  \bibfield  {author} {\bibinfo {author} {\bibfnamefont {T.}~\bibnamefont
  {Andrade}},\ }\href@noop {} {\bibfield  {journal} {\bibinfo  {journal}
  {Journal of Open Source Software}\ }\textbf {\bibinfo {volume} {6}},\
  \bibinfo {pages} {3703} (\bibinfo {year} {2021})}\BibitemShut {NoStop}%
\bibitem [{\citenamefont {Peterson}\ \emph {et~al.}(2023)\citenamefont
  {Peterson}, \citenamefont {Willcox},\ and\ \citenamefont
  {Mösta}}]{Peterson_2023}%
  \BibitemOpen
  \bibfield  {author} {\bibinfo {author} {\bibfnamefont {A.~J.}\ \bibnamefont
  {Peterson}}, \bibinfo {author} {\bibfnamefont {D.}~\bibnamefont {Willcox}}, \
  and\ \bibinfo {author} {\bibfnamefont {P.}~\bibnamefont {Mösta}},\ }\href
  {\doibase 10.1088/1361-6382/ad0b37} {\bibfield  {journal} {\bibinfo
  {journal} {Classical and Quantum Gravity}\ }\textbf {\bibinfo {volume}
  {40}},\ \bibinfo {pages} {245013} (\bibinfo {year} {2023})}\BibitemShut
  {NoStop}%
\bibitem [{\citenamefont {Yamamoto}\ \emph {et~al.}(2008)\citenamefont
  {Yamamoto}, \citenamefont {Shibata},\ and\ \citenamefont
  {Taniguchi}}]{Yamamoto2009}%
  \BibitemOpen
  \bibfield  {author} {\bibinfo {author} {\bibfnamefont {T.}~\bibnamefont
  {Yamamoto}}, \bibinfo {author} {\bibfnamefont {M.}~\bibnamefont {Shibata}}, \
  and\ \bibinfo {author} {\bibfnamefont {K.}~\bibnamefont {Taniguchi}},\ }\href
  {\doibase 10.1103/PhysRevD.78.064054} {\bibfield  {journal} {\bibinfo
  {journal} {Phys. Rev. D}\ }\textbf {\bibinfo {volume} {78}},\ \bibinfo
  {pages} {064054} (\bibinfo {year} {2008})}\BibitemShut {NoStop}%
\bibitem [{\citenamefont {Kiuchi}\ \emph {et~al.}(2017)\citenamefont {Kiuchi},
  \citenamefont {Kawaguchi}, \citenamefont {Kyutoku}, \citenamefont
  {Sekiguchi}, \citenamefont {Shibata},\ and\ \citenamefont
  {Taniguchi}}]{Kiuchi2017}%
  \BibitemOpen
  \bibfield  {author} {\bibinfo {author} {\bibfnamefont {K.}~\bibnamefont
  {Kiuchi}}, \bibinfo {author} {\bibfnamefont {K.}~\bibnamefont {Kawaguchi}},
  \bibinfo {author} {\bibfnamefont {K.}~\bibnamefont {Kyutoku}}, \bibinfo
  {author} {\bibfnamefont {Y.}~\bibnamefont {Sekiguchi}}, \bibinfo {author}
  {\bibfnamefont {M.}~\bibnamefont {Shibata}}, \ and\ \bibinfo {author}
  {\bibfnamefont {K.}~\bibnamefont {Taniguchi}},\ }\href {\doibase
  10.1103/PhysRevD.96.084060} {\bibfield  {journal} {\bibinfo  {journal} {Phys.
  Rev. D}\ }\textbf {\bibinfo {volume} {96}},\ \bibinfo {pages} {084060}
  (\bibinfo {year} {2017})}\BibitemShut {NoStop}%
\bibitem [{\citenamefont {Kidder}\ \emph {et~al.}(2000)\citenamefont {Kidder},
  \citenamefont {Scheel}, \citenamefont {Teukolsky}, \citenamefont {Carlson},\
  and\ \citenamefont {Cook}}]{Kidder2000}%
  \BibitemOpen
  \bibfield  {author} {\bibinfo {author} {\bibfnamefont {L.~E.}\ \bibnamefont
  {Kidder}}, \bibinfo {author} {\bibfnamefont {M.~A.}\ \bibnamefont {Scheel}},
  \bibinfo {author} {\bibfnamefont {S.~A.}\ \bibnamefont {Teukolsky}}, \bibinfo
  {author} {\bibfnamefont {E.~D.}\ \bibnamefont {Carlson}}, \ and\ \bibinfo
  {author} {\bibfnamefont {G.~B.}\ \bibnamefont {Cook}},\ }\href {\doibase
  10.1103/PhysRevD.62.084032} {\bibfield  {journal} {\bibinfo  {journal} {Phys.
  Rev. D}\ }\textbf {\bibinfo {volume} {62}},\ \bibinfo {pages} {084032}
  (\bibinfo {year} {2000})}\BibitemShut {NoStop}%
\bibitem [{\citenamefont {Haas}\ \emph {et~al.}(2016)\citenamefont {Haas},
  \citenamefont {Ott}, \citenamefont {Szilagyi}, \citenamefont {Kaplan},
  \citenamefont {Lippuner}, \citenamefont {Scheel}, \citenamefont {Barkett},
  \citenamefont {Muhlberger}, \citenamefont {Dietrich}, \citenamefont {Duez},
  \citenamefont {Foucart}, \citenamefont {Pfeiffer}, \citenamefont {Kidder},\
  and\ \citenamefont {Teukolsky}}]{Haas2016}%
  \BibitemOpen
  \bibfield  {author} {\bibinfo {author} {\bibfnamefont {R.}~\bibnamefont
  {Haas}}, \bibinfo {author} {\bibfnamefont {C.~D.}\ \bibnamefont {Ott}},
  \bibinfo {author} {\bibfnamefont {B.}~\bibnamefont {Szilagyi}}, \bibinfo
  {author} {\bibfnamefont {J.~D.}\ \bibnamefont {Kaplan}}, \bibinfo {author}
  {\bibfnamefont {J.}~\bibnamefont {Lippuner}}, \bibinfo {author}
  {\bibfnamefont {M.~A.}\ \bibnamefont {Scheel}}, \bibinfo {author}
  {\bibfnamefont {K.}~\bibnamefont {Barkett}}, \bibinfo {author} {\bibfnamefont
  {C.~D.}\ \bibnamefont {Muhlberger}}, \bibinfo {author} {\bibfnamefont
  {T.}~\bibnamefont {Dietrich}}, \bibinfo {author} {\bibfnamefont {M.~D.}\
  \bibnamefont {Duez}}, \bibinfo {author} {\bibfnamefont {F.}~\bibnamefont
  {Foucart}}, \bibinfo {author} {\bibfnamefont {H.~P.}\ \bibnamefont
  {Pfeiffer}}, \bibinfo {author} {\bibfnamefont {L.~E.}\ \bibnamefont
  {Kidder}}, \ and\ \bibinfo {author} {\bibfnamefont {S.~A.}\ \bibnamefont
  {Teukolsky}},\ }\href {\doibase 10.1103/PhysRevD.93.124062} {\bibfield
  {journal} {\bibinfo  {journal} {Phys. Rev. D}\ }\textbf {\bibinfo {volume}
  {93}},\ \bibinfo {pages} {124062} (\bibinfo {year} {2016})}\BibitemShut
  {NoStop}%
\bibitem [{\citenamefont {{Rosswog}}\ and\ \citenamefont
  {{Diener}}(2021)}]{Rosswog2021}%
  \BibitemOpen
  \bibfield  {author} {\bibinfo {author} {\bibfnamefont {S.}~\bibnamefont
  {{Rosswog}}}\ and\ \bibinfo {author} {\bibfnamefont {P.}~\bibnamefont
  {{Diener}}},\ }\href {\doibase 10.1088/1361-6382/abee65} {\bibfield
  {journal} {\bibinfo  {journal} {Classical and Quantum Gravity}\ }\textbf
  {\bibinfo {volume} {38}},\ \bibinfo {eid} {115002} (\bibinfo {year}
  {2021})}\BibitemShut {NoStop}%
\bibitem [{\citenamefont {Ruchlin}\ \emph {et~al.}(2018)\citenamefont
  {Ruchlin}, \citenamefont {Etienne},\ and\ \citenamefont
  {Baumgarte}}]{Ruchlin2018}%
  \BibitemOpen
  \bibfield  {author} {\bibinfo {author} {\bibfnamefont {I.}~\bibnamefont
  {Ruchlin}}, \bibinfo {author} {\bibfnamefont {Z.~B.}\ \bibnamefont
  {Etienne}}, \ and\ \bibinfo {author} {\bibfnamefont {T.~W.}\ \bibnamefont
  {Baumgarte}},\ }\href {\doibase 10.1103/PhysRevD.97.064036} {\bibfield
  {journal} {\bibinfo  {journal} {Phys. Rev. D}\ }\textbf {\bibinfo {volume}
  {97}},\ \bibinfo {pages} {064036} (\bibinfo {year} {2018})}\BibitemShut
  {NoStop}%
\bibitem [{\citenamefont {{Palenzuela}}\ \emph {et~al.}(2018)\citenamefont
  {{Palenzuela}}, \citenamefont {{Mi{\~n}ano}}, \citenamefont {{Vigan{\`o}}},
  \citenamefont {{Arbona}}, \citenamefont {{Bona-Casas}}, \citenamefont
  {{Rigo}}, \citenamefont {{Bezares}}, \citenamefont {{Bona}},\ and\
  \citenamefont {{Mass{\'o}}}}]{Palenzuela2018}%
  \BibitemOpen
  \bibfield  {author} {\bibinfo {author} {\bibfnamefont {C.}~\bibnamefont
  {{Palenzuela}}}, \bibinfo {author} {\bibfnamefont {B.}~\bibnamefont
  {{Mi{\~n}ano}}}, \bibinfo {author} {\bibfnamefont {D.}~\bibnamefont
  {{Vigan{\`o}}}}, \bibinfo {author} {\bibfnamefont {A.}~\bibnamefont
  {{Arbona}}}, \bibinfo {author} {\bibfnamefont {C.}~\bibnamefont
  {{Bona-Casas}}}, \bibinfo {author} {\bibfnamefont {A.}~\bibnamefont
  {{Rigo}}}, \bibinfo {author} {\bibfnamefont {M.}~\bibnamefont {{Bezares}}},
  \bibinfo {author} {\bibfnamefont {C.}~\bibnamefont {{Bona}}}, \ and\ \bibinfo
  {author} {\bibfnamefont {J.}~\bibnamefont {{Mass{\'o}}}},\ }\href {\doibase
  10.1088/1361-6382/aad7f6} {\bibfield  {journal} {\bibinfo  {journal}
  {Classical and Quantum Gravity}\ }\textbf {\bibinfo {volume} {35}},\ \bibinfo
  {eid} {185007} (\bibinfo {year} {2018})}\BibitemShut {NoStop}%
\bibitem [{\citenamefont {{Palenzuela}}\ \emph {et~al.}(2021)\citenamefont
  {{Palenzuela}}, \citenamefont {{Mi{\~n}ano}}, \citenamefont {{Arbona}},
  \citenamefont {{Bona-Casas}}, \citenamefont {{Bona}},\ and\ \citenamefont
  {{Mass{\'o}}}}]{Palenzuela2021}%
  \BibitemOpen
  \bibfield  {author} {\bibinfo {author} {\bibfnamefont {C.}~\bibnamefont
  {{Palenzuela}}}, \bibinfo {author} {\bibfnamefont {B.}~\bibnamefont
  {{Mi{\~n}ano}}}, \bibinfo {author} {\bibfnamefont {A.}~\bibnamefont
  {{Arbona}}}, \bibinfo {author} {\bibfnamefont {C.}~\bibnamefont
  {{Bona-Casas}}}, \bibinfo {author} {\bibfnamefont {C.}~\bibnamefont
  {{Bona}}}, \ and\ \bibinfo {author} {\bibfnamefont {J.}~\bibnamefont
  {{Mass{\'o}}}},\ }\href {\doibase 10.1016/j.cpc.2020.107675} {\bibfield
  {journal} {\bibinfo  {journal} {Computer Physics Communications}\ }\textbf
  {\bibinfo {volume} {259}},\ \bibinfo {eid} {107675} (\bibinfo {year}
  {2021})}\BibitemShut {NoStop}%
\bibitem [{\citenamefont {Fernando}\ \emph {et~al.}(2019)\citenamefont
  {Fernando}, \citenamefont {Neilsen}, \citenamefont {Lim}, \citenamefont
  {Hirschmann},\ and\ \citenamefont {Sundar}}]{Milinda2019}%
  \BibitemOpen
  \bibfield  {author} {\bibinfo {author} {\bibfnamefont {M.}~\bibnamefont
  {Fernando}}, \bibinfo {author} {\bibfnamefont {D.}~\bibnamefont {Neilsen}},
  \bibinfo {author} {\bibfnamefont {H.}~\bibnamefont {Lim}}, \bibinfo {author}
  {\bibfnamefont {E.}~\bibnamefont {Hirschmann}}, \ and\ \bibinfo {author}
  {\bibfnamefont {H.}~\bibnamefont {Sundar}},\ }\href {\doibase
  10.1137/18M1196972} {\bibfield  {journal} {\bibinfo  {journal} {SIAM Journal
  on Scientific Computing}\ }\textbf {\bibinfo {volume} {41}},\ \bibinfo
  {pages} {C97} (\bibinfo {year} {2019})}\BibitemShut {NoStop}%
\bibitem [{\citenamefont {{Fernando}}\ \emph {et~al.}(2023)\citenamefont
  {{Fernando}}, \citenamefont {{Neilsen}}, \citenamefont {{Zlochower}},
  \citenamefont {{Hirschmann}},\ and\ \citenamefont {{Sundar}}}]{Milinda2023}%
  \BibitemOpen
  \bibfield  {author} {\bibinfo {author} {\bibfnamefont {M.}~\bibnamefont
  {{Fernando}}}, \bibinfo {author} {\bibfnamefont {D.}~\bibnamefont
  {{Neilsen}}}, \bibinfo {author} {\bibfnamefont {Y.}~\bibnamefont
  {{Zlochower}}}, \bibinfo {author} {\bibfnamefont {E.~W.}\ \bibnamefont
  {{Hirschmann}}}, \ and\ \bibinfo {author} {\bibfnamefont {H.}~\bibnamefont
  {{Sundar}}},\ }\href {\doibase 10.1103/PhysRevD.107.064035} {\bibfield
  {journal} {\bibinfo  {journal} {\prd}\ }\textbf {\bibinfo {volume} {107}},\
  \bibinfo {eid} {064035} (\bibinfo {year} {2023})}\BibitemShut {NoStop}%
\bibitem [{\citenamefont {{Cordero-Carri{\'o}n}}\ \emph
  {et~al.}(2008)\citenamefont {{Cordero-Carri{\'o}n}}, \citenamefont
  {{Ib{\'a}{\~n}ez}}, \citenamefont {{Gourgoulhon}}, \citenamefont
  {{Jaramillo}},\ and\ \citenamefont {{Novak}}}]{Cordero-Carrion2008}%
  \BibitemOpen
  \bibfield  {author} {\bibinfo {author} {\bibfnamefont {I.}~\bibnamefont
  {{Cordero-Carri{\'o}n}}}, \bibinfo {author} {\bibfnamefont {J.~M.}\
  \bibnamefont {{Ib{\'a}{\~n}ez}}}, \bibinfo {author} {\bibfnamefont
  {E.}~\bibnamefont {{Gourgoulhon}}}, \bibinfo {author} {\bibfnamefont {J.~L.}\
  \bibnamefont {{Jaramillo}}}, \ and\ \bibinfo {author} {\bibfnamefont
  {J.}~\bibnamefont {{Novak}}},\ }\href {\doibase 10.1103/PhysRevD.77.084007}
  {\bibfield  {journal} {\bibinfo  {journal} {Phys. Rev. D}\ }\textbf {\bibinfo
  {volume} {77}},\ \bibinfo {eid} {084007} (\bibinfo {year}
  {2008})}\BibitemShut {NoStop}%
\bibitem [{\citenamefont {Cordero-Carrión}\ \emph {et~al.}(2010)\citenamefont
  {Cordero-Carrión}, \citenamefont {Cerdá-Durán},\ and\ \citenamefont
  {Ibáñez}}]{Cordero-Carrion2010}%
  \BibitemOpen
  \bibfield  {author} {\bibinfo {author} {\bibfnamefont {I.}~\bibnamefont
  {Cordero-Carrión}}, \bibinfo {author} {\bibfnamefont {P.}~\bibnamefont
  {Cerdá-Durán}}, \ and\ \bibinfo {author} {\bibfnamefont {J.~M.}\
  \bibnamefont {Ibáñez}},\ }\href {\doibase 10.1088/1742-6596/228/1/012055}
  {\bibfield  {journal} {\bibinfo  {journal} {Journal of Physics: Conference
  Series}\ }\textbf {\bibinfo {volume} {228}},\ \bibinfo {pages} {012055}
  (\bibinfo {year} {2010})}\BibitemShut {NoStop}%
\bibitem [{\citenamefont {Levy}\ \emph {et~al.}(1999)\citenamefont {Levy},
  \citenamefont {Puppo},\ and\ \citenamefont {Russo}}]{Levy1999}%
  \BibitemOpen
  \bibfield  {author} {\bibinfo {author} {\bibfnamefont {D.}~\bibnamefont
  {Levy}}, \bibinfo {author} {\bibfnamefont {G.}~\bibnamefont {Puppo}}, \ and\
  \bibinfo {author} {\bibfnamefont {G.}~\bibnamefont {Russo}},\ }\href
  {\doibase 10.1051/m2an:1999152} {\bibfield  {journal} {\bibinfo  {journal}
  {Mathematical Modelling and Numerical Analysis}\ }\textbf {\bibinfo {volume}
  {33}},\ \bibinfo {pages} {547 – 571} (\bibinfo {year} {1999})}\BibitemShut
  {NoStop}%
\bibitem [{\citenamefont {Levy}\ \emph {et~al.}(2000)\citenamefont {Levy},
  \citenamefont {Puppo},\ and\ \citenamefont {Russo}}]{Levy2000}%
  \BibitemOpen
  \bibfield  {author} {\bibinfo {author} {\bibfnamefont {D.}~\bibnamefont
  {Levy}}, \bibinfo {author} {\bibfnamefont {G.}~\bibnamefont {Puppo}}, \ and\
  \bibinfo {author} {\bibfnamefont {G.}~\bibnamefont {Russo}},\ }\href
  {\doibase 10.1016/S0168-9274(99)00108-7} {\bibfield  {journal} {\bibinfo
  {journal} {Applied Numerical Mathematics}\ }\textbf {\bibinfo {volume}
  {33}},\ \bibinfo {pages} {415 – 421} (\bibinfo {year} {2000})}\BibitemShut
  {NoStop}%
\bibitem [{\citenamefont {Levy}\ \emph {et~al.}(2001)\citenamefont {Levy},
  \citenamefont {Puppo},\ and\ \citenamefont {Russo}}]{Levy2001}%
  \BibitemOpen
  \bibfield  {author} {\bibinfo {author} {\bibfnamefont {D.}~\bibnamefont
  {Levy}}, \bibinfo {author} {\bibfnamefont {G.}~\bibnamefont {Puppo}}, \ and\
  \bibinfo {author} {\bibfnamefont {G.}~\bibnamefont {Russo}},\ }\href
  {\doibase 10.1137/S1064827599359461} {\bibfield  {journal} {\bibinfo
  {journal} {SIAM Journal on Scientific Computing}\ }\textbf {\bibinfo {volume}
  {22}},\ \bibinfo {pages} {656 – 672} (\bibinfo {year} {2001})}\BibitemShut
  {NoStop}%
\bibitem [{\citenamefont {{Balsara}}\ \emph {et~al.}(2023)\citenamefont
  {{Balsara}}, \citenamefont {{Bhoriya}}, \citenamefont {{Shu}},\ and\
  \citenamefont {{Kumar}}}]{Balsara2023}%
  \BibitemOpen
  \bibfield  {author} {\bibinfo {author} {\bibfnamefont {D.~S.}\ \bibnamefont
  {{Balsara}}}, \bibinfo {author} {\bibfnamefont {D.}~\bibnamefont
  {{Bhoriya}}}, \bibinfo {author} {\bibfnamefont {C.-W.}\ \bibnamefont
  {{Shu}}}, \ and\ \bibinfo {author} {\bibfnamefont {H.}~\bibnamefont
  {{Kumar}}},\ }\href {\doibase 10.48550/arXiv.2303.17672} {\bibfield
  {journal} {\bibinfo  {journal} {arXiv e-prints}\ ,\ \bibinfo {eid}
  {arXiv:2303.17672}} (\bibinfo {year} {2023})}\BibitemShut {NoStop}%
\bibitem [{\citenamefont {{Balsara}}\ \emph
  {et~al.}(2024{\natexlab{a}})\citenamefont {{Balsara}}, \citenamefont
  {{Bhoriya}}, \citenamefont {{Shu}},\ and\ \citenamefont
  {{Kumar}}}]{Balsara2024b}%
  \BibitemOpen
  \bibfield  {author} {\bibinfo {author} {\bibfnamefont {D.~S.}\ \bibnamefont
  {{Balsara}}}, \bibinfo {author} {\bibfnamefont {D.}~\bibnamefont
  {{Bhoriya}}}, \bibinfo {author} {\bibfnamefont {C.-W.}\ \bibnamefont
  {{Shu}}}, \ and\ \bibinfo {author} {\bibfnamefont {H.}~\bibnamefont
  {{Kumar}}},\ }\href {\doibase 10.48550/arXiv.2403.01266} {\bibfield
  {journal} {\bibinfo  {journal} {arXiv e-prints}\ ,\ \bibinfo {eid}
  {arXiv:2403.01266}} (\bibinfo {year} {2024}{\natexlab{a}})}\BibitemShut
  {NoStop}%
\bibitem [{\citenamefont {{Balsara}}\ \emph
  {et~al.}(2024{\natexlab{b}})\citenamefont {{Balsara}}, \citenamefont
  {{Bhoriya}}, \citenamefont {{Zanotti}},\ and\ \citenamefont
  {{Dumbser}}}]{Balsara2024c}%
  \BibitemOpen
  \bibfield  {author} {\bibinfo {author} {\bibfnamefont {D.}~\bibnamefont
  {{Balsara}}}, \bibinfo {author} {\bibfnamefont {D.}~\bibnamefont
  {{Bhoriya}}}, \bibinfo {author} {\bibfnamefont {O.}~\bibnamefont
  {{Zanotti}}}, \ and\ \bibinfo {author} {\bibfnamefont {M.}~\bibnamefont
  {{Dumbser}}},\ }\href {\doibase 10.3847/1538-4365/ad7d0d} {\bibfield
  {journal} {\bibinfo  {journal} {The Astrophysical Journal Supplement Series}\
  }\textbf {\bibinfo {volume} {275}},\ \bibinfo {eid} {18} (\bibinfo {year}
  {2024}{\natexlab{b}})}\BibitemShut {NoStop}%
\bibitem [{\citenamefont {Shu}\ and\ \citenamefont {Osher}(1988)}]{shuosher1}%
  \BibitemOpen
  \bibfield  {author} {\bibinfo {author} {\bibfnamefont {C.}~\bibnamefont
  {Shu}}\ and\ \bibinfo {author} {\bibfnamefont {S.}~\bibnamefont {Osher}},\
  }\href@noop {} {\bibfield  {journal} {\bibinfo  {journal} {Journal of
  Computational Physics}\ }\textbf {\bibinfo {volume} {77}},\ \bibinfo {pages}
  {439} (\bibinfo {year} {1988})}\BibitemShut {NoStop}%
\bibitem [{\citenamefont {Shu}\ and\ \citenamefont {Osher}(1989)}]{shuosher2}%
  \BibitemOpen
  \bibfield  {author} {\bibinfo {author} {\bibfnamefont {C.}~\bibnamefont
  {Shu}}\ and\ \bibinfo {author} {\bibfnamefont {S.}~\bibnamefont {Osher}},\
  }\href@noop {} {\bibfield  {journal} {\bibinfo  {journal} {Journal of
  Computational Physics}\ }\textbf {\bibinfo {volume} {83}},\ \bibinfo {pages}
  {32} (\bibinfo {year} {1989})}\BibitemShut {NoStop}%
\bibitem [{\citenamefont {Jiang}\ and\ \citenamefont
  {Shu}(1996)}]{shu_efficient_weno}%
  \BibitemOpen
  \bibfield  {author} {\bibinfo {author} {\bibfnamefont {G.}~\bibnamefont
  {Jiang}}\ and\ \bibinfo {author} {\bibfnamefont {C.}~\bibnamefont {Shu}},\
  }\href@noop {} {\bibfield  {journal} {\bibinfo  {journal} {Journal of
  Computational Physics}\ }\textbf {\bibinfo {volume} {126}},\ \bibinfo {pages}
  {202} (\bibinfo {year} {1996})}\BibitemShut {NoStop}%
\bibitem [{\citenamefont {Castro}\ \emph {et~al.}(2011)\citenamefont {Castro},
  \citenamefont {Costa},\ and\ \citenamefont {Don}}]{WENOZ}%
  \BibitemOpen
  \bibfield  {author} {\bibinfo {author} {\bibfnamefont {M.}~\bibnamefont
  {Castro}}, \bibinfo {author} {\bibfnamefont {B.}~\bibnamefont {Costa}}, \
  and\ \bibinfo {author} {\bibfnamefont {W.}~\bibnamefont {Don}},\ }\href@noop
  {} {\bibfield  {journal} {\bibinfo  {journal} {Journal of Computational
  Physics}\ }\textbf {\bibinfo {volume} {230}},\ \bibinfo {pages} {1766}
  (\bibinfo {year} {2011})}\BibitemShut {NoStop}%
\bibitem [{\citenamefont {Cravero}\ \emph {et~al.}(2018)\citenamefont
  {Cravero}, \citenamefont {Puppo}, \citenamefont {Semplice},\ and\
  \citenamefont {Visconti}}]{cravero2018cweno}%
  \BibitemOpen
  \bibfield  {author} {\bibinfo {author} {\bibfnamefont {I.}~\bibnamefont
  {Cravero}}, \bibinfo {author} {\bibfnamefont {G.}~\bibnamefont {Puppo}},
  \bibinfo {author} {\bibfnamefont {M.}~\bibnamefont {Semplice}}, \ and\
  \bibinfo {author} {\bibfnamefont {G.}~\bibnamefont {Visconti}},\ }\href@noop
  {} {\bibfield  {journal} {\bibinfo  {journal} {Mathematics of Computation}\
  }\textbf {\bibinfo {volume} {87}},\ \bibinfo {pages} {1689} (\bibinfo {year}
  {2018})}\BibitemShut {NoStop}%
\bibitem [{\citenamefont {Balsara}\ \emph {et~al.}(2016)\citenamefont
  {Balsara}, \citenamefont {Garain},\ and\ \citenamefont {Shu}}]{AOWENO1}%
  \BibitemOpen
  \bibfield  {author} {\bibinfo {author} {\bibfnamefont {D.}~\bibnamefont
  {Balsara}}, \bibinfo {author} {\bibfnamefont {S.}~\bibnamefont {Garain}}, \
  and\ \bibinfo {author} {\bibfnamefont {C.}~\bibnamefont {Shu}},\ }\href@noop
  {} {\bibfield  {journal} {\bibinfo  {journal} {Journal of Computational
  Physics}\ }\textbf {\bibinfo {volume} {326}},\ \bibinfo {pages} {780}
  (\bibinfo {year} {2016})}\BibitemShut {NoStop}%
\bibitem [{\citenamefont {Balsara}\ \emph {et~al.}(2020)\citenamefont
  {Balsara}, \citenamefont {Garain}, \citenamefont {Florinski},\ and\
  \citenamefont {Boscheri}}]{AOWENO2}%
  \BibitemOpen
  \bibfield  {author} {\bibinfo {author} {\bibfnamefont {D.}~\bibnamefont
  {Balsara}}, \bibinfo {author} {\bibfnamefont {S.}~\bibnamefont {Garain}},
  \bibinfo {author} {\bibfnamefont {V.}~\bibnamefont {Florinski}}, \ and\
  \bibinfo {author} {\bibfnamefont {W.}~\bibnamefont {Boscheri}},\ }\href@noop
  {} {\bibfield  {journal} {\bibinfo  {journal} {Journal of Computational
  Physics}\ }\textbf {\bibinfo {volume} {404}},\ \bibinfo {pages} {109062}
  (\bibinfo {year} {2020})}\BibitemShut {NoStop}%
\bibitem [{\citenamefont {Balsara}\ \emph {et~al.}(2023)\citenamefont
  {Balsara}, \citenamefont {Bhoria}, \citenamefont {Shu},\ and\ \citenamefont
  {Kumar}}]{AOWENO3}%
  \BibitemOpen
  \bibfield  {author} {\bibinfo {author} {\bibfnamefont {D.}~\bibnamefont
  {Balsara}}, \bibinfo {author} {\bibfnamefont {D.}~\bibnamefont {Bhoria}},
  \bibinfo {author} {\bibfnamefont {C.}~\bibnamefont {Shu}}, \ and\ \bibinfo
  {author} {\bibfnamefont {H.}~\bibnamefont {Kumar}},\ }\href@noop {}
  {\bibfield  {journal} {\bibinfo  {journal} {Communications on Applied
  Mathematics and Computation}\ } (\bibinfo {year} {2023})},\ \bibinfo {note}
  {dOI: 10.1007/s42967-023-00275-9}\BibitemShut {NoStop}%
\bibitem [{\citenamefont {{Gourgoulhon}}(2012)}]{Gourgoulhon2012}%
  \BibitemOpen
  \bibfield  {author} {\bibinfo {author} {\bibfnamefont {E.}~\bibnamefont
  {{Gourgoulhon}}},\ }\href {\doibase 10.1007/978-3-642-24525-1} {\emph
  {\bibinfo {title} {3+1 Formalism in General Relativity}}},\ Vol.\ \bibinfo
  {volume} {846}\ (\bibinfo {year} {2012})\BibitemShut {NoStop}%
\bibitem [{\citenamefont {Rezzolla}\ and\ \citenamefont
  {Zanotti}(2013)}]{Rezzolla_book:2013}%
  \BibitemOpen
  \bibfield  {author} {\bibinfo {author} {\bibfnamefont {L.}~\bibnamefont
  {Rezzolla}}\ and\ \bibinfo {author} {\bibfnamefont {O.}~\bibnamefont
  {Zanotti}},\ }\href@noop {} {\emph {\bibinfo {title} {{Relativistic
  hydrodynamics}}}}\ (\bibinfo  {publisher} {Oxford University Press},\
  \bibinfo {year} {2013})\BibitemShut {NoStop}%
\bibitem [{\citenamefont {{Takami}}\ \emph {et~al.}(2014)\citenamefont
  {{Takami}}, \citenamefont {{Rezzolla}},\ and\ \citenamefont
  {{Baiotti}}}]{Takami2014}%
  \BibitemOpen
  \bibfield  {author} {\bibinfo {author} {\bibfnamefont {K.}~\bibnamefont
  {{Takami}}}, \bibinfo {author} {\bibfnamefont {L.}~\bibnamefont
  {{Rezzolla}}}, \ and\ \bibinfo {author} {\bibfnamefont {L.}~\bibnamefont
  {{Baiotti}}},\ }\href {\doibase 10.1103/PhysRevLett.113.091104} {\bibfield
  {journal} {\bibinfo  {journal} {\prl}\ }\textbf {\bibinfo {volume} {113}},\
  \bibinfo {eid} {091104} (\bibinfo {year} {2014})}\BibitemShut {NoStop}%
\bibitem [{\citenamefont {{Dexheimer}}\ \emph {et~al.}(2019)\citenamefont
  {{Dexheimer}}, \citenamefont {{Constantinou}}, \citenamefont {{Most}},
  \citenamefont {{Papenfort}}, \citenamefont {{Hanauske}}, \citenamefont
  {{Schramm}}, \citenamefont {{Stoecker}},\ and\ \citenamefont
  {{Rezzolla}}}]{Dexheimer2019}%
  \BibitemOpen
  \bibfield  {author} {\bibinfo {author} {\bibfnamefont {V.}~\bibnamefont
  {{Dexheimer}}}, \bibinfo {author} {\bibfnamefont {C.}~\bibnamefont
  {{Constantinou}}}, \bibinfo {author} {\bibfnamefont {E.~R.}\ \bibnamefont
  {{Most}}}, \bibinfo {author} {\bibfnamefont {L.~J.}\ \bibnamefont
  {{Papenfort}}}, \bibinfo {author} {\bibfnamefont {M.}~\bibnamefont
  {{Hanauske}}}, \bibinfo {author} {\bibfnamefont {S.}~\bibnamefont
  {{Schramm}}}, \bibinfo {author} {\bibfnamefont {H.}~\bibnamefont
  {{Stoecker}}}, \ and\ \bibinfo {author} {\bibfnamefont {L.}~\bibnamefont
  {{Rezzolla}}},\ }\href {\doibase 10.3390/universe5050129} {\bibfield
  {journal} {\bibinfo  {journal} {Universe}\ }\textbf {\bibinfo {volume} {5}},\
  \bibinfo {eid} {129} (\bibinfo {year} {2019})}\BibitemShut {NoStop}%
\bibitem [{\citenamefont {{Weih}}\ \emph {et~al.}(2019)\citenamefont {{Weih}},
  \citenamefont {{Most}},\ and\ \citenamefont {{Rezzolla}}}]{Weigh2019}%
  \BibitemOpen
  \bibfield  {author} {\bibinfo {author} {\bibfnamefont {L.~R.}\ \bibnamefont
  {{Weih}}}, \bibinfo {author} {\bibfnamefont {E.~R.}\ \bibnamefont {{Most}}},
  \ and\ \bibinfo {author} {\bibfnamefont {L.}~\bibnamefont {{Rezzolla}}},\
  }\href {\doibase 10.3847/1538-4357/ab2edd} {\bibfield  {journal} {\bibinfo
  {journal} {\apj}\ }\textbf {\bibinfo {volume} {881}},\ \bibinfo {eid} {73}
  (\bibinfo {year} {2019})}\BibitemShut {NoStop}%
\bibitem [{\citenamefont {Dumbser}\ \emph {et~al.}(2017)\citenamefont
  {Dumbser}, \citenamefont {Boscheri}, \citenamefont {Semplice},\ and\
  \citenamefont {Russo}}]{dumbser2017central}%
  \BibitemOpen
  \bibfield  {author} {\bibinfo {author} {\bibfnamefont {M.}~\bibnamefont
  {Dumbser}}, \bibinfo {author} {\bibfnamefont {W.}~\bibnamefont {Boscheri}},
  \bibinfo {author} {\bibfnamefont {M.}~\bibnamefont {Semplice}}, \ and\
  \bibinfo {author} {\bibfnamefont {G.}~\bibnamefont {Russo}},\ }\href@noop {}
  {\bibfield  {journal} {\bibinfo  {journal} {SIAM Journal on Scientific
  Computing}\ }\textbf {\bibinfo {volume} {39}},\ \bibinfo {pages} {A2564}
  (\bibinfo {year} {2017})}\BibitemShut {NoStop}%
\bibitem [{\citenamefont {Castro}\ \emph {et~al.}(2006)\citenamefont {Castro},
  \citenamefont {Gallardo},\ and\ \citenamefont {Par{\'e}s}}]{Castro2006}%
  \BibitemOpen
  \bibfield  {author} {\bibinfo {author} {\bibfnamefont {M.}~\bibnamefont
  {Castro}}, \bibinfo {author} {\bibfnamefont {J.}~\bibnamefont {Gallardo}}, \
  and\ \bibinfo {author} {\bibfnamefont {C.}~\bibnamefont {Par{\'e}s}},\
  }\href@noop {} {\bibfield  {journal} {\bibinfo  {journal} {Mathematics of
  computation}\ }\textbf {\bibinfo {volume} {75}},\ \bibinfo {pages} {1103}
  (\bibinfo {year} {2006})}\BibitemShut {NoStop}%
\bibitem [{\citenamefont {Par{\'e}s}(2006)}]{Pares2006}%
  \BibitemOpen
  \bibfield  {author} {\bibinfo {author} {\bibfnamefont {C.}~\bibnamefont
  {Par{\'e}s}},\ }\href@noop {} {\bibfield  {journal} {\bibinfo  {journal}
  {SIAM Journal on Numerical Analysis}\ }\textbf {\bibinfo {volume} {44}},\
  \bibinfo {pages} {300} (\bibinfo {year} {2006})}\BibitemShut {NoStop}%
\bibitem [{\citenamefont {Castro}\ \emph {et~al.}(2008)\citenamefont {Castro},
  \citenamefont {Gallardo}, \citenamefont {L\'opez},\ and\ \citenamefont
  {Par\'es}}]{Castro2008}%
  \BibitemOpen
  \bibfield  {author} {\bibinfo {author} {\bibfnamefont {M.}~\bibnamefont
  {Castro}}, \bibinfo {author} {\bibfnamefont {J.}~\bibnamefont {Gallardo}},
  \bibinfo {author} {\bibfnamefont {J.}~\bibnamefont {L\'opez}}, \ and\
  \bibinfo {author} {\bibfnamefont {C.}~\bibnamefont {Par\'es}},\ }\href@noop
  {} {\bibfield  {journal} {\bibinfo  {journal} {SIAM Journal on Numerical
  Analysis}\ }\textbf {\bibinfo {volume} {46}},\ \bibinfo {pages} {1012}
  (\bibinfo {year} {2008})}\BibitemShut {NoStop}%
\bibitem [{\citenamefont {Castro}\ \emph {et~al.}(2010)\citenamefont {Castro},
  \citenamefont {Pardo}, \citenamefont {Par\'es},\ and\ \citenamefont
  {Toro}}]{CastroPardoPares}%
  \BibitemOpen
  \bibfield  {author} {\bibinfo {author} {\bibfnamefont {M.}~\bibnamefont
  {Castro}}, \bibinfo {author} {\bibfnamefont {A.}~\bibnamefont {Pardo}},
  \bibinfo {author} {\bibfnamefont {C.}~\bibnamefont {Par\'es}}, \ and\
  \bibinfo {author} {\bibfnamefont {E.}~\bibnamefont {Toro}},\ }\href@noop {}
  {\bibfield  {journal} {\bibinfo  {journal} {Math. Comput.}\ }\textbf
  {\bibinfo {volume} {79}},\ \bibinfo {pages} {1427} (\bibinfo {year}
  {2010})}\BibitemShut {NoStop}%
\bibitem [{\citenamefont {Gottlieb}\ and\ \citenamefont {Shu}(1998)}]{shu2}%
  \BibitemOpen
  \bibfield  {author} {\bibinfo {author} {\bibfnamefont {S.}~\bibnamefont
  {Gottlieb}}\ and\ \bibinfo {author} {\bibfnamefont {C.}~\bibnamefont {Shu}},\
  }\href@noop {} {\bibfield  {journal} {\bibinfo  {journal} {Mathematics of
  Computation}\ }\textbf {\bibinfo {volume} {67}},\ \bibinfo {pages} {73}
  (\bibinfo {year} {1998})}\BibitemShut {NoStop}%
\bibitem [{\citenamefont {{Hemberger}}\ \emph {et~al.}(2013)\citenamefont
  {{Hemberger}}, \citenamefont {{Scheel}}, \citenamefont {{Kidder}},
  \citenamefont {{Szil{\'a}gyi}}, \citenamefont {{Lovelace}}, \citenamefont
  {{Taylor}},\ and\ \citenamefont {{Teukolsky}}}]{Hemberger2013}%
  \BibitemOpen
  \bibfield  {author} {\bibinfo {author} {\bibfnamefont {D.~A.}\ \bibnamefont
  {{Hemberger}}}, \bibinfo {author} {\bibfnamefont {M.~A.}\ \bibnamefont
  {{Scheel}}}, \bibinfo {author} {\bibfnamefont {L.~E.}\ \bibnamefont
  {{Kidder}}}, \bibinfo {author} {\bibfnamefont {B.}~\bibnamefont
  {{Szil{\'a}gyi}}}, \bibinfo {author} {\bibfnamefont {G.}~\bibnamefont
  {{Lovelace}}}, \bibinfo {author} {\bibfnamefont {N.~W.}\ \bibnamefont
  {{Taylor}}}, \ and\ \bibinfo {author} {\bibfnamefont {S.~A.}\ \bibnamefont
  {{Teukolsky}}},\ }\href {\doibase 10.1088/0264-9381/30/11/115001} {\bibfield
  {journal} {\bibinfo  {journal} {Classical and Quantum Gravity}\ }\textbf
  {\bibinfo {volume} {30}},\ \bibinfo {eid} {115001} (\bibinfo {year}
  {2013})}\BibitemShut {NoStop}%
\bibitem [{\citenamefont {Zhang}\ \emph {et~al.}(2025)\citenamefont {Zhang},
  \citenamefont {Li}, \citenamefont {Weinzierl},\ and\ \citenamefont
  {Barrera-Hinojosa}}]{Zhang2025}%
  \BibitemOpen
  \bibfield  {author} {\bibinfo {author} {\bibfnamefont {H.}~\bibnamefont
  {Zhang}}, \bibinfo {author} {\bibfnamefont {B.}~\bibnamefont {Li}}, \bibinfo
  {author} {\bibfnamefont {T.}~\bibnamefont {Weinzierl}}, \ and\ \bibinfo
  {author} {\bibfnamefont {C.}~\bibnamefont {Barrera-Hinojosa}},\ }\href@noop
  {} {\bibfield  {journal} {\bibinfo  {journal} {Computer Physics
  Communications}\ }\textbf {\bibinfo {volume} {307}} (\bibinfo {year}
  {2025})}\BibitemShut {NoStop}%
\bibitem [{\citenamefont {Pareschi}\ and\ \citenamefont
  {Rey}(2017)}]{PareschiRey}%
  \BibitemOpen
  \bibfield  {author} {\bibinfo {author} {\bibfnamefont {L.}~\bibnamefont
  {Pareschi}}\ and\ \bibinfo {author} {\bibfnamefont {T.}~\bibnamefont {Rey}},\
  }\href@noop {} {\bibfield  {journal} {\bibinfo  {journal} {Computers \&
  Fluids}\ }\textbf {\bibinfo {volume} {156}},\ \bibinfo {pages} {329}
  (\bibinfo {year} {2017})}\BibitemShut {NoStop}%
\bibitem [{\citenamefont {Berberich}\ \emph {et~al.}(2021)\citenamefont
  {Berberich}, \citenamefont {Chandrashekar},\ and\ \citenamefont
  {Klingenberg}}]{berberich2021high}%
  \BibitemOpen
  \bibfield  {author} {\bibinfo {author} {\bibfnamefont {J.~P.}\ \bibnamefont
  {Berberich}}, \bibinfo {author} {\bibfnamefont {P.}~\bibnamefont
  {Chandrashekar}}, \ and\ \bibinfo {author} {\bibfnamefont {C.}~\bibnamefont
  {Klingenberg}},\ }\href@noop {} {\bibfield  {journal} {\bibinfo  {journal}
  {Computers \& Fluids}\ ,\ \bibinfo {pages} {104858}} (\bibinfo {year}
  {2021})}\BibitemShut {NoStop}%
\bibitem [{\citenamefont {Bermudez}\ and\ \citenamefont
  {V\'azquez-Cend\'on}(1994)}]{Bermudez1994}%
  \BibitemOpen
  \bibfield  {author} {\bibinfo {author} {\bibfnamefont {A.}~\bibnamefont
  {Bermudez}}\ and\ \bibinfo {author} {\bibfnamefont {M.}~\bibnamefont
  {V\'azquez-Cend\'on}},\ }\href@noop {} {\bibfield  {journal} {\bibinfo
  {journal} {Computers \& Fluids}\ }\textbf {\bibinfo {volume} {23}},\ \bibinfo
  {pages} {1049} (\bibinfo {year} {1994})}\BibitemShut {NoStop}%
\bibitem [{\citenamefont {LeVeque}(1998)}]{leveque1998balancing}%
  \BibitemOpen
  \bibfield  {author} {\bibinfo {author} {\bibfnamefont {R.}~\bibnamefont
  {LeVeque}},\ }\href@noop {} {\bibfield  {journal} {\bibinfo  {journal}
  {Journal of Computational Physics}\ }\textbf {\bibinfo {volume} {146}},\
  \bibinfo {pages} {346} (\bibinfo {year} {1998})}\BibitemShut {NoStop}%
\bibitem [{\citenamefont {Gosse}(2001)}]{gosse2001well}%
  \BibitemOpen
  \bibfield  {author} {\bibinfo {author} {\bibfnamefont {L.}~\bibnamefont
  {Gosse}},\ }\href@noop {} {\bibfield  {journal} {\bibinfo  {journal}
  {Mathematical Models and Methods in Applied Sciences}\ }\textbf {\bibinfo
  {volume} {11}},\ \bibinfo {pages} {339} (\bibinfo {year} {2001})}\BibitemShut
  {NoStop}%
\bibitem [{\citenamefont {Audusse}\ \emph {et~al.}(2004)\citenamefont
  {Audusse}, \citenamefont {Bouchut}, \citenamefont {Bristeau}, \citenamefont
  {Klein},\ and\ \citenamefont {Perthame}}]{audusse2004fast}%
  \BibitemOpen
  \bibfield  {author} {\bibinfo {author} {\bibfnamefont {E.}~\bibnamefont
  {Audusse}}, \bibinfo {author} {\bibfnamefont {F.}~\bibnamefont {Bouchut}},
  \bibinfo {author} {\bibfnamefont {M.}~\bibnamefont {Bristeau}}, \bibinfo
  {author} {\bibfnamefont {R.}~\bibnamefont {Klein}}, \ and\ \bibinfo {author}
  {\bibfnamefont {B.}~\bibnamefont {Perthame}},\ }\href@noop {} {\bibfield
  {journal} {\bibinfo  {journal} {SIAM Journal on Scientific Computing}\
  }\textbf {\bibinfo {volume} {25}},\ \bibinfo {pages} {2050} (\bibinfo {year}
  {2004})}\BibitemShut {NoStop}%
\bibitem [{\citenamefont {Botta}\ \emph {et~al.}(2004)\citenamefont {Botta},
  \citenamefont {Klein}, \citenamefont {Langenberg},\ and\ \citenamefont
  {L{\"u}tzenkirchen}}]{BottaKlein}%
  \BibitemOpen
  \bibfield  {author} {\bibinfo {author} {\bibfnamefont {N.}~\bibnamefont
  {Botta}}, \bibinfo {author} {\bibfnamefont {R.}~\bibnamefont {Klein}},
  \bibinfo {author} {\bibfnamefont {S.}~\bibnamefont {Langenberg}}, \ and\
  \bibinfo {author} {\bibfnamefont {S.}~\bibnamefont {L{\"u}tzenkirchen}},\
  }\href@noop {} {\bibfield  {journal} {\bibinfo  {journal} {Journal of
  Computational Physics}\ }\textbf {\bibinfo {volume} {196}},\ \bibinfo {pages}
  {539} (\bibinfo {year} {2004})}\BibitemShut {NoStop}%
\bibitem [{\citenamefont {{Alcubierre}}\ \emph {et~al.}(2004)\citenamefont
  {{Alcubierre}}, \citenamefont {{Allen}},\ and\ \citenamefont {{Bona et
  al.}}}]{Alcubierre2004}%
  \BibitemOpen
  \bibfield  {author} {\bibinfo {author} {\bibfnamefont {M.}~\bibnamefont
  {{Alcubierre}}}, \bibinfo {author} {\bibfnamefont {G.}~\bibnamefont
  {{Allen}}}, \ and\ \bibinfo {author} {\bibfnamefont {C.}~\bibnamefont {{Bona
  et al.}}},\ }\href {\doibase 10.1088/0264-9381/21/2/019} {\bibfield
  {journal} {\bibinfo  {journal} {Classical and Quantum Gravity}\ }\textbf
  {\bibinfo {volume} {21}},\ \bibinfo {pages} {589} (\bibinfo {year}
  {2004})}\BibitemShut {NoStop}%
\bibitem [{\citenamefont {Babiuc}\ \emph {et~al.}(2008)\citenamefont {Babiuc},
  \citenamefont {Husa}, \citenamefont {Alic}, \citenamefont {Hinder},
  \citenamefont {Lechner}, \citenamefont {Schnetter}, \citenamefont
  {Szilágyi}, \citenamefont {Zlochower}, \citenamefont {Dorband},
  \citenamefont {Pollney},\ and\ \citenamefont {Winicour}}]{Babiuc_2008}%
  \BibitemOpen
  \bibfield  {author} {\bibinfo {author} {\bibfnamefont {M.~C.}\ \bibnamefont
  {Babiuc}}, \bibinfo {author} {\bibfnamefont {S.}~\bibnamefont {Husa}},
  \bibinfo {author} {\bibfnamefont {D.}~\bibnamefont {Alic}}, \bibinfo {author}
  {\bibfnamefont {I.}~\bibnamefont {Hinder}}, \bibinfo {author} {\bibfnamefont
  {C.}~\bibnamefont {Lechner}}, \bibinfo {author} {\bibfnamefont
  {E.}~\bibnamefont {Schnetter}}, \bibinfo {author} {\bibfnamefont
  {B.}~\bibnamefont {Szilágyi}}, \bibinfo {author} {\bibfnamefont
  {Y.}~\bibnamefont {Zlochower}}, \bibinfo {author} {\bibfnamefont
  {N.}~\bibnamefont {Dorband}}, \bibinfo {author} {\bibfnamefont
  {D.}~\bibnamefont {Pollney}}, \ and\ \bibinfo {author} {\bibfnamefont
  {J.}~\bibnamefont {Winicour}},\ }\href {\doibase
  10.1088/0264-9381/25/12/125012} {\bibfield  {journal} {\bibinfo  {journal}
  {Classical and Quantum Gravity}\ }\textbf {\bibinfo {volume} {25}},\ \bibinfo
  {pages} {125012} (\bibinfo {year} {2008})}\BibitemShut {NoStop}%
\bibitem [{\citenamefont {{Bona}}\ \emph {et~al.}(2004)\citenamefont {{Bona}},
  \citenamefont {{Ledvinka}}, \citenamefont {{Palenzuela}},\ and\ \citenamefont
  {{{\v{Z}}{\'a}{\v{c}}ek}}}]{Bona:2003qn}%
  \BibitemOpen
  \bibfield  {author} {\bibinfo {author} {\bibfnamefont {C.}~\bibnamefont
  {{Bona}}}, \bibinfo {author} {\bibfnamefont {T.}~\bibnamefont {{Ledvinka}}},
  \bibinfo {author} {\bibfnamefont {C.}~\bibnamefont {{Palenzuela}}}, \ and\
  \bibinfo {author} {\bibfnamefont {M.}~\bibnamefont
  {{{\v{Z}}{\'a}{\v{c}}ek}}},\ }\href {\doibase 10.1103/PhysRevD.69.064036}
  {\bibfield  {journal} {\bibinfo  {journal} {\prd}\ }\textbf {\bibinfo
  {volume} {69}},\ \bibinfo {eid} {064036} (\bibinfo {year}
  {2004})}\BibitemShut {NoStop}%
\bibitem [{\citenamefont {Alic}\ \emph {et~al.}(2012)\citenamefont {Alic},
  \citenamefont {Bona-Casas}, \citenamefont {Bona}, \citenamefont {Rezzolla},\
  and\ \citenamefont {Palenzuela}}]{Alic:2011a}%
  \BibitemOpen
  \bibfield  {author} {\bibinfo {author} {\bibfnamefont {D.}~\bibnamefont
  {Alic}}, \bibinfo {author} {\bibfnamefont {C.}~\bibnamefont {Bona-Casas}},
  \bibinfo {author} {\bibfnamefont {C.}~\bibnamefont {Bona}}, \bibinfo {author}
  {\bibfnamefont {L.}~\bibnamefont {Rezzolla}}, \ and\ \bibinfo {author}
  {\bibfnamefont {C.}~\bibnamefont {Palenzuela}},\ }\href@noop {} {\bibfield
  {journal} {\bibinfo  {journal} {Physical Review D}\ }\textbf {\bibinfo
  {volume} {85}},\ \bibinfo {pages} {064040} (\bibinfo {year}
  {2012})}\BibitemShut {NoStop}%
\bibitem [{\citenamefont {{Tichy}}(2009)}]{Tichy2009}%
  \BibitemOpen
  \bibfield  {author} {\bibinfo {author} {\bibfnamefont {W.}~\bibnamefont
  {{Tichy}}},\ }\href {\doibase 10.1103/PhysRevD.80.104034} {\bibfield
  {journal} {\bibinfo  {journal} {Phys. Rev. D}\ }\textbf {\bibinfo {volume}
  {80}},\ \bibinfo {eid} {104034} (\bibinfo {year} {2009})}\BibitemShut
  {NoStop}%
\bibitem [{\citenamefont {Babiuc}\ \emph {et~al.}(2006)\citenamefont {Babiuc},
  \citenamefont {Szil\'agyi},\ and\ \citenamefont {Winicour}}]{Babiuc2006}%
  \BibitemOpen
  \bibfield  {author} {\bibinfo {author} {\bibfnamefont {M.~C.}\ \bibnamefont
  {Babiuc}}, \bibinfo {author} {\bibfnamefont {B.}~\bibnamefont {Szil\'agyi}},
  \ and\ \bibinfo {author} {\bibfnamefont {J.}~\bibnamefont {Winicour}},\
  }\href {\doibase 10.1103/PhysRevD.73.064017} {\bibfield  {journal} {\bibinfo
  {journal} {Phys. Rev. D}\ }\textbf {\bibinfo {volume} {73}},\ \bibinfo
  {pages} {064017} (\bibinfo {year} {2006})}\BibitemShut {NoStop}%
\bibitem [{\citenamefont {{Cowling}}(1941)}]{Cowling41}%
  \BibitemOpen
  \bibfield  {author} {\bibinfo {author} {\bibfnamefont {T.~G.}\ \bibnamefont
  {{Cowling}}},\ }\href@noop {} {\bibfield  {journal} {\bibinfo  {journal}
  {Mon. Not. R. Astron. Soc.}\ }\textbf {\bibinfo {volume} {101}},\ \bibinfo
  {pages} {367} (\bibinfo {year} {1941})}\BibitemShut {NoStop}%
\bibitem [{\citenamefont {{Mignone}}\ and\ \citenamefont
  {{Bodo}}(2005)}]{Mignone2005}%
  \BibitemOpen
  \bibfield  {author} {\bibinfo {author} {\bibfnamefont {A.}~\bibnamefont
  {{Mignone}}}\ and\ \bibinfo {author} {\bibfnamefont {G.}~\bibnamefont
  {{Bodo}}},\ }\href {\doibase 10.1111/j.1365-2966.2005.09546.x} {\bibfield
  {journal} {\bibinfo  {journal} {Mon. Not. R. Astron. Soc.}\ }\textbf
  {\bibinfo {volume} {364}},\ \bibinfo {pages} {126} (\bibinfo {year}
  {2005})}\BibitemShut {NoStop}%
\bibitem [{\citenamefont {Radice}\ and\ \citenamefont
  {Rezzolla}(2012)}]{Radice2012a}%
  \BibitemOpen
  \bibfield  {author} {\bibinfo {author} {\bibfnamefont {D.}~\bibnamefont
  {Radice}}\ and\ \bibinfo {author} {\bibfnamefont {L.}~\bibnamefont
  {Rezzolla}},\ }\href@noop {} {\bibfield  {journal} {\bibinfo  {journal}
  {Astronomy \& Astrophysics}\ }\textbf {\bibinfo {volume} {547}},\ \bibinfo
  {pages} {A26} (\bibinfo {year} {2012})}\BibitemShut {NoStop}%
\bibitem [{\citenamefont {Zhang}\ and\ \citenamefont
  {MacFadyen}(2006)}]{Zhang2006}%
  \BibitemOpen
  \bibfield  {author} {\bibinfo {author} {\bibfnamefont {W.}~\bibnamefont
  {Zhang}}\ and\ \bibinfo {author} {\bibfnamefont {A.}~\bibnamefont
  {MacFadyen}},\ }\href {http://iopscience.iop.org/0067-0049/164/1/255}
  {\bibfield  {journal} {\bibinfo  {journal} {The Astrophysical Journal
  Supplement Series}\ }\textbf {\bibinfo {volume} {164}},\ \bibinfo {pages}
  {255} (\bibinfo {year} {2006})}\BibitemShut {NoStop}%
\bibitem [{\citenamefont {Ant{\'o}n}\ \emph {et~al.}(2006)\citenamefont
  {Ant{\'o}n}, \citenamefont {Zanotti}, \citenamefont {Miralles}, \citenamefont
  {Mart{\'\i}}, \citenamefont {Ib{\'a}{\~n}ez}, \citenamefont {Font},\ and\
  \citenamefont {Pons}}]{Anton06}%
  \BibitemOpen
  \bibfield  {author} {\bibinfo {author} {\bibfnamefont {L.}~\bibnamefont
  {Ant{\'o}n}}, \bibinfo {author} {\bibfnamefont {O.}~\bibnamefont {Zanotti}},
  \bibinfo {author} {\bibfnamefont {J.~A.}\ \bibnamefont {Miralles}}, \bibinfo
  {author} {\bibfnamefont {J.~M.}\ \bibnamefont {Mart{\'\i}}}, \bibinfo
  {author} {\bibfnamefont {J.~M.}\ \bibnamefont {Ib{\'a}{\~n}ez}}, \bibinfo
  {author} {\bibfnamefont {J.~A.}\ \bibnamefont {Font}}, \ and\ \bibinfo
  {author} {\bibfnamefont {J.~A.}\ \bibnamefont {Pons}},\ }\href@noop {}
  {\bibfield  {journal} {\bibinfo  {journal} {The Astrophysical Journal}\
  }\textbf {\bibinfo {volume} {637}},\ \bibinfo {pages} {296} (\bibinfo {year}
  {2006})}\BibitemShut {NoStop}%
\bibitem [{\citenamefont {Del~Zanna}\ \emph {et~al.}(2007)\citenamefont
  {Del~Zanna}, \citenamefont {Zanotti}, \citenamefont {Bucciantini},\ and\
  \citenamefont {Londrillo}}]{DelZanna2007}%
  \BibitemOpen
  \bibfield  {author} {\bibinfo {author} {\bibfnamefont {L.}~\bibnamefont
  {Del~Zanna}}, \bibinfo {author} {\bibfnamefont {O.}~\bibnamefont {Zanotti}},
  \bibinfo {author} {\bibfnamefont {N.}~\bibnamefont {Bucciantini}}, \ and\
  \bibinfo {author} {\bibfnamefont {P.}~\bibnamefont {Londrillo}},\ }\href@noop
  {} {\bibfield  {journal} {\bibinfo  {journal} {Astronomy \& Astrophysics}\
  }\textbf {\bibinfo {volume} {473}},\ \bibinfo {pages} {11} (\bibinfo {year}
  {2007})}\BibitemShut {NoStop}%
\bibitem [{\citenamefont {Tolman}(1939)}]{Tolman}%
  \BibitemOpen
  \bibfield  {author} {\bibinfo {author} {\bibfnamefont {R.~C.}\ \bibnamefont
  {Tolman}},\ }\href@noop {} {\bibfield  {journal} {\bibinfo  {journal}
  {Physical Review}\ }\textbf {\bibinfo {volume} {55}},\ \bibinfo {pages} {364}
  (\bibinfo {year} {1939})}\BibitemShut {NoStop}%
\bibitem [{\citenamefont {Oppenheimer}\ and\ \citenamefont
  {Volkoff}(1939)}]{Oppenheimer39b}%
  \BibitemOpen
  \bibfield  {author} {\bibinfo {author} {\bibfnamefont {J.~R.}\ \bibnamefont
  {Oppenheimer}}\ and\ \bibinfo {author} {\bibfnamefont {G.~M.}\ \bibnamefont
  {Volkoff}},\ }\href@noop {} {\bibfield  {journal} {\bibinfo  {journal}
  {Physical Review}\ }\textbf {\bibinfo {volume} {55}},\ \bibinfo {pages} {374}
  (\bibinfo {year} {1939})}\BibitemShut {NoStop}%
\bibitem [{\citenamefont {{Camenzind}}(2007)}]{Camenzind2007}%
  \BibitemOpen
  \bibfield  {author} {\bibinfo {author} {\bibfnamefont {M.}~\bibnamefont
  {{Camenzind}}},\ }\href {\doibase 10.1007/978-3-540-49912-1} {\emph {\bibinfo
  {title} {{Compact objects in astrophysics : white dwarfs, neutron stars, and
  black holes}}}}\ (\bibinfo {year} {2007})\BibitemShut {NoStop}%
\bibitem [{\citenamefont {{Font}}\ \emph {et~al.}(2002)\citenamefont {{Font}},
  \citenamefont {{Goodale}}, \citenamefont {{Iyer}}, \citenamefont {{Miller}},
  \citenamefont {{Rezzolla}}, \citenamefont {{Seidel}}, \citenamefont
  {{Stergioulas}}, \citenamefont {{Suen}},\ and\ \citenamefont
  {{Tobias}}}]{Font2002}%
  \BibitemOpen
  \bibfield  {author} {\bibinfo {author} {\bibfnamefont {J.~A.}\ \bibnamefont
  {{Font}}}, \bibinfo {author} {\bibfnamefont {T.}~\bibnamefont {{Goodale}}},
  \bibinfo {author} {\bibfnamefont {S.}~\bibnamefont {{Iyer}}}, \bibinfo
  {author} {\bibfnamefont {M.}~\bibnamefont {{Miller}}}, \bibinfo {author}
  {\bibfnamefont {L.}~\bibnamefont {{Rezzolla}}}, \bibinfo {author}
  {\bibfnamefont {E.}~\bibnamefont {{Seidel}}}, \bibinfo {author}
  {\bibfnamefont {N.}~\bibnamefont {{Stergioulas}}}, \bibinfo {author}
  {\bibfnamefont {W.-M.}\ \bibnamefont {{Suen}}}, \ and\ \bibinfo {author}
  {\bibfnamefont {M.}~\bibnamefont {{Tobias}}},\ }\href {\doibase
  10.1103/PhysRevD.65.084024} {\bibfield  {journal} {\bibinfo  {journal}
  {Physical Review D}\ }\textbf {\bibinfo {volume} {65}},\ \bibinfo {eid}
  {084024} (\bibinfo {year} {2002})}\BibitemShut {NoStop}%
\bibitem [{\citenamefont {Cipolletta}\ \emph {et~al.}(2020)\citenamefont
  {Cipolletta}, \citenamefont {Kalinani}, \citenamefont {Giacomazzo},\ and\
  \citenamefont {Ciolfi}}]{Cipolletta_2020}%
  \BibitemOpen
  \bibfield  {author} {\bibinfo {author} {\bibfnamefont {F.}~\bibnamefont
  {Cipolletta}}, \bibinfo {author} {\bibfnamefont {J.~V.}\ \bibnamefont
  {Kalinani}}, \bibinfo {author} {\bibfnamefont {B.}~\bibnamefont
  {Giacomazzo}}, \ and\ \bibinfo {author} {\bibfnamefont {R.}~\bibnamefont
  {Ciolfi}},\ }\href {\doibase 10.1088/1361-6382/ab8be8} {\bibfield  {journal}
  {\bibinfo  {journal} {Classical and Quantum Gravity}\ }\textbf {\bibinfo
  {volume} {37}},\ \bibinfo {pages} {135010} (\bibinfo {year}
  {2020})}\BibitemShut {NoStop}%
\bibitem [{\citenamefont {{Ng}}\ \emph {et~al.}(2024)\citenamefont {{Ng}},
  \citenamefont {{Jiang}}, \citenamefont {{Musolino}}, \citenamefont {{Ecker}},
  \citenamefont {{Tootle}},\ and\ \citenamefont {{Rezzolla}}}]{Ng2024}%
  \BibitemOpen
  \bibfield  {author} {\bibinfo {author} {\bibfnamefont {H.~H.-Y.}\
  \bibnamefont {{Ng}}}, \bibinfo {author} {\bibfnamefont {J.-L.}\ \bibnamefont
  {{Jiang}}}, \bibinfo {author} {\bibfnamefont {C.}~\bibnamefont {{Musolino}}},
  \bibinfo {author} {\bibfnamefont {C.}~\bibnamefont {{Ecker}}}, \bibinfo
  {author} {\bibfnamefont {S.~D.}\ \bibnamefont {{Tootle}}}, \ and\ \bibinfo
  {author} {\bibfnamefont {L.}~\bibnamefont {{Rezzolla}}},\ }\href {\doibase
  10.1103/PhysRevD.109.064061} {\bibfield  {journal} {\bibinfo  {journal}
  {Phys. Rev. D}\ }\textbf {\bibinfo {volume} {109}},\ \bibinfo {eid} {064061}
  (\bibinfo {year} {2024})},\ \Eprint {http://arxiv.org/abs/2312.11358}
  {arXiv:2312.11358 [gr-qc]} \BibitemShut {NoStop}%
\bibitem [{\citenamefont {Brandt}\ and\ \citenamefont
  {Br\"ugmann}(1997)}]{Brandt1997}%
  \BibitemOpen
  \bibfield  {author} {\bibinfo {author} {\bibfnamefont {S.}~\bibnamefont
  {Brandt}}\ and\ \bibinfo {author} {\bibfnamefont {B.}~\bibnamefont
  {Br\"ugmann}},\ }\href {\doibase 10.1103/PhysRevLett.78.3606} {\bibfield
  {journal} {\bibinfo  {journal} {Phys. Rev. Lett.}\ }\textbf {\bibinfo
  {volume} {78}},\ \bibinfo {pages} {3606} (\bibinfo {year}
  {1997})}\BibitemShut {NoStop}%
\bibitem [{\citenamefont {Ansorg}\ \emph {et~al.}(2004)\citenamefont {Ansorg},
  \citenamefont {Br{\"u}gmann},\ and\ \citenamefont {Tichy}}]{Ansorg:2004ds}%
  \BibitemOpen
  \bibfield  {author} {\bibinfo {author} {\bibfnamefont {M.}~\bibnamefont
  {Ansorg}}, \bibinfo {author} {\bibfnamefont {B.}~\bibnamefont
  {Br{\"u}gmann}}, \ and\ \bibinfo {author} {\bibfnamefont {W.}~\bibnamefont
  {Tichy}},\ }\href@noop {} {\bibfield  {journal} {\bibinfo  {journal} {Phys.
  Rev. D}\ }\textbf {\bibinfo {volume} {70}},\ \bibinfo {pages} {064011}
  (\bibinfo {year} {2004})}\BibitemShut {NoStop}%
\bibitem [{\citenamefont {Reisswig}\ \emph {et~al.}(2009)\citenamefont
  {Reisswig}, \citenamefont {Bishop}, \citenamefont {Pollney},\ and\
  \citenamefont {Szil\'agyi}}]{Reisswig2009}%
  \BibitemOpen
  \bibfield  {author} {\bibinfo {author} {\bibfnamefont {C.}~\bibnamefont
  {Reisswig}}, \bibinfo {author} {\bibfnamefont {N.~T.}\ \bibnamefont
  {Bishop}}, \bibinfo {author} {\bibfnamefont {D.}~\bibnamefont {Pollney}}, \
  and\ \bibinfo {author} {\bibfnamefont {B.}~\bibnamefont {Szil\'agyi}},\
  }\href {\doibase 10.1103/PhysRevLett.103.221101} {\bibfield  {journal}
  {\bibinfo  {journal} {Phys. Rev. Lett.}\ }\textbf {\bibinfo {volume} {103}},\
  \bibinfo {pages} {221101} (\bibinfo {year} {2009})}\BibitemShut {NoStop}%
\bibitem [{\citenamefont {Hannam}\ \emph {et~al.}(2010)\citenamefont {Hannam},
  \citenamefont {Husa}, \citenamefont {Ohme}, \citenamefont {M\"uller},\ and\
  \citenamefont {Br\"ugmann}}]{Hannam2010}%
  \BibitemOpen
  \bibfield  {author} {\bibinfo {author} {\bibfnamefont {M.}~\bibnamefont
  {Hannam}}, \bibinfo {author} {\bibfnamefont {S.}~\bibnamefont {Husa}},
  \bibinfo {author} {\bibfnamefont {F.}~\bibnamefont {Ohme}}, \bibinfo {author}
  {\bibfnamefont {D.}~\bibnamefont {M\"uller}}, \ and\ \bibinfo {author}
  {\bibfnamefont {B.}~\bibnamefont {Br\"ugmann}},\ }\href {\doibase
  10.1103/PhysRevD.82.124008} {\bibfield  {journal} {\bibinfo  {journal} {Phys.
  Rev. D}\ }\textbf {\bibinfo {volume} {82}},\ \bibinfo {pages} {124008}
  (\bibinfo {year} {2010})}\BibitemShut {NoStop}%
\bibitem [{\citenamefont {Tichy}\ and\ \citenamefont
  {Br\"ugmann}(2004)}]{Tichy2004}%
  \BibitemOpen
  \bibfield  {author} {\bibinfo {author} {\bibfnamefont {W.}~\bibnamefont
  {Tichy}}\ and\ \bibinfo {author} {\bibfnamefont {B.}~\bibnamefont
  {Br\"ugmann}},\ }\href {\doibase 10.1103/PhysRevD.69.024006} {\bibfield
  {journal} {\bibinfo  {journal} {Phys. Rev. D}\ }\textbf {\bibinfo {volume}
  {69}},\ \bibinfo {pages} {024006} (\bibinfo {year} {2004})}\BibitemShut
  {NoStop}%
\end{thebibliography}%

\end{document}